\documentclass[%
 reprint,
superscriptaddress,
 aps,
 pra,
]{revtex4-2}

\usepackage{amsthm,amsmath,amssymb,graphicx,float,subfigure}
\usepackage{graphicx}
\usepackage{dcolumn}
\usepackage{bm}
\usepackage{amsthm,amsmath,amssymb}
\usepackage{mathrsfs}
\usepackage{amsfonts}

\begin{document}

\preprint{APS/123-QED}

\title{Dynamics simulation and numerical analysis of arbitrary time-dependent $\mathcal{PT}$-symmetric system based on density operators\\}
\author{Xiaogang Li}
 \affiliation{State Key Laboratory of Low-Dimensional Quantum Physics and Department of Physics, Tsinghua University, Beijing 100084, China}
\author{Chao Zheng}
 \affiliation{Department of Physics, College of Science, North China University of Technology, Beijing 100144, China}
\author{Jiancun Gao}
 \affiliation{State Key Laboratory of Low-Dimensional Quantum Physics and Department of Physics, Tsinghua University, Beijing 100084, China} \affiliation{Frontier Science Center for Quantum Information, Beijing 100084, China}
\author{Guilu Long}
\email{gllong@mail.tsinghua.edu.cn}
 \affiliation{State Key Laboratory of Low-Dimensional Quantum Physics and Department of Physics, Tsinghua University, Beijing 100084, China}
 \affiliation{Frontier Science Center for Quantum Information, Beijing 100084, China}
 \affiliation{Beijing Academy of Quantum Information Sciences, Beijing 100193, China}
 \affiliation{Beijing National Research Center for Information Science and Technology and School of Information, Tsinghua University,
 Beijing 100084, China}
\date{\today}

\begin{abstract}
$\mathcal{PT}$-symmetric system has attracted extensive attention in recent years because of its unique properties and applications. How to simulate $\mathcal{PT}$-symmetric system in traditional quantum mechanical system has not only fundamental theoretical significance but also practical value. We propose a dynamics simulation scheme of arbitrary time-dependent $\mathcal{PT}$-symmetric system based on density operators, and the results are compatible with previous methods based on pure-state vectors.
Based on the above, we are able to study the influence of quantum noises on the simulation results with the technique of vectorization of density operators and matrixization of superoperators (VDMS), and we show the depolarizing (Dep) noise is the most fatal and should be avoided as much as possible. Meanwhile, we also give a numerical analysis. We find that the problem of chronological product usually has to be solved not only in the numerical calculation, but also even in the experiment, because the dilated higher-dimensional Hamiltonian is usually time-dependent. Through theoretical analysis and numerical calculation, we find that on the premise of meeting the goal of calculation accuracy and saving computing resources, the time step of calculation and the cut-off term of Magnus series have to be carefully balanced.  
\end{abstract}

\maketitle


\section{\label{Introduction}Introduction}

That all physical observables, including Hamiltonians, must be Hermitian operators has long been seen as one of the axioms in \emph{conventional quantum mechanics} (CQM) \cite{Griffiths2018}, because Hermitian operators have real eigenspectrums as we all know. However, Bender et al. found that some non-Hermitian Hamiltonians, which are parity-time ($\mathcal{PT}$)-reversal symmetric, may also have real eigenspectrums in 1998 \cite{Bender1998}, and then established the \emph{$\mathcal{PT}$-symmetric quantum mechanics} ($\mathcal{PT}$-QM) \cite{Bender1999,Bender2002,Bender2004}. In last two decades, $\mathcal{PT}$-symmetry theory has developed rapidly \cite{Mostafazadeh2002b,Mostafazadeh2002a,Mostafazadeh2002,Mostafazadeh2003a,Mostafazadeh2007a,Curtright2007,Brody2013,Zhang2020}, and aroused wide attention \cite{Bender2007,Mostafazadeh2010}. The phenomenon related to  $\mathcal{PT}$-symmetry exists widely, not only in classical systems, such as optical systems\cite{Makris2008,Rueter2010}, microcavities \cite{Peng2014} and circuits \cite{Assawaworrarit2017}, but also in quantum systems, such as strongly correlated many-body systems \cite{Ashida2017}, quantum critical spin chains \cite{Couvreur2017}, and ultracold atoms \cite{li2019observation}. In addition, $\mathcal{PT}$-symmetric system also shows practical value in quantum sensors \cite{Zhang2019a,Chu2020,Yu2020}, which can use the sensitivity of $\mathcal{PT}$-symmetric system near the exceptional points (EPs) to amplify small signals. In particular, it is worth noting that, with the increasing interest in $\mathcal{PT}$-symmetric systems, some new phenomena have emerged \cite{Croke2015,Kawabata2017}, which are impressive because they seem to conflict with theory of conventional quantum mechanics or theory of relativity, such as the instantaneous quantum brachistochrone problem \cite{Mostafazadeh2007,Bender2007a,Guenther2008,Guenther2008a,Mostafazadeh2010,Ramezani2012,ZhengChao2013,Beygi2018,Brody2021}, the discrimination of nonorthogonal quantum states \cite{Bender2013,Wang2020}, and the violation of no-signaling principle \cite{Barnett2002,Lee2014,Brody2016,Tang2016,feng2017non,Beygi2018,Bagchi2020}. However, some anomalies actually come from not knowing how to simulate $\mathcal{PT}$-symmetric system in conventional quantum system, for instance, if the discarded probability during the simulation of $\mathcal{PT}$-symmetric system is considered, the no-signaling principle will still hold \cite{Huang2018}. Therefore, finding a way to simulate $\mathcal{PT}$-symmetric system in conventional quantum system has not only practical value, but also theoretical significance.

At present, there are at least three technical routes to simulate $\mathcal{PT}$-symmetric system, and all of them can be realized in experiments \cite{ZhengChao2013,Yu2020,wu2019observation}. The method of linear combination of unitaries (LCU) \cite{GuiLu2006}, can be used to simulate various non-Hermitian $\mathcal{PT}$-symmetric systems in discrete time, whether they are in unbroken or broken phase \cite{GuiLu2006,ZhengChao2013,Zheng2018,Gao2021}, and even anti-$\mathcal{PT}$-symmetric systems \cite{Zheng2019a}. The method of weak measurement\cite{Huang2019,Yu2020}, can also be used to simulate various time-independent unbroken and broken $\mathcal{PT}$-symmetric systems in restricted continuous time under the condition of weak interaction by a weak measurement. The methods based on embedding \cite{Huang2018,Li2022,wu2019observation}, can be used to simulate the dynamics of unbroken time-independent $\mathcal{PT}$-symmetric system (in pure-states case \cite{Huang2018} or mixed-states case \cite{Li2022}) in unrestricted continuous time with only one qubit as an auxiliary system.

What deserves special attention is that in 2019, Wu et al. proposed a general simulation scheme of dynamics of time-dependent (TD) arbitrary $\mathcal{PT}$-symmetric system based on pure-state vectors through the dilation method, and realized it with a single nitrogen-vacancy center in diamond \cite{wu2019observation}. Specifically, their method is performed by dilating a general TD $\mathcal{PT}$–symmetric Hamiltonian into a higher dimensional TD Hermitian one with the help of an auxiliary qubit system, and evolving the state in the dilated Hermitian system for a period of time, then performing a fixed projection measurement on the auxiliary system, after that, the remained main system is equivalent to going through the evolution process govern by the $\mathcal{PT}$-symmetric Hamiltonian. However, the dilated Hamiltonian is usually time-dependent, which means the system is actually an open quantum system. From one point of view, as we all know, in an open quantum system, the influence of quantum noises in the environment is usually inevitable so that they have to be considered \cite{nielsen2002quantum}. The time evolution of an open quantum system interacting with memoryless environment can be described by the  Lindblad master equation \cite{nielsen2002quantum,Minganti2019}, which is usually based on the density operators (matrices) rather than the state vectors. From another point of view, a pure state will evolve to a mixed state under the quantum noises, while the theory based on pure-state vectors can not conveniently deal with the question related to mixed states, then the tool of pure-state vectors may be failed, in this situation, the density operators will also be a better tool.

In this paper, we first generalize the outstanding work of Wu et al. based on dilation method from the pure-state vectors case to the mixed-state density operators case with the tool of density operators \cite{wu2019observation}, and provide more mathematical and physical completeness. It is worth emphasizing that it is not a trivial process for generalizing the quantum state from pure-state vectors to mixed-state density operators in the simulation of $\mathcal{PT}$-symmetric system, and the difficulty mainly comes from the flexibility of characterizing quantum states in $\mathcal{PT}$-symmetric system and the uncertainty of mapping them to the high-dimensional quantum states in conventional quantum system \cite{Li2022,Ohlsson2021}. Only based on this groundwork can we deal with problems in open quantum systems, and we deal with these problems using the vectorization of density operators and matrixization of superoperators (VDMS) technique. Then we study the influence of quantum noises to the dynamics of time-dependent arbitrary $\mathcal{PT}$-symmetric system, meanwhile, we also give a numerical analysis. Through theoretical analysis and numerical calculation, we find that on the premise of meeting the goal of calculation accuracy and saving computing resources, the time step of calculation and the cut-off term of Magnus series have to be carefully balanced. In the numerical analysis, we also find that the time step $h$ of each numerical calculation step shall be limited to its corresponding critical time $T_c$ of the convergence of Magnus series, especially when the high-order terms of Magnus series are considered. This phenomenon occurs because the dilated higher-dimensional Hamiltonian is usually time-dependent, then the problem of chronological product usually has to be dealt with, and the Magnus series may have to be calculated \cite{Magnus1954,Blanes1998,Blanes2009}, which may diverge when $t\rightarrow T_c$ so that the error may be amplified after the Magnus series is truncated to a high-order term in calculation. Meanwhile, the implemented duration of experimental running is actually bounded by the critical time $T_l$ of the legitimacy of dilation method. This phenomenon occurs because when $t\rightarrow T_l$, the energy may diverge. In fact, the problem of chronological product may have to be solved not only in the numerical calculation, but also even in the experiment, because the dilated $\hat{H}_{AS}(t)$ has to be parameterized in advance by numerically calculating the chronological product caused by $H_S(t)$ needed to be dilated. In addition, when considering the influence of quantum noises, we find that the depolarizing (Dep) noise (channel) is the most fatal to the simulation of $\mathcal{PT}$-symmetric system among three kinds of quantum noises we considered and should be avoided as much as possible. It is worth noting that when the system considered is time-independent and $\mathcal{PT}$-symmetry unbroken, the results of dynamics simulation in this work are consistent with our previous results in Ref.\cite{Li2022}, and when the state considered is the pure state, the results of this work are consistent with the theoretical results given in Ref.\cite{wu2019observation}. In summary, this work provides a general theoretical framework based on density operators to analytically and numerically analyze the dynamics of time-dependent arbitrary $\mathcal{PT}$-symmetric system and the influence of quantum noises.

The rest of this paper is organized as follows. In Sec.\ref{Theoretical_preparations}, we give some necessary basic theories of $\mathcal{PT}$-symmetric system. In Sec.\ref{dilation_method}, we give a universal Hermitian dilation method of non-Hermitian Hamiltonians with density operators, and based on that, we give a universal simulation scheme of the dynamics of arbitrary TD $\mathcal{PT}$-symmetric system. To be able to solve problems in open quantum system, we vectorize density operators and matrixize the Liouvillian superoperators in Sec.\ref{VD_and_ML}. In addition, we specially discuss the numerical calculation methods of time-dependent linear matrix differential equations involved in this paper in Sec.\ref{numerical_calculation}. In Sec.\ref{example}, we give an example of two-dimensional $\mathcal{PT}$-symmetric system, and numerically analyze its dynamics, meanwhile, we also consider the influence of three kinds of quantum noises. In Sec.\ref{conclusions}, we give conclusions and discussions. Further more, we also make the Appendix \ref{appendix_derivation-H_24} to show the details of the dilated Hamiltonians, and the Appendix \ref{chronological product} to introduce the problem of chronological product at the end of this paper.

\section{Theoretical preparations\label{Theoretical_preparations}}
Given a $n$-dimensional non-Hermitian $\mathcal{PT}$-symmetric Hamiltonian $\mathcal{H}$, the parity operator $\mathcal{P}$
and time-reversal operator $\mathcal{T}$, where $\mathcal{T}$ is an anti-linear operator, and $H$, $P$, $T$ denote their matrix representation, respectively. They have the following properties:
\begin{align}
P^2=I, T\overline{T}=&I, PT=T\overline{P}, \nonumber \\
PT\overline{H}=&HPT,
\end{align}
where $\overline{[\cdot]}$ denotes complex conjugate of $[\cdot]$, and it occurs because $\mathcal{T}$ is an anti-linear operator. If $H$ is similar to a real diagonal matrix, $H$ will be $\mathcal{PT}$-symmetric unbroken, otherwise, $H$ is called $PT$-symmetry broken
if and only if it satisfies either of these two conditions \cite{Mostafazadeh2002a,Mostafazadeh2002b,huang2021solvable,Li2022}: (1) it cannot be diagonalized, (2) it has complex eigenvalues that appears in complex conjugate pairs.

For a time-independent $\mathcal{PT}$-symmetry $H$, there is a time-independent operator $\eta$ that satisfies:
\begin{equation}\label{metric1}
\eta H=H^\dag\eta,
\end{equation}
where $\eta$ is called the metric operator of $H$, and it is a reversible operator (usually Hermitian), when $H$ is $\mathcal{PT}$-symmetry unbroken, it can be completely positive. The metric operator is usually not unique, for instance, if $\eta$ is a metric operator of $H$, so is $r\eta$ ($r\in \mathbb{R}$). The above Eq.\eqref{metric1} is also referred to as the pseudo-Hermiticity relation, and $H$ is also referred to as pseudo-Hermitian (Hamiltonian) \cite{Mostafazadeh2002b,Zhang2019}. It is worth mentioning that in 2002, Mostafazadeh pointed out that all the $\mathcal{PT}$-symmetric non-Hermitian Hamiltonians belong to the class of pseudo-Hermitian Hamiltonians \cite{Mostafazadeh2002b}, recently, in 2020, this conclusion was more strictly proved and strengthened by Ruili Zhang et al., i.e., $\mathcal{PT}$-symmetry entails pseudo-Hermiticity regardless of diagonalizability \cite{Zhang2020}. The theory of pseudo-Hermiticity provides more convenience in dealing with questions related to mixed states, so we use the tools usually used in the framework of pseudo-Hermiticity hereafter, such as biorthonormal eigenbasis, metric operators , etc.

The elements of $\mathcal{PT}$-QM in the unbroken phase of $H$ can be represented by biorthogonal basis of $H$: \{$|\chi_n,a\rangle, |\phi_n,a\rangle$\}, and they have the properties as follows \cite{Mostafazadeh2002b}:
\begin{subequations}\label{biorthononal_basis1}
\begin{align}
\langle\chi_m,a|\phi_n,b\rangle=&\delta_{mn}\delta_{ab} \\
H|\phi_n,a\rangle=E_n|\phi_n,a\rangle,\quad &H^\dag|\chi_n,a\rangle=E_n|\chi_n,a\rangle  \\
\sum\limits_n\sum\limits_{a=1}^{d_n}|\chi_n,a\rangle\langle\phi_n,a|=&\sum\limits_n\sum\limits_{a=1}^{d_n}|\phi_n,a\rangle\langle\chi_n,a|=I,\\
  |\chi_n,a\rangle=&\eta|\phi_n,a\rangle,\\
\eta=\sum_n\sum_{a=1}^{d_n}&|\chi_n,a\rangle\langle\chi_n,a|,\\
  \eta^{-1}=\sum_n\sum_{a=1}^{d_n}&|\phi_n,a\rangle\langle\phi_n,a|,
\end{align}
\end{subequations}
where $d_n$ is the degree of degeneracy of the eigenvalue $E_n$ (in the $\mathcal{PT}$-unbroken case, $E_n$ is real), and $a$ and $b$ are degeneracy labels, and $|\phi_k\rangle$s ($|\chi_k\rangle$s) are usually not orthogonal to each other. The possible real coefficients before $\eta$ have been absorbed into the biorthogonal basis. It is worth noting that in the extreme case, when $H$ becomes Hermitian, then the biorthogonal basis will become orthogonal basis because of $H=H^\dag$ and then $\{|\phi_k\rangle\}=\{|\chi_k\rangle\}$. For convenience, we set $d_n=1$ hereafter. If we recorded that $\Phi=[|\phi_1\rangle,...|\phi_i\rangle,...|\phi_n\rangle]$, $\Xi=[|\chi_1\rangle,...|\chi_i\rangle,...|\chi_n\rangle]$, $E=$diag$(E_1,...E_i,...E_n)$, then according to Eqs.\eqref{biorthononal_basis1} we will get \cite{Huang2018,Li2022}:
\begin{equation}\label{jordan_block_H1}
 \begin{split}
  &\Phi^{-1}H\Phi=E, \quad \Xi^{-1}H^\dag\Xi=E \\
  &\eta=\Xi\Xi^\dag, \quad \eta^{-1}=\Phi\Phi^\dag \\
  &\Xi=\eta\Phi, \quad \Xi^\dag\Phi=I_n.
\end{split}
\end{equation}

Through the positive Hermitian metric operator $\eta$, the representations of a quantum observable $\mathcal{O}$ under the framework of CQM and the framework of $\mathcal{PT}$-QM can be connected by a similarity transformation, i.e., the Dyson map \cite{Fring2016,Luiz2020}:
 \begin{align}\label{Dyson_map1}
  O_c=\eta^{\frac{1}{2}}\cdot O_{\mathcal{PT}}\cdot \eta^{-\frac{1}{2}},
 \end{align}
where $O_c$ is the observable in $\mathcal{PT}$-QM framework, $O_{\mathcal{PT}}$ is the corresponding observable in CQM. Similarly, there is a relation between the quantum state $\rho_c$ in CQM and the state $\rho_{\mathcal{PT}}$ in $\mathcal{PT}$-QM:
\begin{equation}\label{rhoc_rhopt}
  \rho_c=\sum_{mn}{\rho_c}_{mn}|m\rangle\langle n| \Leftrightarrow \rho_{\mathcal{PT}}=\sum_{mn}{\rho_c}_{mn}|\phi_m\rangle\langle \chi_n|,
\end{equation}
where $\{|n\rangle\}$ is a mutual orthogonal basis in CQM; ${\rho_c}_{mn}$s are the matrix elements; $\rho_c$ and $\rho_{\mathcal{PT}}$ are connected by the Dyson map given above in Eq.\eqref{Dyson_map1}. Therefore, there is a fixed relation between the quantum state (unnormalized) $\rho_S$ in the framework of CQM and the quantum state $\rho_{\mathcal{PT}}$ in the framework of $\mathcal{PT}$-QM (for simplicity, we only use this conclusion without more derivation, more details can be found in our previous work in Ref.\cite{Li2022}):
\begin{align}\label{rho_simulation1}
  \rho_S&=\rho_{\mathcal{PT}}\cdot\eta^{-1}  \nonumber \\
        &=\sum_{mn}{\rho_c}_{mn}|\phi_m\rangle\langle \phi_n|,
\end{align}
where $\rho_S$ can be normalized by $\rm{Tr}(\eta\rho_S)=1$. Clarifying the relation between the density operators in $\mathcal{PT}$-QM and the density operators in CQM is an important step to extend the dynamics simulation scheme of $\mathcal{PT}$-symmetric system from the pure-state vectors case to the mixed-state density operators case \cite{Li2022}. From the above, we know that this process is not trivial.

Now we introduce the concept of $\eta$-inner product \cite{Mostafazadeh2002b,Li2022}:
\begin{align}\label{eta-inner-product1}
  (|\psi_{1}\rangle,|\psi_{2}\rangle)_{\eta}\equiv\langle\psi_{1} \mid \psi_{2}\rangle_{\eta}:=\left\langle\psi_{1}|\eta| \psi_{2}\right\rangle& \nonumber\\
  \forall\left|\psi_{1}\right\rangle,\left|\psi_{2}\right\rangle \in L(\mathcal{H}),&
\end{align}
where $L(\mathcal{H})$ denotes Hilbert space, and $\eta$ is a reversible Hermitian metric operator of this $\eta$-inner space, especially in the unbroken phase of $\mathcal{PT}$-QM, it can be a positive operator \cite{Mostafazadeh2002b}.

Next we discuss the situation that non-Hermitian $\mathcal{PT}$-symmetric system $H(t)$ is time-dependent. Considering two evolving states $|\psi_1(t)\rangle$ and $|\psi_2(t)\rangle$, we assume that they satisfy the Schr\"{o}dinger-like equation:
\begin{align}
\frac{\mathrm{d}|\psi(t)\rangle}{\mathrm{d}t}=-iH(t)|\psi(t)\rangle,
\end{align}
where we have set $\hbar=1$ here and after. According to the probability conservation in the inner product space defined like in Eq.\eqref{eta-inner-product1}, we can obtain that
\begin{align}\label{probability_conservation2}
  &\frac{\mathrm{d}}{\mathrm{d}t}\langle\psi_{1}(t)|\psi_{2}(t)\rangle_{\eta(t)} \nonumber\\
\equiv& \frac{\mathrm{d}}{\mathrm{d}t}\langle\psi_{1}(t)|\eta(t)|\psi_{2}(t)\rangle \nonumber\\
=&\langle\psi_{1}(t)|-i\eta(t)H(t)+iH^\dag(t) \eta(t)+\eta'(t)| \psi_{2}(t)\rangle \nonumber\\
=&0,
\end{align}
where we have recorded the differential operator $\frac{\mathrm{d}}{\mathrm{d}t}$ as the symbol "$'$". Then we can get the result:
\begin{align}\label{TD-pseudo-Hermiticity-relation}
\eta'(t)=i[\eta(t)H(t)-H^\dag(t) \eta(t)].
\end{align}
The above Eq.\eqref{TD-pseudo-Hermiticity-relation} is referred to as the time-dependent (TD) pseudo-Hermiticity relation, $\eta(t)$ is the time-dependent (TD) metric operator in its corresponding inner product space, i.e., $\eta(t)$-inner product space, which leads to the probability conservation \cite{Fring2016,Luiz2020}. Nothing that $\eta(t)$-inner product space is time-dependent.

The solution of Eq.\eqref{TD-pseudo-Hermiticity-relation} can be obtained as:
\begin{align}\label{eta_TD}
  \eta(t)=\mathbb{T}e^{-i\int_{0}^{t}{H}^\dag(\tau)d\tau}\eta(0)\mathbb{\overline{T}}e^{i\int_{0}^{t}H(\tau)d\tau},
\end{align}
where $\mathbb{T}$ is time-ordering operator and $\mathbb{\overline{T}}$ is the anti-time-ordering operator, moreover, $\eta(0)$ can be arbitrary Hermitian operator. If we take $\eta(0)>0$, then there must exist a period of time $T_t$ make $\eta(t)>0$ during $t\in[0,T_t)$. It is worth noting that when $H$ is time-independent, $\eta(t)$ may be time-independent so that $\eta'=0$ (for instance, when $H$ is $\mathcal{PT}$-symmetric unbroken, $\eta(0)$ can be taken as the metric operator like the one in Eq.\eqref{metric1}), and then the TD pseudo-Hermiticity relation given above in Eq.\eqref{TD-pseudo-Hermiticity-relation} will be reduced to the pseudo-Hermiticity relation given in Eq.\eqref{metric1}.

In addition, according to Eq.\eqref{probability_conservation2}, if we set $|\psi(t)\rangle_1=|\psi(t)\rangle_2=|\psi(t)\rangle$, and $\rho_S(t)=|\psi(t)\rangle\langle\psi(t)|$, then we know:
\begin{align}
\frac{\mathrm{d}\mathrm{Tr}[\rho_S(t)\eta(t)]}{\mathrm{d}t}\equiv\frac{\mathrm{d}\mathrm{Tr}[\rho_{\mathcal{PT}}(t)]}{\mathrm{d}t}\equiv0.
\end{align}
Here $\rho_{\mathcal{PT}}(t)\equiv\rho_S(t)\eta(t)$ can be seen as a quantum state in TD $\mathcal{PT}$-QM, and can be normalized by $\mathrm{Tr}[\rho_{\mathcal{PT}}(t)]=\mathrm{Tr}[\rho_S(t)\eta(t)]\equiv1$. The form of $\rho_{\mathcal{PT}}(t)$ can be easily generalized, and can be mapped to a quantum state $\rho_c(t)$ in (TD) CQM through a time-dependent (TD) similarity transformation, i.e., the TD Dyson map similar to Eq.\eqref{Dyson_map1} \cite{Luiz2020}:
 \begin{align}\label{Dyson_map1-TD}
  O_c(t)=\eta^{\frac{1}{2}}(t)\cdot O_{\mathcal{PT}}(t)\cdot \eta^{-\frac{1}{2}}(t),
 \end{align}
and $\rho_c(t)$ is similar to Eq.\eqref{rhoc_rhopt}:
\begin{align}\label{rhoc_rhopt-TD}
  \rho_c(t)=&\sum_{mn}{\rho_c}_{mn}(t)|m(t)\rangle\langle n(t)| \Leftrightarrow \nonumber\\ \rho_{\mathcal{PT}}(t)=&\sum_{mn}{\rho_c}_{mn}(t)|\phi_m(t)\rangle\langle \chi_n(t)|, \nonumber\\
                      =&\sum_{mn}{\rho_c}_{mn}(t)|\phi_m(t)\rangle\langle \phi_n(t)|\cdot\eta(t),
\end{align}
where $\rho_c(t)$ is a quantum state in CQM with a TD mutual orthogonal basis $\{|m(t)\rangle\}$ in CQM, and $\{|\phi(t)\rangle,|\chi(t)\rangle\}$ is a TD biorthogonal basis in $\mathcal{PT}$-QM, and $|\chi(t)\rangle=\eta(t)|\phi(t)\rangle$. Therefore, if we take $\eta(0)>0$ given in Eq.\eqref{TD-pseudo-Hermiticity-relation}, similar to Eq.\eqref{rhoc_rhopt}, we can obtain that
\begin{align}\label{rho_simulation1-TD}
  \rho_S(t)&=\rho_{\mathcal{PT}}(t)\cdot\eta^{-1}(t)  \nonumber \\
        &=\sum_{mn}{\rho_c}_{mn}(t)|\phi_m(t)\rangle\langle \phi_n(t)|,
\end{align}
which means on the promise that $\eta(t)>0$, we can always find an unnormalized quantum state $\rho_S(t)$ in TD CQM related to a quantum state $\rho_{\mathcal{PT}}(t)$ in TD $\mathcal{PT}$-QM (refer to Appendix B in Ref.\cite{Li2022} for the proof that $\rho_S$ is actually an unnormalized state in (TD) CQM). According to the relation between the unnormalized state of $\rho_S(t)$ in TD CQM and quantum state $\rho_{\mathcal{PT}}(t)$ in TD $\mathcal{PT}$-QM given above in Eq.\eqref{rho_simulation1-TD}, we know that once $\eta(t)$ is known (given, or calculated), for the purpose of simulation, $\rho_S$ can be used to represent $\rho_{\mathcal{PT}}$. It is worth noting that the two are actually different in the physical sense, because they are not satisfied with similarity transformation. Fortunately, for a simulation task, it is not necessary to pursue the absolute equivalence of the two physical meanings, but only to ensure that their form is appropriate and can be realized physically \cite{Li2022,Guenther2008a,Brody2012}. The Eq.\eqref{rho_simulation1-TD} is actually the prerequisite for the implementation of the dilation method based on density operators we will discuss next.

\section{\label{dilation_method}Universal Hermitian dilation method of non-Hermitian Hamiltonians based on density operators}

One of the methods to simulate the dynamics of $\mathcal{PT}$-symmetric system is to find a dilated higher-dimensional Hermitian system (marked by "$AS$", where "$A$", "$S$", "$AS$" represents the auxiliary system, the main system used to generate the dynamics of the non-Hermitian system, and the composite system, respectively), which obey the von Neumann equation (it reduces to the Schr\"{o}dinger equation when restricted to pure-state vectors), to simulate the dynamics of non-Hermitian system, which obey the von Neumann-like equation (Schr\"{o}dinger-like equation in pure-state vectors case). We assume the evolution equation (the von Neumann-like equation) of unnormalized state $\rho_S$ that has been mentioned in Eq.\eqref{rho_simulation1-TD} is (hereafter, we set $\hbar=1$) \cite{Ohlsson2021,Brody2012,Kawabata2017,Xiao2019}:
\begin{align}\label{H_dyn}
  \frac{\mathrm{d}{\rho_S(t)}}{\mathrm{d} t}&=-i[H_S(t), \rho_S(t)]_{\dag}\nonumber\\
                                        &\equiv-i[H_S(t)\rho_S(t)-\rho_S(t){H_S}^\dag(t)],
\end{align}
where $H_S(t)$ is non-Hermitian Hamiltonian in the system $S$ and can be $\mathcal{PT}$-symmetric Hamiltonian, for generality, we assume it is time-dependent, so the time-independent Hamiltonian can be regarded as a special case of it. It is worth noting that because $H(t)$ is non-Hermitian, the Eq.\eqref{H_dyn} cannot usually be realized directly in physics. $\rho_S(t)$ is the unnormalized state in the system $S$ at time $t$s, its normalized form is
\begin{align}\label{rho_S_N}
\rho_{S_\mathrm{N}}(t)=\frac{\rho_S(t)}{\mathrm{Tr}[\rho_S(t)]},
\end{align}
where $\rho_{S_\mathrm{N}}(t)$ is a legal quantum state (i.e., a positive-semidefinite operator with unit trace) actually measured in the experiment, and $P_0(t)\equiv\mathrm{Tr}[\rho_S(t)]$ represents the corresponding measured probability of the state $\rho_{S_\mathrm{N}}(t)$ at the time moment $t$. Because both $P_0(t)$ and $\rho_{S_\mathrm{N}}(t)$ can be measured in an experiment, $\rho_S(t)$ can be used to represent $\rho_{S_\mathrm{N}}(t)$, the probability $P_0(t)$ is actually
absorbed in the unnormalized density matrix $\rho_S(t)$ (see the Ref.\cite{Li2022} and its Appendix B for details), and this practice meet the common usage\cite{Bender2007a,Guenther2008,Guenther2008a,wu2019observation,Li2022}. Without misunderstanding, we assume that they are equivalent and are strictly distinguished only when calculating probability. It should be noted that since $H_S$ is non-Hermitian, $\mathrm{Tr}[\rho_S(t)]$ is usually not constant, and may be greater than one so that loses its physical meaning. Therefore, we set the constraint condition $\mathrm{Tr}[\rho_S(t)]\leqslant1$, which can be satisfied by choosing the appropriate unnormalized initial state $\rho_S(0)$ and duration time $T$ of the evolution. 

However, because $H_S$ is non-Hermitian, the state evolution satisfying von Neumann-like equation like Eq.\eqref{H_dyn} can not be realized directly in CQM. Fortunately, we can map it into the following von Neumann equation. The von Neumann equation related to the above von Neumann-like equation in the Eq.\eqref{H_dyn} is:
\begin{align}\label{H_hat_dyn}
  \frac{\mathrm{d}\rho_{AS}(t)}{\mathrm{d} t}=-i[\hat{H}_{AS}(t), \rho_{AS}(t)],
\end{align}
where $\hat{H}_{AS}(t)$ is a Hermitian Hamiltonian in the system $AS$, so the Eq.\eqref{H_hat_dyn} can be realized directly in physics compared with Eq.\eqref{H_dyn}, and $\rho_{AS}(t)$ is the density operator in this system.

We assume that
\begin{align}\label{rho_rhoS}
  \rho_{AS}(t)&=\begin{bmatrix}
  I_S \\
\xi(t)
\end{bmatrix}\cdot\rho_S(t)\cdot
\begin{bmatrix}
  I_S &\xi^\dag(t)
\end{bmatrix}\nonumber\\
&=\begin{bmatrix}
\rho_S(t) &\rho_S(t)\cdot\xi^\dag(t)\\
\xi(t)\rho_S(t)&\xi(t)\cdot\rho_S(t)\cdot\xi^\dag(t)
\end{bmatrix},
\end{align}
where $\begin{bmatrix}
  I_S \\
\xi(t)
\end{bmatrix}=|0\rangle_A\otimes I_S+|1\rangle_A\otimes\xi(t)$, and we can see that the state $\rho_{AS}(t)$ is usually an entanglement state.

Meanwhile, we assume
\begin{equation}\label{H_dag}
  \hat{H}_{AS}(t)=
  \begin{bmatrix}
  H_1(t) & H_2(t)\\
  H_2^\dag(t) &H_4(t)
  \end{bmatrix},
\end{equation}
where $H_1(t),H_2(t),H_4(t)$ are all operators, and it is obvious that $H_1^\dag(t)=H_1(t),H_4^\dag(t)=H_4(t)$, i.e., they are Hermitian, because $\hat{H}_{AS}(t)$ is Hermitian.

Then according to the probability conservation principle, we know that
\begin{align}\label{probability conservation}
&\frac{\mathrm{d}{\mathrm{Tr}[\rho_{AS}(t)]}}{\mathrm{d}t}\nonumber\\
&=\frac{\mathrm{d}\mathrm{Tr}[\rho_S(t)+\xi(t)\rho_S(t)\xi^\dag(t)]}{\mathrm{d}t}\nonumber\\
&=\frac{\mathrm{d}\mathrm{Tr}[(\xi^\dag(t)\xi(t)+I_S)\rho_S(t)]}{\mathrm{d}t} \nonumber \\
&\equiv\frac{\mathrm{d}\mathrm{Tr}[M(t)\rho_S(t)]}{\mathrm{d}t} \nonumber \\
&=\mathrm{Tr}[M'(t)\rho_S(t)+M(t)\rho_S'(t)] \nonumber \\
&=\mathrm{Tr}\{M'(t)\rho_S(t)-iM(t)[H_S(t)\rho_S(t)-\rho_S(t){H_S}^\dag(t)]\} \nonumber \\
&=\mathrm{Tr}\{[M'(t)-i(M(t)H_S(t)-{H_S}^\dag(t)M(t))]\rho_S(t)\} \nonumber \\
&\equiv0,
\end{align}
where we have made
\begin{equation}\label{M_xi}
  M(t)=\xi^\dag(t)\xi(t)+I_S.
\end{equation}
It is obvious that $M(t)$ is Hermitian and $M(t)>1$, so M(t) is reversible. For the convenience, we set
\begin{align}\label{unit_trace}
\mathrm{Tr}[\rho_{AS}(t)]=\mathrm{Tr}[M(t)\rho_{S}(t)]\equiv1.
\end{align}

Then according to Eq.\eqref{probability conservation} we get the result
\begin{equation}\label{M'_TD}
  M'(t)=-i[{H_S}^\dag(t)M(t)-M(t)H_S(t)].
\end{equation}
It is worth noting that the above Eq.\eqref{M'_TD} has the same form with Eq.\eqref{TD-pseudo-Hermiticity-relation}, so this relation can also be a TD pseudo-Hermiticity relation, which can replace the (time-independent) pseudo-Hermiticity relation like ${H_S}^\dag M= M H_S$ \cite{Mostafazadeh2007}. In general, $M'(t)\neq0$, so ${H_S}^\dag(t)M(t)\neq M(t)H_S(t)$, which means $M(t)$ is not the metric operator of $H_S(t)$, but only the TD metric operator of inner product space (i.e., the $M(t)$-inner product space), which leads to probability conservation. We find that $M(t)$ is actually the metric operator of TD pseudo-Hermitian Hamiltonian $h_S(t)$ that will be introduced next.

Then multiplying left and right sides of the above Eq.\eqref{M'_TD} by $M^{-1}(t)$, we can get
\begin{equation}\label{K_mid}
  iM^{-1}(t)M'(t)M^{-1}(t)=M^{-1}(t){H_S}^\dag(t)-H_S(t)M^{-1}(t).
\end{equation}
If we introduce an operator:
\begin{equation}\label{K}
  K(t)=H_S(t)M^{-1}(t)+\frac{i}{2}M^{-1}(t)M'(t)M^{-1}(t),
\end{equation}
then according to the above Eq.\eqref{K_mid}, we will find $K^\dag(t)=K(t)$, which means $K(t)$ is Hermitian. The introduction of the operator $K(t)$ will be very beneficial to our following work. We then introduce a new quantity $h_S(t)$:
\begin{align}
h_S(t)&\equiv K(t)M(t) \nonumber\\
   &=H_S(t)+\frac{i}{2}M^{-1}(t)M'(t)\nonumber\\
   &=\frac{1}{2}H_S(t)+\frac{1}{2}M^{-1}(t)H_S^\dag(t)M(t),
\end{align}
where the Eq.\eqref{M'_TD} is used in the derivation. Obviously, $h_S(t)$ is non-Hermitian. At the same time, it is easy to verify that
\begin{align}\label{Mh_s}
M(t)h_S(t)=h_S^\dag(t)M(t),
\end{align}
which is exactly pseudo-Hermitian relation mentioned in Eq.\eqref{metric1}, and that means $h_S(t)$ is actually a pseudo-Hermitian quantity, and $M(t)$ is actually its metric operator. As we have known in Eq.\eqref{M_xi}, $M(t)>I_S$, so according to above Eq.\eqref{Mh_s}, there is a similarity transformation make
\begin{align}
h_{\mathrm{phys}}(t)&\equiv M^{\frac{1}{2}}(t)\cdot h_S(t)\cdot M^{-\frac{1}{2}}(t)\nonumber\\
                    &=M^{-\frac{1}{2}}(t)\cdot M(t)h_S(t)\cdot M^{-\frac{1}{2}}(t) \nonumber\\
                    &=h_{\mathrm{phys}}^\dag(t) \nonumber \\
                    &=\frac{1}{2}[M^{\frac{1}{2}}(t)H_S(t)M^{-\frac{1}{2}}(t)\!+\!M^{-\frac{1}{2}}(t)H_S^\dag(t)M^{\frac{1}{2}}(t)],
\end{align}
which means $h_{\mathrm{phys}}(t)$ is a Hermitian operator with real eigenspectrum and can be a legal observable in physics, so $h_S(t)$ is also a legal observable with real eigenspectrum \cite{Bender2003,Bender2007,Mostafazadeh2010,Fring2016}, and can be a physically permissible Hamiltonian. We reveal the physical meaning of $M(t)$, and find the physical observables $h_S(t)$ and $h_{\mathrm{phys}}(t)$ related to it, so provide more physical completeness \cite{wu2019observation}.

By solving the Eq.\eqref{M'_TD}, we obtain that
\begin{align}\label{M_TD}
  M(t)=\mathbb{T}e^{-i\int_{0}^{t}{H_S}^\dag(\tau)d\tau}M(0)\mathbb{\overline{T}}e^{i\int_{0}^{t}H_S(\tau)d\tau},
\end{align}
where $M(0)$ can be any Hermitian operator satisfying the condition $M(0)>I_S$. The above Eq.\eqref{M_TD} is also obtained in Ref.\cite{wu2019observation}. It is worth noting that if we take $M(0)=\eta(0)>1$ given in Eq.\eqref{eta_TD}, then $M(t)$ will be $\eta(t)$. This means that $M(t)$ is actually an TD metric operator like $\eta(t)$ given in Eq.\eqref{eta_TD}.
In addition, when $H_S$ is given, some eigenvalues of $M(t)$ may be decreases with time in some cases, such as in the $\mathcal{PT}$-symmetry broken phase of $H_S$, so we can defined a critical time $T_l$ of the legitimacy that make $M(t)$ legal, i.e., $M(t)>I_S$ when $t\in[0, T_l)$, while when $t=T_l$, at least one of eigenvalues of $M(T_l)$ become one. It should be noted that $T_l$ may be infinite in some cases, such as in the $\mathcal{PT}$-symmetry unbroken phase of $H_S$. In addition, in the cases that $T_l\neq \infty$, $T_l$ will depend on the initial setting of $M(0)$, in general, $T_l$ will increase with the increase of eigenvalues of $M(0)$, a big enough $T_l$ can always be obtained by scaling $M(0)$ to meet the requirement of experiment or numerical calculation tasks.

It is worth noting that when the system is time-independent and $\mathcal{PT}$-symmetry unbroken, there must be a metric operator $\eta$ that satisfies $\eta>1$, then $M(0)$ can be chosen as $M(0)=\eta$, after that
\begin{align}\label{Mt-eta}
  M(t)&=e^{-i\int_{0}^{t}H^\dag \mathrm{d}\tau}\eta e^{i\int_{0}^{t}H\mathrm{d}\tau} \nonumber \\
      &=\eta\cdot e^{-i\int_{0}^{t}H\mathrm{d}\tau}\cdot e^{i\int_{0}^{t}H \mathrm{d}\tau} \nonumber \\
      &=\eta,
\end{align}
which means $M(t)$ will be the metric operator and is time-independent.

By the above Eq.\eqref{rho_rhoS}, we have established a map between $\rho_S(t)$ and $\rho_{AS}(t)$, now we try to establish the map between $H_S(t)$ and $\hat{H}_{AS}(t)$, specifically, we need to find the solutions of $H_1(t), H_2(t), H_4(t)$ in Eq.\eqref{H_dag}.

 Substituting the Eqs.\eqref{rho_rhoS} and \eqref{H_dyn} into the Eq.\eqref{H_hat_dyn}, we obtain that
 \begin{widetext}
 \begin{align}\label{H_hat_map_H}
  \frac{\mathrm{d}{\rho_{AS}(t)}}{\mathrm{d}t}&=
  \begin{bmatrix}
  0 \\ \xi'(t)
\end{bmatrix}\cdot\rho_S(t)\cdot
\begin{bmatrix}
  I_S &\xi^\dag(t)
\end{bmatrix}+
 \begin{bmatrix}
  I_S \\
   \xi(t)
\end{bmatrix}\cdot\rho'_0\cdot
\begin{bmatrix}
  I_S &\xi^\dag(t)
\end{bmatrix}+
 \begin{bmatrix}
  I_S \\ \xi(t)
\end{bmatrix}\cdot\rho_S\cdot
\begin{bmatrix}
  0 &\xi'^\dag(t)
\end{bmatrix} \nonumber\\
&=-i\left\{\begin{bmatrix}
  H_S(t) \\
  \xi(t)H_S(t)+i\xi'(t)
\end{bmatrix}\cdot\rho_S(t)\cdot
\begin{bmatrix}
  I_S &\xi^\dag(t)
\end{bmatrix}-
\begin{bmatrix}
   I_S \\
   \xi(t)
\end{bmatrix}\cdot\rho_S(t)\cdot
\begin{bmatrix}
  {H_S}^\dag(t) &{H_S}^\dag(t)\xi^\dag(t)-i\xi'^\dag(t)
\end{bmatrix}\right\} \nonumber \\
&=-i[\hat H_S(t)\rho_{AS}(t)-\rho_{AS}(t)\hat H_S(t)] \nonumber\\
&=-i\left\{\begin{bmatrix}
  H_1(t) & H_2(t)\\
  H_2^\dag(t) &H_4(t)
  \end{bmatrix}\cdot
    \begin{bmatrix}
  I_S \\
  \xi(t)
  \end{bmatrix}\cdot \rho_S\cdot
  \begin{bmatrix}
  I_S &\xi^\dag(t)
  \end{bmatrix}-
  \begin{bmatrix}
  I_S \\
  \xi(t)
  \end{bmatrix}\cdot \rho_S\cdot
  \begin{bmatrix}
  I_S &\xi^\dag(t)
  \end{bmatrix}\cdot
  \begin{bmatrix}
  H_1(t) & H_2(t)\\
  H_2^\dag(t) &H_4(t)
  \end{bmatrix}\right\} \nonumber\\
 &=-i\left\{\begin{bmatrix}
 H_1(t)+H_2(t)\xi(t) \\
 H_2^\dag(t)+H_4(t)\xi(t)
 \end{bmatrix} \cdot \rho_S\cdot
  \begin{bmatrix}
  I_S &\xi^\dag(t)
  \end{bmatrix}-
  \begin{bmatrix}
  I_S \\
   \xi(t)
  \end{bmatrix}\cdot\rho_S(t)\cdot
\begin{bmatrix}
 H_1(t)+\xi^\dag(t)H_2^\dag(t) &H_2(t)+\xi^\dag(t)H_4
 \end{bmatrix}\right\}.
\end{align}
\end{widetext}
According to the above Eq.\eqref{H_hat_map_H}, we can get the following relation:
\begin{subequations}\label{relation1}
\begin{align}
 H_1(t)+H_2(t)\xi(t)&=H_S(t)  \label{eq1}\\
 H_2(t)+H_4(t)\xi(t)&=i\xi'(t)+\xi(t)H_S(t). \label{eq2}
\end{align}
\end{subequations}
Observing the above equations, we know that the solutions of $H_1(t), H_2(t), H_4(t)$ are not unique, and $H_1$ can be chosen as the unique variable. Therefore, we can add a gauge in order to obtain unique solutions. However, the value of $H_1$ is artificially assigned in Ref.\cite{wu2019observation}, so some unique properties of the resulting dilated $\hat{H}_{AS}$ may be masked, we will see that in our Sec.\ref{example}, and especially in Fig.\ref{magnus3_eigenvalue} we will see the eigenspectrum of $\hat{H}_{AS}$ actually has symmetric property, and $\hat{H}_{AS}$ obtained in Ref.\cite{wu2019observation} is actually the result of the application of a symmetric gauge. We then provide more mathematical completeness.

By observing the Eq.\eqref{H_dag}, we know that the space of $\hat{H}_{AS}(t)$ is $2n$-dimensional, however, the space of $\rho_S$ defined in the Eq.\eqref{rho_rhoS} is just $n$-dimensional, consequently, similar to the Eq.\eqref{rho_rhoS}, we can assume that
\begin{align}\label{rho_rhoS-2}
  \rho_{AS}^\bot(t)&=
\begin{bmatrix}
  -\xi^\dag(t) \\
  I_S
\end{bmatrix}\cdot\rho_S(t)\cdot
\begin{bmatrix}
  -\xi(t) &I_S
\end{bmatrix}\nonumber\\
&=\begin{bmatrix}
\xi^\dag(t)\cdot\rho_S(t)\cdot\xi(t) &-\xi^\dag(t)\cdot\rho_S(t)\\
-\rho_S(t)\xi(t)&\rho_S(t)
\end{bmatrix}.
\end{align}
It is easy to check that $\mathrm{Tr}[\rho_{AS}(t)\rho^\bot(t)]\equiv0$, which means that the space of $\rho$ and $\rho^\bot$ are mutually orthogonal, then we know that $\rho$ and $\rho^\bot$ are located in two different orthogonal subspaces in the space where $\hat{H}_{AS}(t)$ is located. Therefore, we can adopt the following symmetric gauge (there are also some other valid gauges\cite{Zhang2019,Huang2019,Luiz2020}):
\begin{equation}\label{gauge}
  \frac{\mathrm{d}{\rho_{AS}^\bot(t)}}{\mathrm{d}t}=-i[\hat{H}_{AS}(t), \rho_{AS}^\bot(t)].
\end{equation}
With this gauge, according to the probability conservation principle again, similar to Eq.\eqref{probability conservation}, we can also derive another relation between $M(t)$ and $\xi(t)$ as follows:
\begin{equation}\label{M_xi-2}
  M(t)=\xi(t)\xi^\dag(t)+I_S,
\end{equation}
where $M(t)$ has been given in Eq.\eqref{M_TD}. Comparing the Eq.\eqref{M_xi-2} with the Eq.\eqref{M_xi}, we get
\begin{equation}\label{xi_xi}
  \xi(t)\xi^\dag(t)=\xi^\dag(t)\xi(t),
\end{equation}
which means $\xi(t)$ will be a normal operator under the symmetric gauge in Eq.\eqref{gauge}, therefore, for convenience, we take $\xi(t)$ Hermitian, and then
 \begin{equation}\label{xi}
   \xi(t)=[M(t)-I_S]^\frac{1}{2}.
 \end{equation}

At the same time, by adopting the similar method with Eq.\eqref{H_hat_map_H}, we can obtain the following relation:
\begin{subequations}\label{relation2}
\begin{align}
 -H_1(t)\xi^\dag(t)+H_2(t)&=-i\xi^\dag{'}(t)-\xi^\dag(t)H_S(t)  \label{eq3}\\
 -H_2^\dag(t)\xi^\dag(t)+H_4(t)&=H_S(t). \label{eq4}
\end{align}
\end{subequations}
Then multiplying Eq.\eqref{eq1} right by $\xi(t)$ and substituting it to Eq.\eqref{eq3}, we can obtain
\begin{align}\label{H_2}
  H_2(t)&=[-i\xi'(t)+H_S(t)\xi(t)-\xi(t)H_S(t)]M^{-1}(t) \nonumber \\
        &=K(t)\xi(t)-\xi(t)K(t)-\frac{i}{2}[\xi'(t)M^{-1}(t)+\nonumber\\
        &\quad M^{-1}(t)\xi'(t)].
\end{align}
It is obvious that $H_2(t)$ is anti-Hermitian. The details of the derivation is given in  Appendix \ref{appendix_derivation-H_24}.

Then in a similar way like above, multiplying Eq.\eqref{eq2} right by $\xi(t)$ and substituting it into Eq.\eqref{eq4}, we can obtain
\begin{align}\label{H_4}
  H_4(t)=&[i\xi'(t)\xi(t)+\xi(t)H_S(t)\xi(t)+H_S(t)]M^{-1}(t) \nonumber \\
        =&K(t)\!+\!\xi(t)K(t)\xi(t)\!+\!\frac{i}{2}[\xi'(t)\xi(t)M^{-1}(t)\!-\nonumber\\
         &M^{-1}(t)\xi(t)\xi'(t)],
\end{align}
$H_4(t)$ is Hermitian, obviously. The details of the derivation is also given in  Appendix \ref{appendix_derivation-H_24}.

Next, according to Eq.\eqref{H_2} and Eq.\eqref{eq1}, the result as follows will be obtained:
\begin{align}\label{H_1}
  H_1(t)=&H_S(t)-H_2(t)\xi(t) \nonumber \\
        =&H_S(t)-[-i\xi'(t)+H_S(t)\xi(t)-\xi(t)H_S(t)]\cdot \nonumber\\
         &M^{-1}(t)\cdot\xi(t) \nonumber \\
        =&[i\xi'(t)\xi(t)+\xi(t)H_S(t)\xi(t)+H_S(t)]M^{-1}(t) \nonumber \\
        =&K(t)+\xi(t)K(t)\xi(t)+\frac{i}{2}[\xi'(t)\xi(t)M^{-1}-\nonumber\\
         &M^{-1}\xi(t)\xi'(t)] \nonumber \\
        =&H_4(t).
\end{align}
Consequently, according to Eq.\eqref{H_1} and Eq.\eqref{H_2} we can easily get
\begin{small}
\begin{widetext}
\begin{subequations}\label{H1_add_iH2}
\begin{align}
H_1(t)+iH_2(t)&=[I_S-i\xi(t)]H_S(t)M^{-1}(t)[I_S+i\xi(t)]+\xi'(t)M^{-1}(t) \nonumber \\
              &=[I_S-i\xi(t)]K(t)[I_S+i\xi(t)]+\frac{1}{2}[\xi'(t)M(t)^{-1}[I_S+i\xi(t)]+[I_S-i\xi(t)]M(t)^{-1}\xi'(t)] \nonumber \\
              &=[I_S-i\xi(t)]\left\{K(t)+\frac{1}{2}[I_S-i\xi(t)]^{-1}[\xi'(t)[I_S-i\xi(t)]^{-1}+[I_S+i\xi(t)]^{-1}\xi'(t)][I_S+i\xi(t)]^{-1}\right\}[I_S+i\xi(t)], \\
H_1(t)-iH_2(t)&=[I_S+i\xi(t)]H_S(t)M^{-1}(t)[I_S-i\xi(t)]-\xi'(t)M^{-1}(t) \nonumber \\
              &=[I_S+i\xi(t)]K(t)[I_S-i\xi(t)]-\frac{1}{2}[\xi'(t)M(t)^{-1}[I_S-i\xi(t)]+[I_S+i\xi(t)]M(t)^{-1}\xi'(t)] \nonumber \\
              &=[I_S+i\xi(t)]\left\{K(t)-\frac{1}{2}[I_S+i\xi(t)]^{-1}[\xi'(t)[I_S+i\xi(t)]^{-1}+[I_S-i\xi(t)]^{-1}\xi'(t)][I_S-i\xi(t)]^{-1}\right\}[I_S-i\xi(t)],
\end{align}
\end{subequations}
\end{widetext}
\end{small}
where $M(t)=\xi^2(t)+I_S=[I_S\pm i\xi(t)]\cdot[I_S\mp i\xi(t)]$. Both $H_1(t)+iH_2(t)$ and $H_1(t)-iH_2(t)$ are both Hermitian.

Finally, according to Eq.\eqref{H_dag}, we obtain the final form of $\hat{H}_{AS}(t)$:
\begin{align}\label{H_hat_t}
 \hat{H}_{AS}(t)=&I_S\otimes H_1(t)+i\sigma_y\otimes H_2(t) \nonumber\\
                =&|+_y\rangle\langle+_y|\otimes[H_1(t)+iH_2(t)]\nonumber\\
                +&|-_y\rangle\langle-_y|\otimes[H_1(t)-iH_2(t)],
\end{align}
where $|+_y\rangle=\frac{1}{\sqrt{2}}(|0\rangle_A+i|1\rangle_A), |-_y\rangle=\frac{1}{\sqrt{2}}(|0\rangle_A-i|1\rangle_A)$ are the eigenstates of $\sigma_y$ corresponding to its eigenvalues of $+1, -1$, respectively. The equation above has a similar form to the result of pure-state case given in Ref.\cite{wu2019observation} (see Eqs.(19)-(21) in their Supplementary Materials), while the differences are mainly caused by they actually use the basis $\{|-_y\rangle,-i|+_y\rangle\}$, while we use the basis $\{|0\rangle,|1\rangle\}$. According to Eq.\eqref{H_hat_t}, we can find $\hat{H}(t)$ is highly symmetric under the symmetric gauge of Eq.\eqref{gauge}, and use only one single qubit as its auxiliary system. We have to point out that, whether in the experiments or numerical calculations, as long as we use a time-dependent $\hat{H}_{AS}(t)$ given above in Eq.\eqref{H_hat_t}, it may be inevitable to solve the problem of chronological product. The reason for the numerical calculation situation is obvious, while in an experiment, $\hat{H}_{AS}(t)$ usually has to be parameterized in advance by numerically calculating $M(t)$ given in Eq.\eqref{M_TD}, which needs to numerically computing the chronological product caused by $H_S(t)$, so it usually needs to deal with the problem of chronological product unless any two moments of $\hat{H}_{AS}(t)$ are commute to each other (such as the case of $H_S$ is time-independent and $\mathcal{PT}$-symmetry unbroken, then $M(0)$ can be chosen as its metric operator $\eta$ just like the case in Eq.\eqref{Mt-eta} \cite{wu2019observation}. We will see the impact of the chronological product on simulation accuracy more clearly in Fig.\ref{Magnus_expansion—figs} and Fig.\ref{Magnus2-fig} in our example given in Sec.\ref{example}.

For the convenience of expression, we define the "$\circ$" operation as $C\circ[\cdot]=C[\cdot]C^\dag$. After that, looking back Eqs.\eqref{H_dyn} and \eqref{H_hat_dyn}, their solution can be expressed as:
\begin{subequations}\label{H_hat_H_dyn-solutions}
\begin{align}
\rho_S(t)&=\mathcal{U}_S\circ\rho_S(0)\equiv\mathbb{T}e^{-i\int^t_0 H_S(\tau)\mathrm{d}\tau}\circ\rho_S(0),\label{H_hat_H_dyn-solutions1}\\
\rho_{AS}(t)&=U_{AS}\circ\rho_{AS}(0)\equiv\mathbb{T}e^{-i\int^t_0 \hat{H}_{AS}(\tau)\mathrm{d}\tau}\circ\rho_{AS}(0), \label{H_hat_H_dyn-solutions2}
\end{align}
\end{subequations}
where $\hat{H}_{AS}(t)$ has been obtained in Eq.\eqref{H_hat_t}, and $\mathcal{U}_S$ is the non-unitary evolution operator related to the non-Hermitian Hamiltonian $H_S$, while $U_{AS}$ is the unitary evolution operator related to the Hermitian Hamiltonian $H_{AS}$. We will call the method of obtaining $\rho_{AS}(t)$ through Eq.\eqref{H_hat_H_dyn-solutions2} the dilation method, while call the method of first obtaining $\rho_S(t)$ according to Eq.\eqref{H_hat_H_dyn-solutions1} and then combining it into $\rho_{AS}(t)$ by Eq.\eqref{rho_rhoS} the combination method. The difference between two methods is that the delated Hamiltonian $\hat{H}_{AS}$ containing the term $\xi'(t)$ is avoid to be calculated for latter, which may increase the error in numerical calculation, while only need to calculate $M(t)$ given in Eq.\eqref{M_TD} and $\xi(t)$ given in Eq.\eqref{xi}. The error of the numerical calculation between the dilation method and the combination method can be defined as follows:
\begin{align}\label{norm_Delta}
\Delta_\rho(t)=\|{\rho_{AS}}_\mathrm{dilation}(t)-{\rho_{AS}}_\mathrm{combination}(t)\|_F,
\end{align}
where ${\rho_{AS}}_\mathrm{dilation}(t)$ denotes the result calculated by the dilation method, while ${\rho_{AS}}_\mathrm{combination}(t)$ denotes the result calculated by the combination method, and the symbol "$F$" denotes Frobenius norm.

We define the measurement operator: $\Pi_k=|k\rangle_A\langle k|\otimes I_S, k\in \{0,1\}$, and the map $\mathcal{M}_k$: $\mathcal{M}_k[\rho_{AS}]=\mathrm{Tr}_A[\Pi_k\circ\rho_{AS}]$. Therefore, the Eq.\eqref{H_hat_H_dyn-solutions2} can be mapped to the Eq.\eqref{H_hat_H_dyn-solutions1} by the map $\mathcal{M}_0$, experimentally, by performing a projection measurement $\Pi_0\equiv|0\rangle_A\langle0|$ on the auxiliary qubit:
\begin{align}\label{M_Map}
\mathcal{M}_0[\rho_{AS}(t)]=\rho_S(t)=\mathbb{T}e^{-i\int^t_0 H_S(\tau)\mathrm{d}\tau}\circ\rho_S(0),
\end{align}
which means we can simulate the non-unitary evolution (the dynamics) of non-Hermitian system $H_S$ in the higher-dimensional system $\hat{H}_{AS}$. It is worth noting that the map $\mathcal{M}_0$ is realized by a fixed projection measurement (post-selection) $\Pi_0\equiv|0\rangle_A\langle0|$, which is time-independent, so it will be easy to be realized in experiment \cite{wu2019observation}. However, the success of this process $\mathcal{M}_0$ is probabilistic, and the corresponding success probability $P_0(t)$ is
\begin{align}
P_0(t)=\mathrm{Tr}[\rho_S(t)].
\end{align}
Obviously, in general, $P_0(t)<1$. After the measurement $\Pi_0$, the state in main system $S$ will be ${\rho_S}_\mathrm{N}(t)$ in Eq.\eqref{rho_S_N} with success probability $P_0(t)$ given above. It is worth mentioning that the success probability may be optimized by technical means, such as using the local-operations-and-classical-communication (LOCC) protocol scheme proposed in Ref.\cite{Li2022}, which may be of significance for experiment.

In particular, in the situation that $H_S$ is $\mathcal{PT}$-unbroken and time-independent, according to the result of Eq.\eqref{Mt-eta}, we can take the operator $M(t)\equiv \eta$, where $\eta$ is a positive metric operator and $\eta>1$, and then according to the Eqs.\eqref{K}, \eqref{H_1}, \eqref{H_2} and \eqref{H_4}, $K, H_1, H_2, H_4$ all will be time-independent, specifically, according to Eqs.\eqref{jordan_block_H1}, they will become \cite{Huang2018,Huang2019,Li2022}:
\begin{subequations} \label{H124_TID}
\begin{align}
 M(t)&\equiv\eta\Rightarrow\xi=(\eta-I_S)^\frac{1}{2},\\
   K&\equiv H_S\cdot\eta^{-1}=\Phi\cdot E_S\cdot\Phi^\dag, \\
 H_1&=H_S\eta^{-1}+\xi H_S\eta^{-1}\xi \nonumber\\
    &=\Phi E_S\Phi^\dag+\xi\cdot\Phi E_S\Phi^\dag\cdot\xi   \\
 H_2&=H_S\eta^{-1}\xi-\xi H_S\eta^{-1}, \nonumber   \\
    &=\Phi E_S\Phi^\dag\cdot\xi-\xi\cdot\Phi E_S\Phi^\dag \nonumber \\
    &=-H_2^\dag, \\
 H_4&=H_S\eta^{-1}+\xi H_S\eta^{-1}\xi \nonumber  \\
    &=\xi^{-1}(H_S\eta^{-1}+\eta H_S-H_S-H_S^\dag)\xi^{-1}+H_S\eta^{-1} \nonumber \\
    &=H_1.
\end{align}
\end{subequations}
They are the same as the results we obtained in Ref.\cite{Li2022}. Therefore, according to Eq.\eqref{H1_add_iH2}, $H_1\pm iH_2$ will also become time-independent:
\begin{subequations}
\begin{align}
H_1+iH_2&=(I_S-i\xi)\Phi\circ E_S=\Phi_{+}\circ E_S, \\
H_1-iH_2&=(I_S+i\xi)\Phi\circ E_S=\Phi_{-}\circ E_S,
\end{align}
\end{subequations}
where $\Phi_{\pm}=(I_S\mp i\xi)$.
At the same time, the dilated higher-dimensional system $\hat{H}_{AS}$ will also become time-independent, and according to the result of Eq.\eqref{H_hat_t}:
\begin{align}\label{H_hat_unique_TID}
  \hat H_{AS}&=I_A\otimes H_1+i\sigma_y\otimes H_2 \nonumber\\
        &=\!I_A\!\!\otimes\!(\!H_S\eta^{-1}\!\!+\!\!\xi H_S\eta^{-1}\xi)\!+\!i\sigma_y\!\!\otimes\!(H_S\eta^{-1}\xi\!-\!\xi H_S\eta^{-1}\!)\nonumber \\
        &=V_{AS}\circ [I_A\otimes E_S],
\end{align}
where
\begin{equation*}\label{V}
 \begin{split}
  V_{AS}&=\frac{1}{\sqrt{2}}[(I_A+i\sigma_x)\otimes I_S-i(\sigma_y+\sigma_z)\otimes\xi]\cdot I_A\otimes\Phi \\
   &=\frac{1}{\sqrt{2}}
   \begin{bmatrix}
   \Phi_{+}  &i\Phi_{-} \\
   i\Phi_{+} &\Phi_{-}
   \end{bmatrix} \\
   &=\frac{1}{\sqrt{2}}
   \begin{bmatrix}
   (I_S-i\xi)\Phi  &i(I_S+i\xi)\Phi \\
   i(I_S-i\xi)\Phi &(I_S+i\xi)\Phi
   \end{bmatrix}
 \end{split}
\end{equation*}
is an unitary operator, while $V_{AS}$ is not unique.  This result is also the same as that in Ref.\cite{Li2022}. From the above Eq.\eqref{H_hat_unique_TID}, it can be found that the degeneracy of higher-dimensional dilated system $\hat{H}_{AS}$ is twice that of lower-dimensional $\mathcal{PT}$-symmetric system $H_S$ in this situation.

\section{Vectorization of density operators and matrixization of Liouvillian superoperators in open quantum system\label{VD_and_ML}}
The evolution equation of an open quantum system with a Markovian approximation (i.e., memoryless) can be expressed by Lindblad master equation \cite{Minganti2019}:
 \begin{equation}\label{master equation}
   \frac{\mathrm{d}\rho_{AS}(t)}{\mathrm{d} t}=\mathcal{L}\rho_{AS}(t)=-i[\hat H_{AS}(t),\rho_{AS}(t)]+\sum_{\mu}\mathcal{D}[\Gamma_\mu]\rho_{AS}(t),
 \end{equation}
where $\rho_{AS}(t)$ is the density operator of the system, $\Gamma_\mu$ is the jump operator, $\mathcal{L}$ is the Liouvillian superoperator, and $\mathcal{D}[\Gamma_\mu]$ is the dissipator related to the $\Gamma_\mu$, which is used to describe the dissipation:
\begin{equation}\label{Lindblad}
  \mathcal{D}\left[\Gamma_\mu\right] \rho_{AS}(t)=\Gamma_\mu \rho_{AS}(t) \Gamma_\mu^\dag-\frac{\Gamma_\mu^\dag \Gamma_\mu}{2} \rho_{AS}(t)-\rho_{AS}(t) \frac{\Gamma_\mu^\dag \Gamma_\mu}{2}.
\end{equation}
The operator-sum representation is a convenient tool to describe the open system, various models of decoherence and dissipation in open quantum systems have been widely studied, such as amplitude damping (AD) channel model, phase damping (PD) channel model, and depolarizing (Dep) channel. In some cases of simple decoherence (such as the case of Hamiltonian $H=0$), these models have a concise form in the framework of operator-sum representation, however, when the Hamiltonian gets complicated just like the one in Eq.\eqref{H_hat_t}, the application of the operator-sum representation will be indirect and inconvenient. Therefore, the vectorization of density operators and matrixization of superoperators (VDMS) technique may be a more convenient tool (see the Appendix of Ref.\cite{Minganti2019} for details).

We adopt VDMS technique to carry out Kraus decomposition of density operators in matrix basis $\{|\alpha\rangle\langle\beta|, \alpha, \beta=1,2,\cdots,n\}$ \cite{andersson2007finding}. In this VDMS technique, a matrix $A$ can be mapped to a vector $\vec{A}$ by stacking all the rows of the matrix $A$ to a column in order:
\begin{align*}
A=
 \begin{bmatrix}
 a_1 &a_2\\
 a_3 &a_4
 \end{bmatrix}\rightarrow
\vec{A}=
 \begin{bmatrix}
 a_1\\a_2\\a_3\\a_4
 \end{bmatrix},
\end{align*}
In the similar way, the density operator $\rho$ will be mapped to the vector $\vec\rho$, and the superoperators $\mathcal{A}[\cdot]=A[\cdot]B$ will be mapped to a matrix, which can be recorded as $\overline{\overline{\mathcal{A}}}=A\otimes B^{\mathrm{T}}$. so
$\overrightarrow {{\cal A}[ \cdot ]}  = \overrightarrow {A[ \cdot ]B}  = \overline {\overline {\mathcal{A} }} \vec{[ \cdot ]}  = A \otimes {B^{\rm{T}}}\vec{[ \cdot ]}$, where the superscript "$\mathrm{T}$" denotes the transpose operation..

Meanwhile, give a matrix
\begin{align*}
B=
 \begin{bmatrix}
 b_1 &b_2 \\
 b_3 &b_4
 \end{bmatrix},
\end{align*}
in the VDMS technique, the Hilbert-Schmidt inner product can be introduced as follows:
\begin{align}
\langle B|A\rangle=
 \begin{bmatrix}
 b_1^*&b_2^*&b_3^*&b_4^*
 \end{bmatrix}
 \begin{bmatrix}
 a_1\\a_2\\a_3\\a_4
 \end{bmatrix}
 =\mathrm{Tr}[B^\dag A],
\end{align}
where $|A\rangle, \langle B|$ are actually the $\vec{A}, \vec{B}^\dag$ mentioned above. From this point of view, it is easy to find that the vector of $\rho_{AS}$ given in Eq.\eqref{rho_rhoS} and $\rho^\bot_{AS}$ given in Eq.\eqref{rho_rhoS-2} will still be mutually orthogonal.

Therefore, the Lindbald superoperator $\mathcal{L}$ in Eq.\eqref{master equation} will be mapped to
\begin{align}\label{Lindblad operator_matrix}
  \overline{\overline{\mathcal{L}}}(t)=&-i[\hat{H}_{AS}(t)\otimes I_{AS}-I_{AS}\otimes\hat{H}^{\mathrm{T}}_{AS}(t)]\nonumber\\
                                    &+\sum_\mu{ [\Gamma_\mu\otimes({\Gamma^\dag_\mu})^\mathrm{T}-\frac{\Gamma^\dag_\mu\Gamma_\mu\otimes I_{AS}}{2}-\frac{I_{AS}\otimes({\Gamma^\dag_\mu\Gamma_\mu})^\mathrm{T}}{2} ] },
\end{align}
After that the Eq.\eqref{H_hat_dyn} will be mapped to
\begin{equation}\label{Lindblad equation vec}
  \frac{\mathrm{d}\vec{\rho}_{AS}(t)}{\mathrm{d} t}=\overline{\overline{\mathcal{L}}}(t)\vec{\rho}_{AS}(t),
\end{equation}
and then we can get the solution of the above equation:
\begin{equation}\label{solution_master_equation}
  \vec{\rho}_{AS}(t)=\mathbb{T}e^{\int_0^t \overline{\overline{\mathcal{L}}}(\tau)\mathrm{d}\tau }\vec{\rho}_{AS}(0),
\end{equation}
where $\mathbb{T}$ is the time-ordering operator mentioned above in Eq.\eqref{eta_TD}, and $\vec{\rho}_{AS}(0)$ is the vector representation of the initial density operator $\rho_{AS}(0)$. In general, the dilated $\hat{H}_{AS}(t)$ is time-dependent, so $\overline{\overline{\mathcal{L}}}$ will also be time-dependent, then the problem of chronological product may also have to be dealt with (see more details in Appendix \ref{chronological product} and in our example given in Sec.\ref{example}).

\section{Numerical calculation of the relevant linear time-dependent matrix differential equations\label{numerical_calculation}}
As we all know, when the system $H_S$ and $H_{AS}$ is time-independent, the related calculations including Eq.\eqref{M_TD}, Eq.\eqref{H_hat_t}, Eqs.\eqref{H_hat_H_dyn-solutions} and et.al. are trivial. However, when they are time-dependent, the problem of chronological product may have to be considered. According to Magnus's theory \cite{Magnus1954,Blanes2009}, the solution of a linear matrix differential equation (include time-dependent Schr\"{o}dinger equation):
\begin{align}\label{matrix-differential-equation2}
Y^{\prime}(t)=A(t) Y(t), \quad Y\left(t_{0}\right)=Y_{0},
\end{align}
can be expressed by the form of exponential matrix like:
\begin{align}\label{Magnus2}
 Y(t)=\exp \left(\Omega\left(t, t_{0}\right)\right) Y_{0},
\end{align}
where $Y_{0}$ is the initial vector (state), and
\begin{align}\label{Omega2}
\Omega(t,t_0)=\sum_{k=1}^{\infty} \Omega_{k}(t,t_0)
\end{align}
is the matrix series: so-called Magnus expansion or Magnus series, and $\Omega_n(t,t_0)$ is the $n$-th term (see more details in Appendix \ref{chronological product}).

However, for an arbitrary TD linear operator (matrix) $A(t)$, the above Magnus series may diverge \cite{Blanes2009}, and a sufficient but unnecessary condition for it to converge in $t\in[t_0, t_f)$ is:
\begin{align}\label{convergence2}
\int_{t_0}^{t_f}\|A(s)\|_{2} \mathrm{d}s<\pi,
\end{align}
where $\|A\|_{2}$ denotes 2-norm of $A$. The above integral converges completely in the interval $t\in[t_0, t_0+T_c)$ (see more details in Appendix \ref{chronological product}), where $T_c$ is defined as the critical time of convergence that makes the above integral take $\pi$, and it only depends on the operator $A$ itself. This brings some restrictions to the time step $h$ in numerical calculation, because when $h<T_c$, the results can be completely trusted, while when $h>T_c$, the results becomes untrustworthy and the error may be amplified after Magnus series is truncated to a high-order term $\Omega_n$. The higher the order of items $\Omega_n$, the more obvious it is.

An alternative method to overcome the above finite convergence interval is to divide the interval into $N$ segments so that the Magnus series in each segment $[t_k,t_{k+1}]$ meets the above convergence conditions Eq.\eqref{convergence2}, i.e., $t_{k+1}-t_k<T_c$, then the Eq.\eqref{Magnus2} can be replaced by (see Eq.(240) in Ref.\cite{Blanes2009}):
\begin{align}\label{Magnus_product}
Y\left(t_N\right)=\prod_{k=0}^{N-1} \exp \left(\Omega\left(t_{k+1}, t_{k}\right)\right) Y_0,
\end{align}
where $\Omega\left(t_{k+1}, t_{k}\right)$ can be appropriately truncated to $\Omega_n\left(t_{k+1}, t_{k}\right)$ in practical use. 

Suppose that a fixed step size $h$ is adopted in computation, according to the analysis in Ref.\cite{Blanes2009}) (see around Eq.(66) and Eq.(242) for details), the error of computations is $O(h^3)$ when the computations up to first term of Magnus series $\Omega_1$ are carried out, and $O(h^5)$ when the computations up to second term $\Omega_2$ are carried  out. It is worth noting that, for the purpose of numerical calculation, we usually have to consider the computational complexity of the exponential matrix $\exp \left(\Omega\left(t, t_{0}\right)\right)$, because the computation cost of the exponential matrix is usually very expensive, especially when the matrix becomes large, so in this case, we should try to avoid its frequent computation. Therefore, on the premise of meeting the goal of calculation accuracy and saving computing resources, the time step of calculation and the cut-off term of Magnus series $\Omega_n$ have to be carefully balanced, specifically, we can reduce the calculation times of the exponential matrix by selecting a big enough $h$, and compensate for the loss of calculation accuracy by calculating up to higher order terms of Magnus series in Eq.\eqref{Magnus_product}, and we will see this compensation in the comparison between Fig.\ref{Magnus_expansion—figs} and Fig.\ref{Magnus2-fig} in next section.

\section{An example: 2-dimensional $\mathcal{PT}$-symmetric system\label{example}}

In this section, we analyze an example: 2-dimensional $\mathcal{PT}$-symmetric system:
\begin{equation}\label{example_H}H_S=
 \begin{bmatrix}
  re^{i\theta}&s  \\
   s          &re^{-i\theta}
 \end{bmatrix}, r, s\in\mathbb{R},\theta\in[-\pi/2, \pi/2],
\end{equation}
$\theta$ can be understood as the parameter representing the degree of non-Hermiticity of the Hamiltonian $H_S$, and the degree of non-Hermiticity will increase with $|\theta|$ (when $\theta=0$, $H_S$ will be Hermitian, when $\theta=\pi/2$, $H_S$ will be anti-Hermitian, see Sec.V in Ref.\cite{Li2022} for details). The eigenvalues of $H_S$ are $E_\pm=r\cos{\theta}\pm\sqrt{s^2-r^2\sin^2{\theta}}$, and when $s^2-r^2\sin^2{\theta}>0$, $H$ is $\mathcal{PT}$-symmetry unbroken, otherwise, when $s^2-r^2\sin^2{\theta}<0$, $H$ is $\mathcal{PT}$-symmetry broken, then the two eigenvalues become complex conjugate.

The results of pure-state vectors case has been given in Ref.\cite{wu2019observation}, in order to facilitate comparison with the results in Ref.\cite{wu2019observation}, we set $\theta=\pi/2,s=1$, which makes $H_S$ have the same form as the one in Ref.\cite{wu2019observation}, and under this parameter configuration, the system is in the $\mathcal{PT}$-symmetry unbroken phase, then the eigenvalues of $H_S$ will be $E_\pm=\pm\sqrt{1-r^2}$. At the same time, we set $M(0)=5I_S$ given in Eq.\eqref{M_TD}. We set the initial density operator of pure-state case as $\rho_S(0)=\frac{1}{5}|0\rangle_S\langle0|$, which correspondings to the initial state $|0\rangle_S$ in Ref.\cite{wu2019observation}, and the coefficient $1/5$ is required by $\mathrm{Tr}[\rho_{AS}]\equiv1$ according to Eq.\eqref{unit_trace}. We set the initial density operator of mixed-state case as ${\rho_S}_{\mathrm{mixed}}(0)=\frac{1}{30}\begin{pmatrix}
              4 & 1 \\
              1 & 2
            \end{pmatrix}$, which can not be described by the pure-state vectors used in Ref.\cite{wu2019observation}. 

\subsection{The effectiveness of the density operators tool and the eigenspectrum of Hamiltonian before and after dilation}
\begin{figure}
  \centering
  \includegraphics[width=0.95\linewidth]{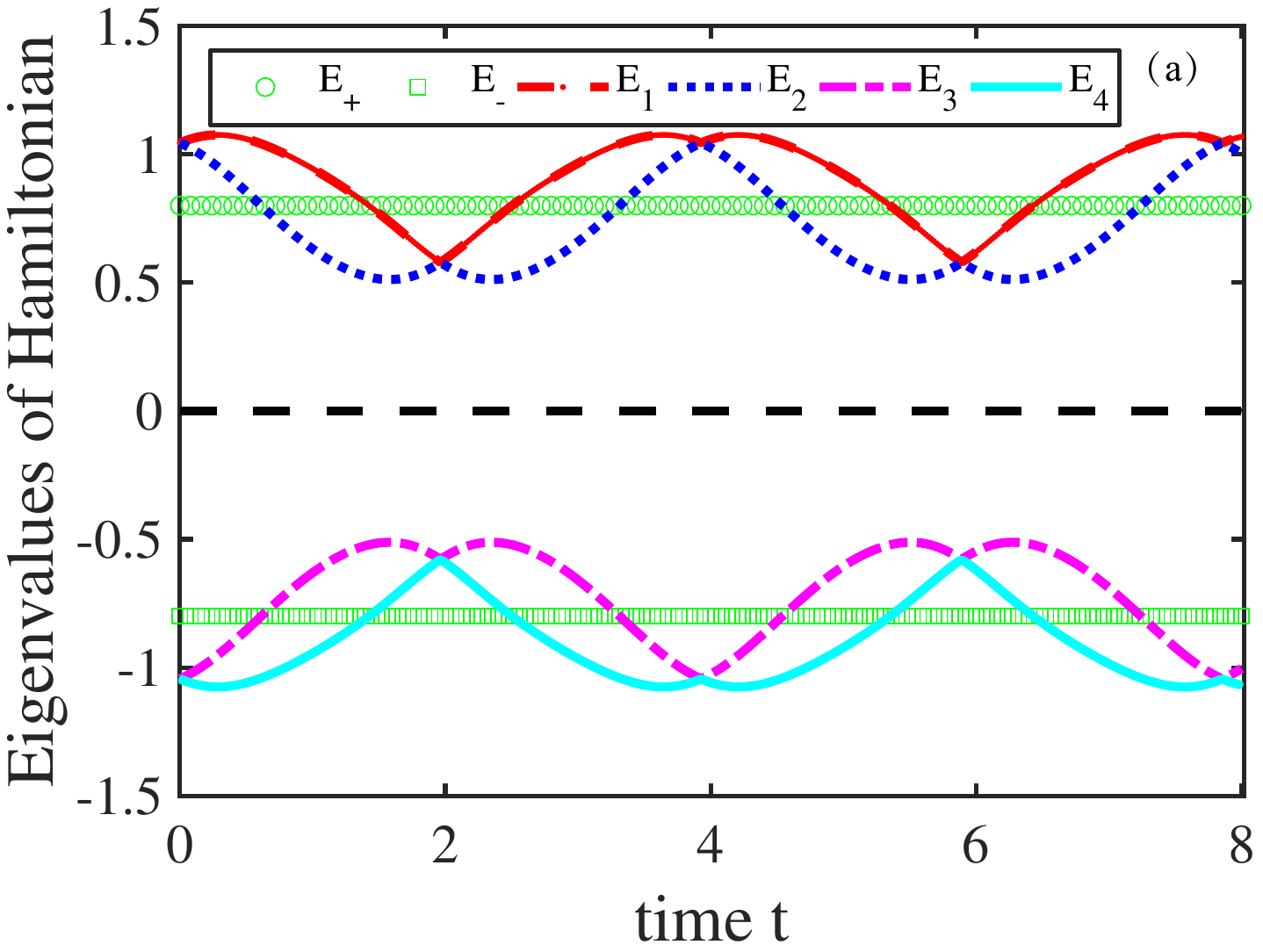}\label{unbroken1_magnus3_eigenvalue}\\
  \includegraphics[width=0.95\linewidth]{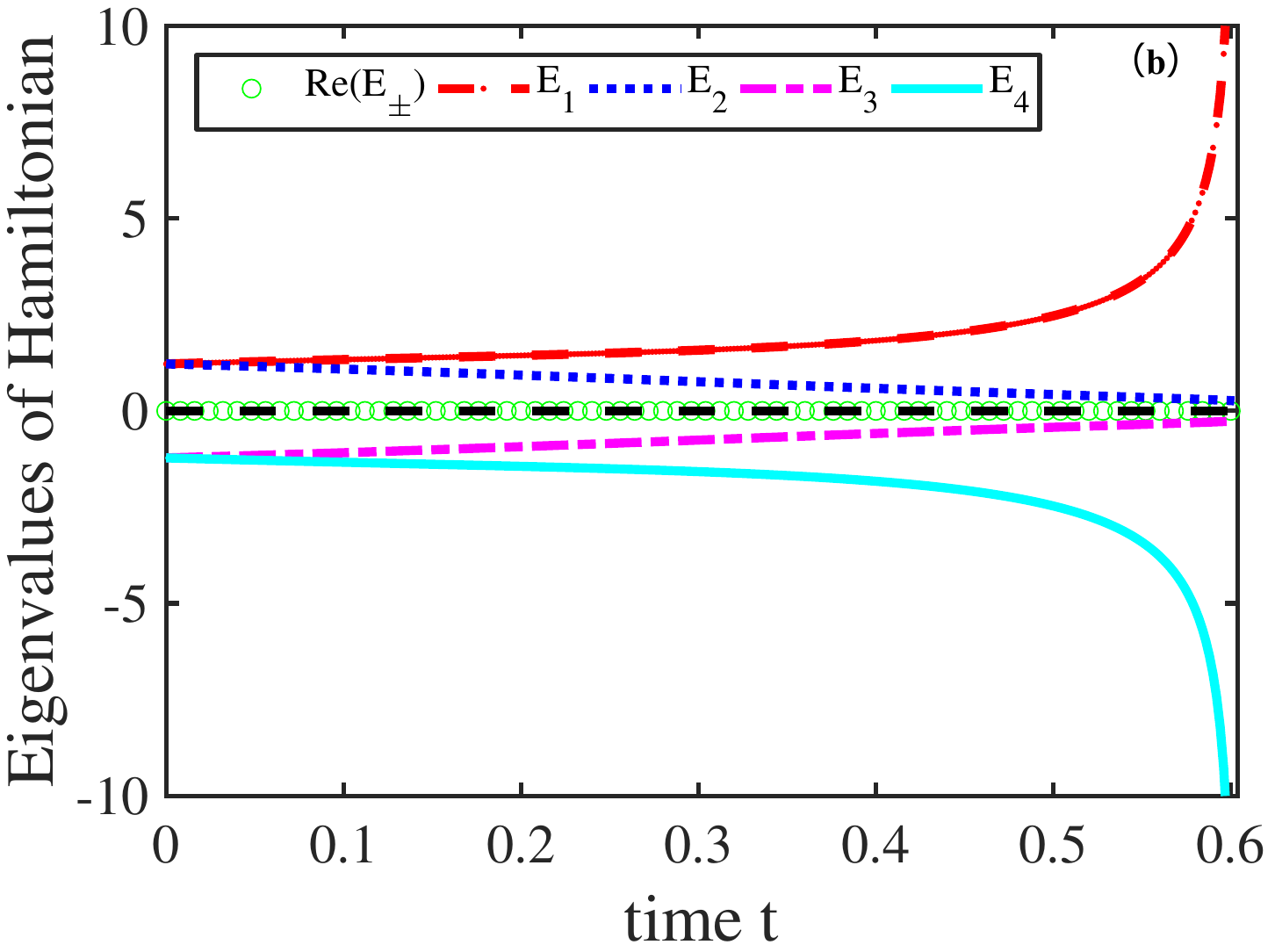}\label{broken1_magnus3_eigenvalue}
  \caption{The eigenvalues of $\mathcal{PT}$-symmetric Hamiltonian $H_S$ (green curves with small circles or boxes, $E_\pm=\pm0.8$) and the eigenvalues of its dilated Hermitian Hamiltonian $\hat{H}_{AS}$ (curves with other colors, $E_1,E_2,E_3,E_4$). (a) $\mathcal{PT}$-symmetry unbroken phase ($r=0.6$, $T_l=\infty$). (b) $\mathcal{PT}$-symmetry broken phase, $E_\pm=\pm0.98i$ ($r=1.4$, $T_l\approx0.604$).}\label{magnus3_eigenvalue}
\end{figure}
The eigenvalues of $\mathcal{PT}$-symmetric Hamiltonian $H_S$ ($E_\pm$) and the eigenvalues of its dilated Hermitian Hamiltonian $\hat{H}_{AS}$ ($E_1,E_2,E_3,E_4$, arranged in descending order) are given in Fig.\ref{magnus3_eigenvalue}. Fig.\ref{magnus3_eigenvalue}(a) is about the $\mathcal{PT}$-symmetric unbroken phase ($r=0.6$, $E_\pm=\pm\sqrt{1-0.6^2}=\pm0.8$), while Fig.\ref{magnus3_eigenvalue}(b) is about the broken phase ($r=1.4,E_\pm=\pm\sqrt{1-1.4^2}=\pm0.98i$ will be conjugate complex numbers, which is exactly what $\mathcal{PT}$-symmetry theory predicts \cite{Mostafazadeh2002b}). From Fig.\ref{magnus3_eigenvalue}(a), we can see $E_\pm$ remain unchanged, while $E_1$ and $E_2$  oscillate around $E_+$, $E_3$ and $E_4$  oscillate around $E_-$ and change periodically with time $t$ (in this case, the critical time $T_l$ of the legitimacy given below Eq.\eqref{convergence2} under the setting $r=0.6$ is infinite, i.e., $T_l=\infty$). Meanwhile, $E_1$ and $E_4$, $E_2$ and $E_3$ are symmetric about $E=0$ (black dashed line). From Fig.\ref{magnus3_eigenvalue}(b), we can see that $E_k$s ($k=1,2,3,4$) are also symmetric about $E=0$ (black dashed line), which is similar as the case of unbroken phase. In fact, the symmetry in the case of unbroken and broken phase is caused by the symmetric gauge adopted in Eq.\eqref{gauge}. However, the periodicity oscillation of $E_k$s are destroyed, and $E_k$s increase (decrease) monotonically with time $t$. Especially, when $t\rightarrow T_l$ (the critical time $T_l$ of the legitimacy under the setting $r=1.4$ is about 0.604), $E_1$ ($E_4$) will increases (decreases) sharply to infinity, which is caused by one of eigenvalues of $M(t)$ given in Eq.\eqref{M_TD} tend to one. At this moment, the energy of system $\hat{H}_{AS}$ may diverge, so can not be realized by an experiment. In a summary, the critical time $T_l$ of the legitimacy actually bounds the duration of implemented experimental running.

We also characterize the evolution using the renormalized population ${P_\mathrm{N}}_0(t)$ of state $\rho_S(0)$ in main system $S$ \cite{wu2019observation}. In this situation, the renormalized population ${P_\mathrm{N}}_0(t)$ can be obtained by
\begin{align}\label{P_N0}
{P_\mathrm{N}}_0(t)=\frac{\mathrm{Tr}[|0\rangle_S\langle0|\cdot\rho_S(t)]}{\mathrm{Tr}[\rho_S(t)]}=\frac{\langle0|\rho_S(t)|0\rangle_S}{P_0(t)},
\end{align}
where $P_0(t)=\mathrm{Tr}[\rho_S(t)]$ has been given below Eq.\eqref{rho_S_N}. The result based on the pure-state vectors has been given analytically in Ref.\cite{wu2019observation} as follows:
\begin{widetext}
\begin{align}\label{reference}
{P_\mathrm{N}}_0(t)=
\begin{cases}
\frac{\left|e^{t \sqrt{r^{2}-1}}\left(r+\sqrt{r^{2}-1}\right)-e^{-t \sqrt{r^{2}-1}}\left(r-\sqrt{r^{2}-1}\right)\right|^{2}}{\left|e^{t \sqrt{r^{2}-1}}\left(r+\sqrt{r^{2}-1}\right)-e^{-t \sqrt{r^{2}-1}}\left(r-\sqrt{r^{2}-1}\right)\right|^{2}+\left|i e^{-t \sqrt{r^{2}-1}}-i e^{t \sqrt{r^{2}-1}}\right|^{2}},  &r \neq 1 \\
\frac{(t+1)^{2}}{(t+1)^{2}+t^{2}}, &r=1.
\end{cases}
\end{align}
\end{widetext}

\begin{figure}
  \centering
  \includegraphics[width=0.95\linewidth]{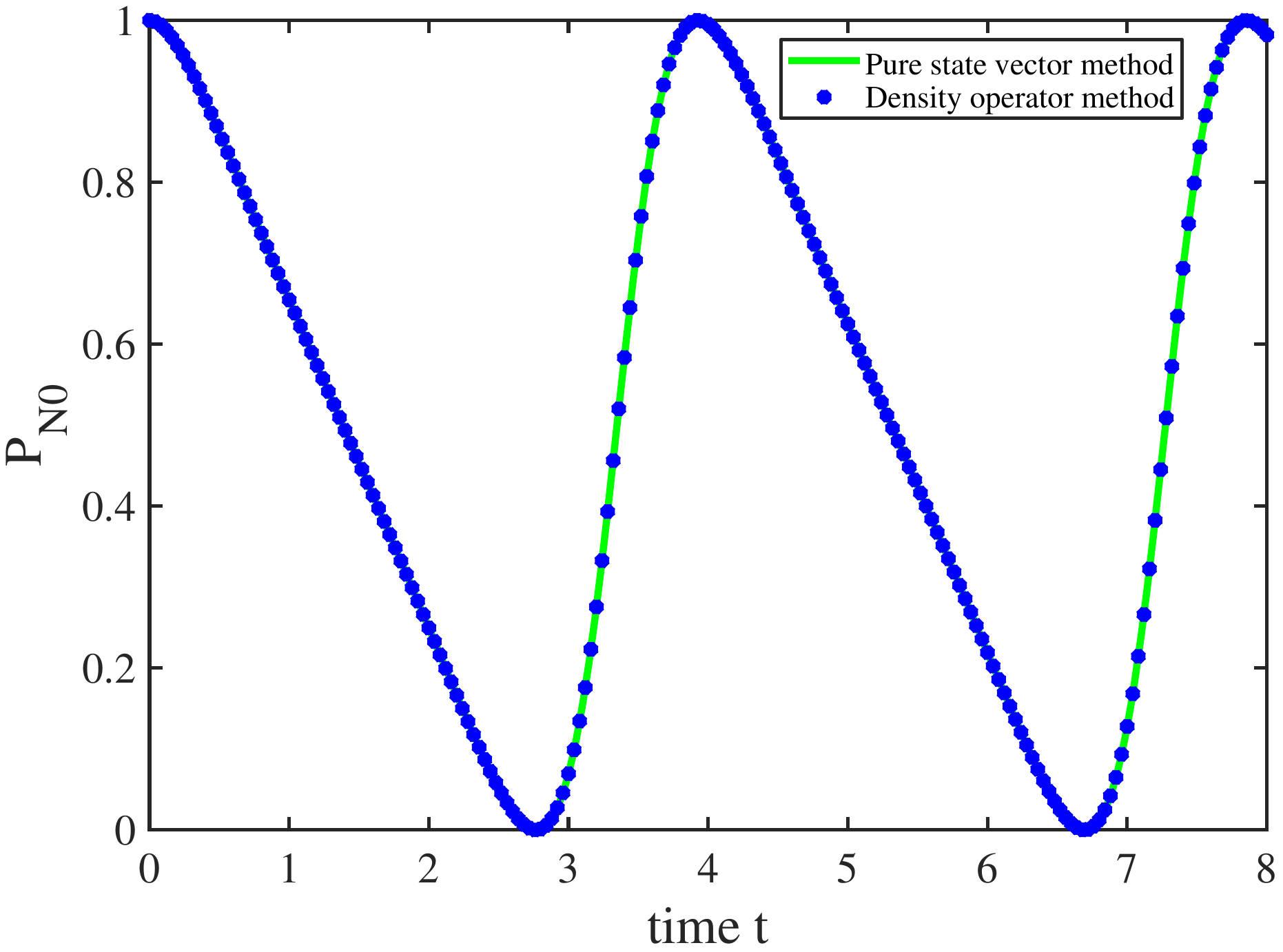}
  \caption{Comparisons of the results for 2-dimensional $\mathcal{PT}$-symmetric dynamics simulation (unbroken phase) between the method based on pure-state vectors (green curve) and the method based on density operator (blue dotted curve) ($\theta=\pi/2,s=1,r=0.6$). Both $X$ and $Y$ axes are dimensionless. In this case, the system is in $\mathcal{PT}$-symmetry unbroken phase. }\label{comparisons}
\end{figure}

According to Eq.\eqref{H_hat_H_dyn-solutions1} and above Eq.\eqref{P_N0}, we can calculate the renormalized population ${P_\mathrm{N}}_0(t)$ under the initial density operator $\rho_S(0)$. To verify the effectiveness of density operators method in Sec.\ref{dilation_method}, we take $r=0.6$ and draw the Fig.\ref{comparisons}. The green curve is drawn according to the analytical results based on the pure-state vector method given in the above Eq.\eqref{reference}, while the blue dotted curve is drawn according to Eqs.\eqref{H_hat_H_dyn-solutions1} and \eqref{P_N0}, and we can find their results are completely the same, which means the density operator method are compatible with pure-state vector method, just as we intuitively think. It should be pointed out that the Hamiltonian $H_S$ in this example is time-independent, so the problem of chronological product can be avoid in the calculation of $M(t)$ according to Eq.\eqref{M_TD}.

\subsection{The influence of three kinds of quantum noises generated in main system}
In the experimental simulation of $\mathcal{PT}$-symmetric system dynamics, the quantum state will inevitably be disturbed by quantum noises, especially when the state is entangled just like the state in Eq.\eqref{rho_rhoS}, the damage of quantum noise to entanglement may be so fatal that has to be studied carefully. In this section, we introduce three common quantum noise (channel) models and using the VDMS technique mentioned in Sec.\ref{VD_and_ML} to characterize the Lindblad master equation corresponding to them. We only consider the situation where the noises act on the main system space (similarly, the situation where the noises act on the auxiliary system can also be studied). Based on that, we analysis their effects on the simulation of $\mathcal{PT}$-symmetric system dynamics.

In AD channel and PD channel mentioned blow Eq.\eqref{Lindblad}, there is only one jump operator (for convenience, all parameters of decay rate have been set to $\gamma$): $\Gamma_{\mathrm{AD}}=\sqrt{\gamma}\sigma^S_{-}=\sqrt{\gamma}|0\rangle_S\langle1|$, $\Gamma_{\mathrm{PD}}=\sqrt{\gamma}\sigma^S_z$; while in Dep channel, there are three jump operators:
${\Gamma_{\mathrm{Dep}}}_k=\sqrt{\gamma}\sigma^S_k$, where $k=x,y,z$, $\{\sigma^S_k\}$ are Pauli operators, and the subscript "$S$" indicates it acts on the main system, and the complete form of any operator $X$ with different superscripts is $X^S=I_A\otimes X^S, X^A=X^A\otimes I_S$.

Considering the system $\hat{H}_{AS}(t)$ used to simulate the $\mathcal{PT}$-symmetric system dynamics in Eq.\eqref{H_hat_t}, and substituting jump operators into Eq.\eqref{Lindblad operator_matrix} we obtain that
\begin{widetext}
\begin{subequations}\label{AD PD_Dep}
\begin{align}
  \overline{\overline{\mathcal{L}}}_{\mathrm{AD}}(t)&=-i[\hat{H}_{AS}(t)\otimes I_{AS}-I_{AS}\otimes\hat{H}^{\mathrm{T}}_{AS}(t)]+\gamma[ |0\rangle_S\langle1|\otimes(|1\rangle_S\langle0|)^\mathrm{T}-\frac{|1\rangle_S\langle1|}{2}\otimes I_S-I_S\otimes\frac{(|1\rangle_S\langle1|)^\mathrm{T}}{2} ],\label{AD channel}\\
  \overline{\overline{\mathcal{L}}}_{\mathrm{PD}}(t)&=-i[\hat{H}_{AS}(t)\otimes I_{AS}-I_{AS}\otimes\hat{H}^{\mathrm{T}}_{AS}(t)]+\gamma[ \sigma^S_z\otimes({\sigma^S_z})^\mathrm{T}-I_S\otimes I_S],\label{PD channel}\\
  \overline{\overline{\mathcal{L}}}_{\mathrm{Dep}}(t)&=-i[\hat{H}_{AS}(t)\otimes I_{AS}-I_{AS}\otimes\hat{H}^{\mathrm{T}}_{AS}(t)]+\gamma\sum_{k\in\{x,y,z\}}[ \sigma^S_k\otimes({\sigma}^S_k)^\mathrm{T}-I_S\otimes I_S],\label{Dep channel}
\end{align}
\end{subequations}
\end{widetext}
where $\hat{H}_{AS}(t)$ is the dilated Hamiltonian in Eq.\eqref{H_hat_t}, $I_{AS}$ and $I_S$ represent the identity operator in the composite system $AS$ and the main system $S$ respectively, and they are the same when written in complete form. We calculate Eqs.\eqref{AD PD_Dep} and Eqs.\eqref{H_hat_H_dyn-solutions}  by Magnus series given in Eq.\eqref{Magnus2}, and only the first two terms of Magnus series, i.e., $\Omega_1,\Omega_2$ are considered. Specifically, in the calculation, the operator $A(t)$ given in Eq.\eqref{matrix-differential-equation2} can be replaced by $-i\hat{H}_{AS}(t)$ obtained in Eq.\eqref{H_hat_t}, and $\overline{\overline{\mathcal{L}}}_{\mathrm{AD}}(t), \overline{\overline{\mathcal{L}}}_{\mathrm{PD}}(t), \overline{\overline{\mathcal{L}}}_{\mathrm{Dep}}(t)$ given in Eqs.\eqref{AD PD_Dep}.

\begin{figure*}
  \centering
    {\includegraphics[width=0.630\linewidth]{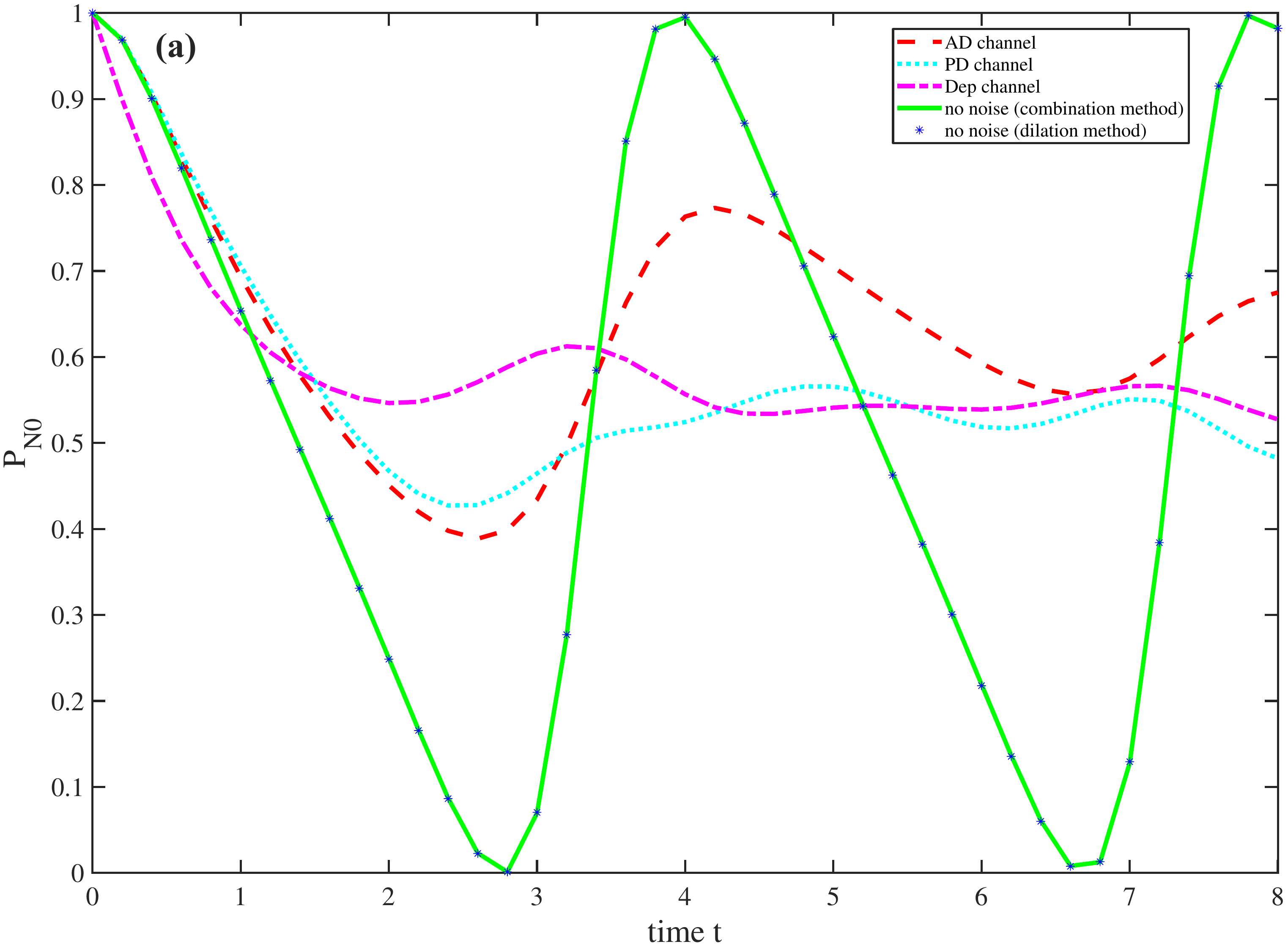}}\label{unbroken1_magnus1}
    {\includegraphics[width=0.222\linewidth]{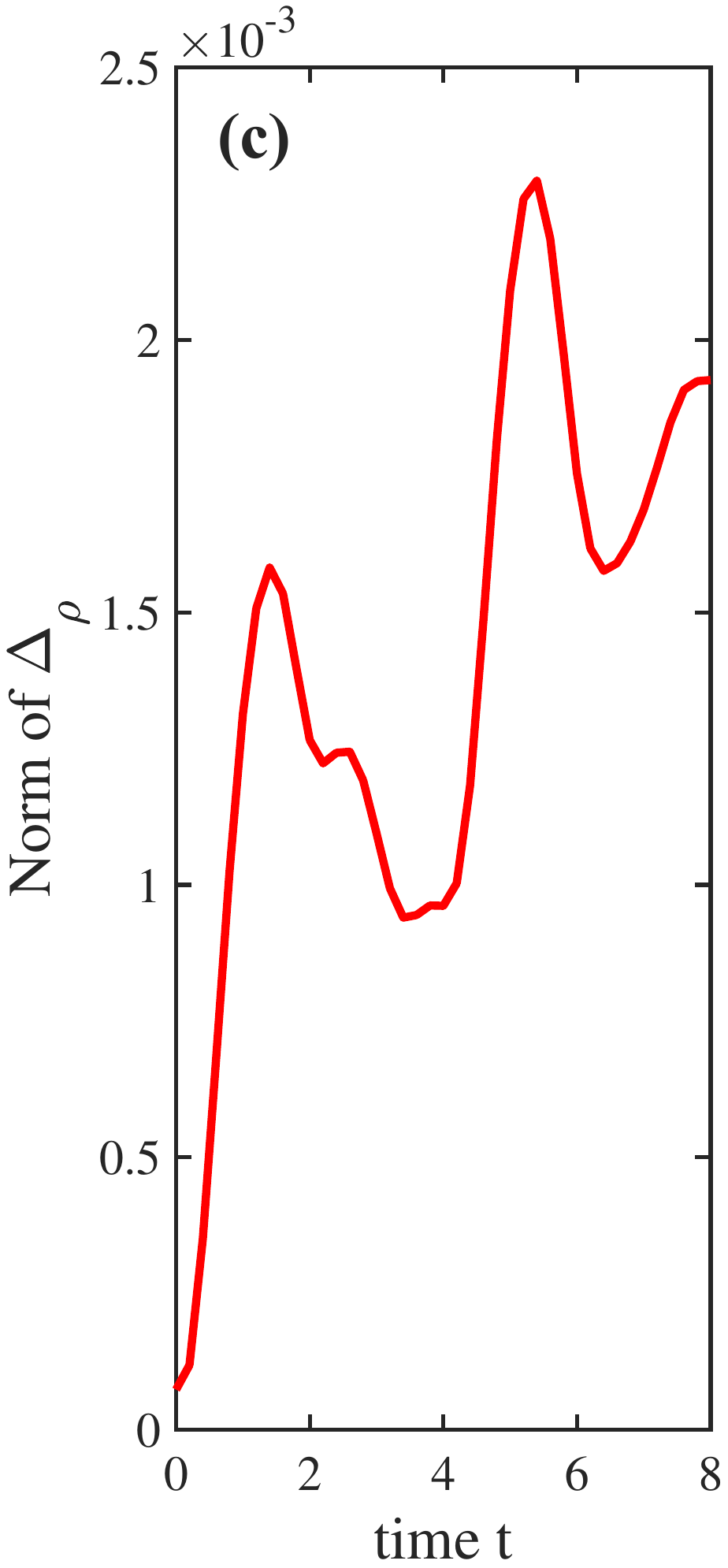}}\label{unbroken1_magnus1_norm}\\
    {\includegraphics[width=0.630\linewidth]{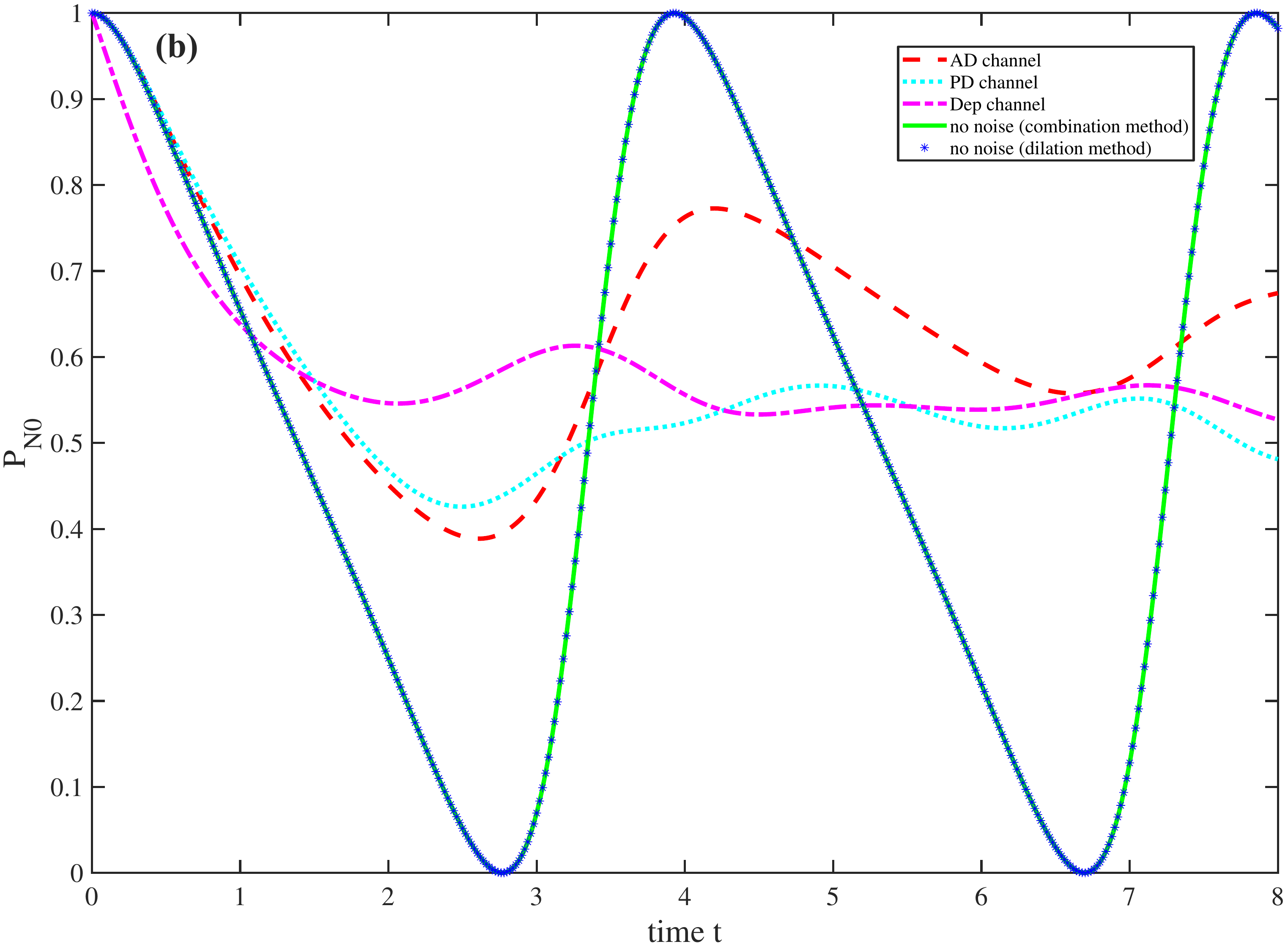}}\label{unbroken1_magnus2}
    {\includegraphics[width=0.222\linewidth]{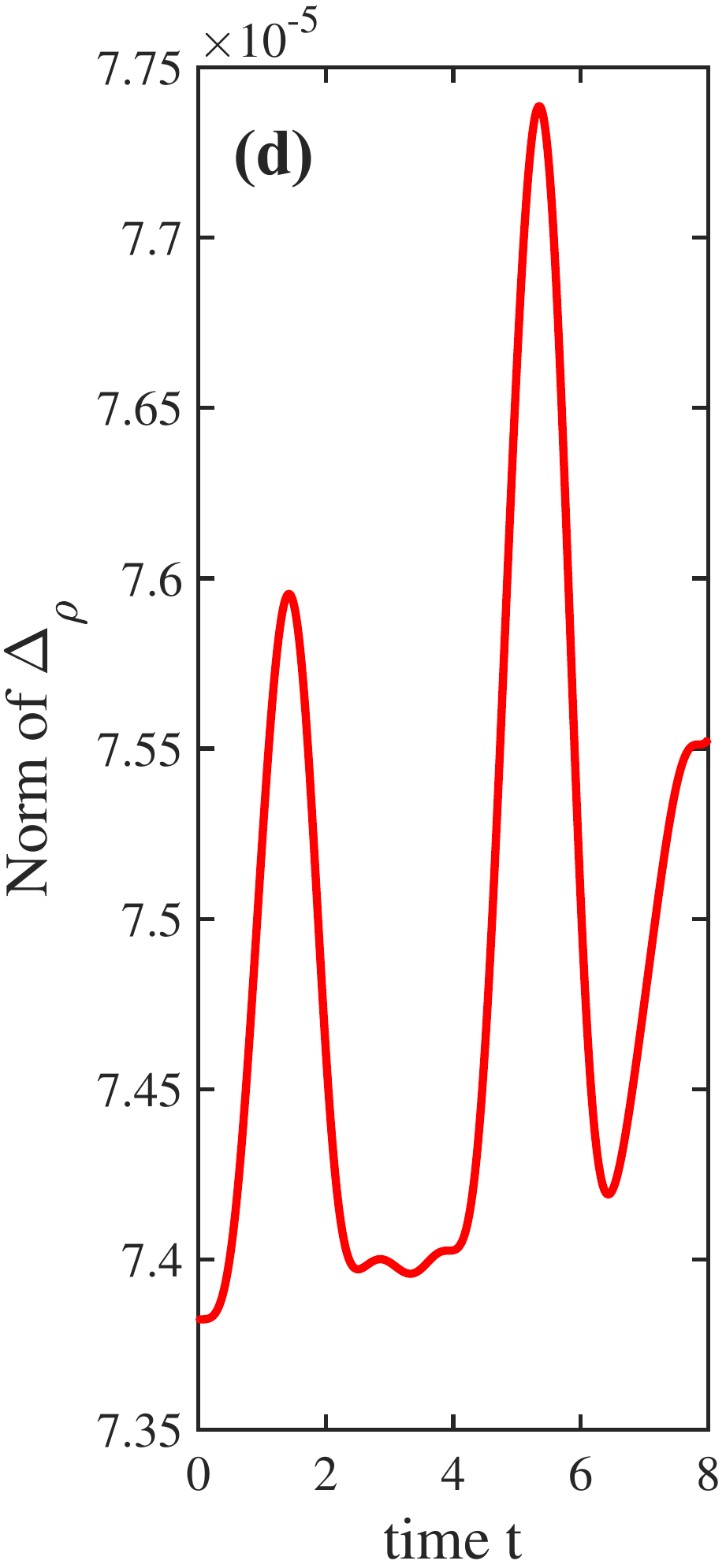}}\label{unbroken1_magnus2_norm}
  \caption{The effects of quantum noises on the dynamics simulation of $\mathcal{PT}$-symmetry unbroken system under the initial state of pure-state density operator ($\gamma=0.25,\theta=\pi/2,s=1,r=0.6$). Both $X$ and $Y$ axes are dimensionless. (a-b) Renormalized population ${P_\mathrm{N}}_0$ with Magnus series are calculated to the first term $\Omega_1$ in time steps $h=0.2$ and $h=0.02$ respectively. The linestyles and colors related to each noise (including no noise) are represented differently. (c-d) The errors between the green curves and the blue dotted curves corresponding to subfigures (a-b) respectively. }
\label{Magnus_expansion—figs}
\end{figure*}

\begin{figure*}
  \centering
    {\includegraphics[width=0.630\linewidth]{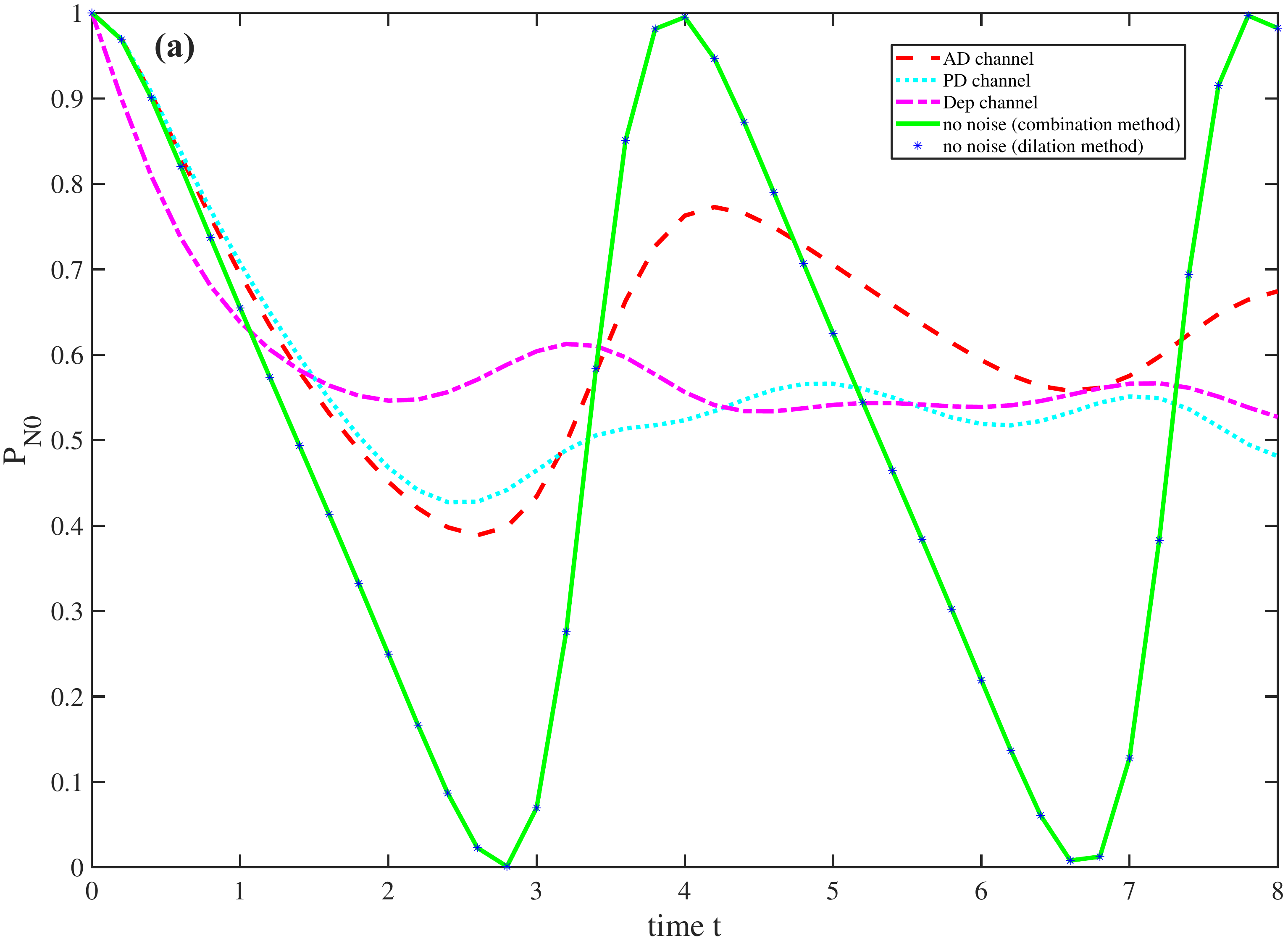}}\label{unbroken1_magnus3}
    {\includegraphics[width=0.222\linewidth]{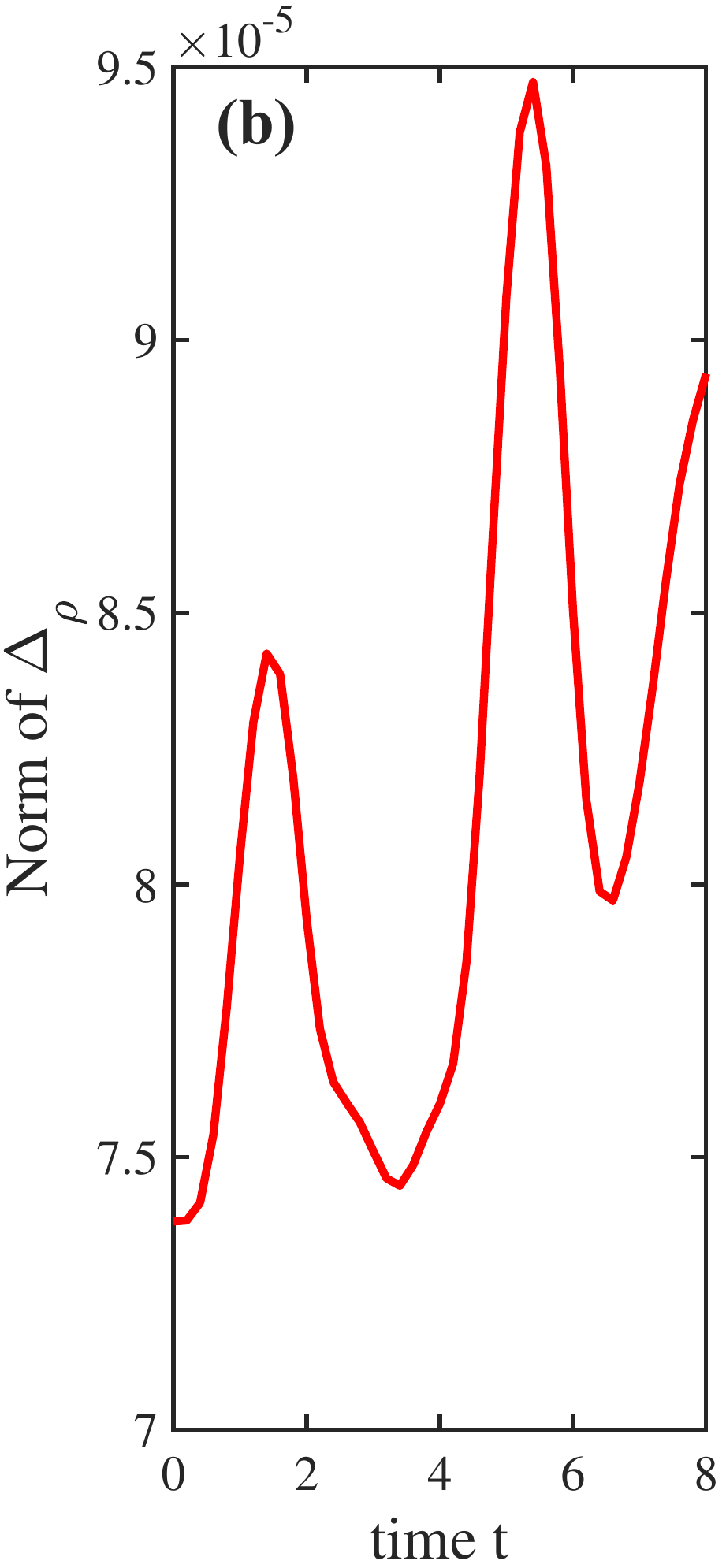}}\label{unbroken1_magnus3_norm}
  \caption{The effects of quantum noises on the dynamics simulation of $\mathcal{PT}$-symmetry unbroken system under the initial state of pure-state density operator ($\gamma=0.25,\theta=\pi/2,s=1,r=0.6$). Both $X$ and $Y$ axes are dimensionless. (a) Renormalized population ${P_\mathrm{N}}_0$ with Magnus series are calculated to the second term $\Omega_2$ in time step $h=0.2$. The linestyles and colors related to each noise (including no noise) are represented differently. (b) The errors between the green curves and the blue dotted curves corresponding to subfigure (a). } 
\label{Magnus2-fig}
\end{figure*}

\begin{figure*}
  \centering
    {\includegraphics[width=0.65\linewidth]{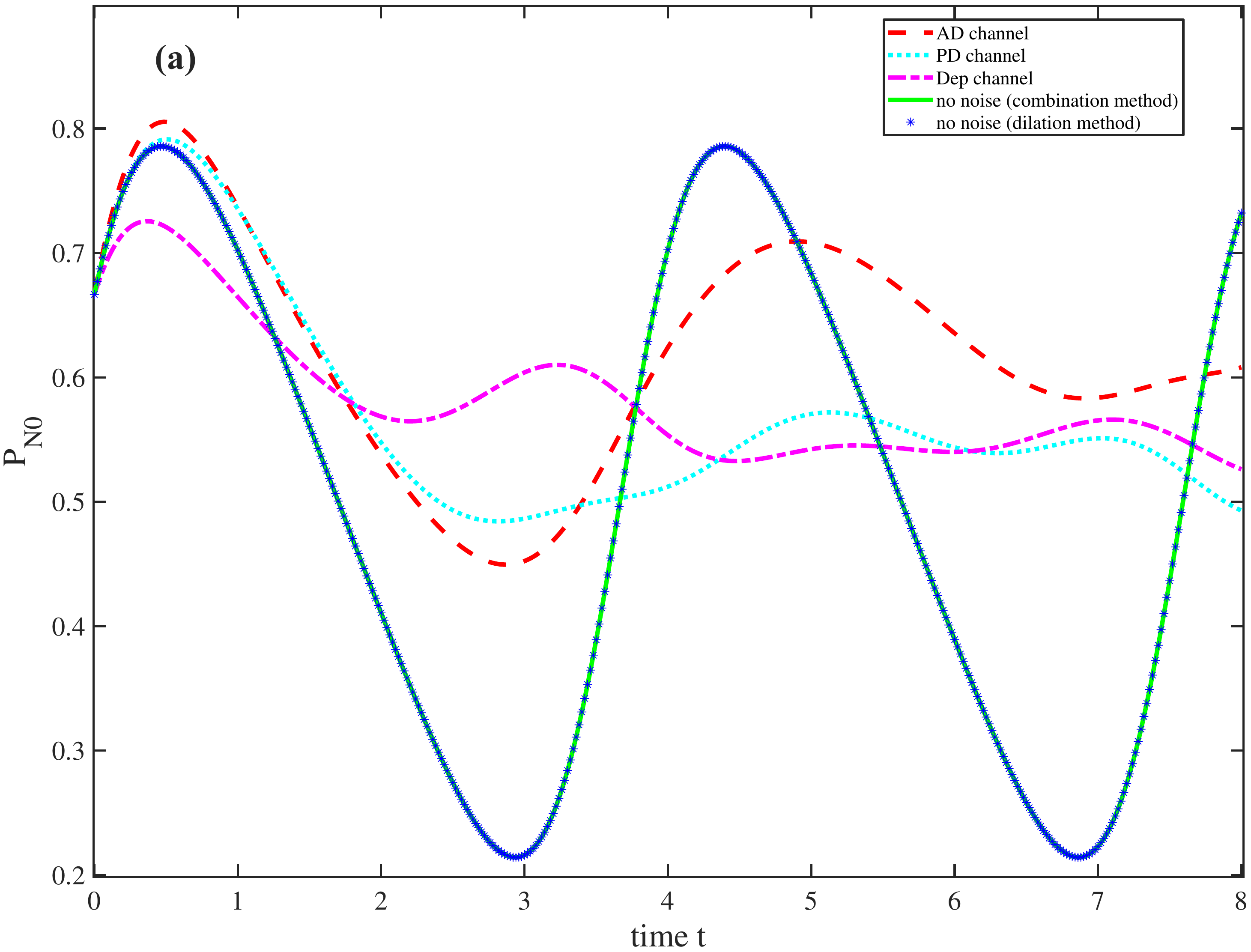}}\label{mixed_unbroken1_magnus1}
    {\includegraphics[width=0.27\linewidth]{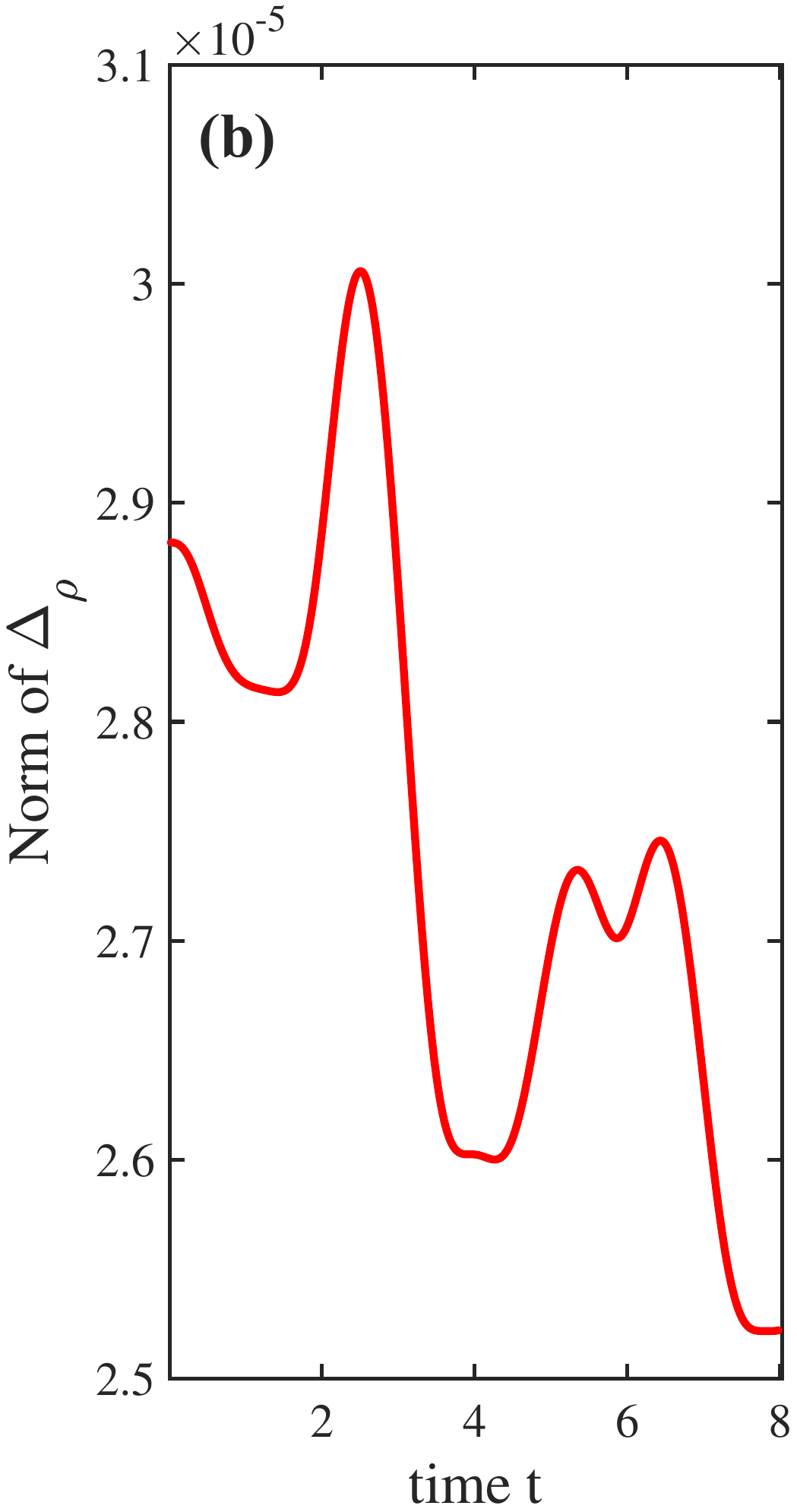}}\label{mixed_unbroken1_magnus1_norm}\\
  \caption{The effects of quantum noises on the dynamics simulation of $\mathcal{PT}$-symmetry unbroken system ($\gamma=0.25,\theta=\pi/2,s=1,r=0.6$, and set time step =0.02) under the initial state of mixed-state density operator. Both $X$ and $Y$ axes are dimensionless. (a) Renormalized population ${P_\mathrm{N}}_0$ under the first term of Magnus series $\Omega_1$ according to Eq.\eqref{Magnus_product}. The linestyles (dotted curves) and colors related to each noise (including no noise) are represented differently. (b) The error between the green curve and the blue dotted curve corresponding to (a).}
\label{Magnus_expansion—figs_mixed}
\end{figure*}

Based on the density operators, we are able to consider the effect of quantum noise to the simulation of the dynamics of $\mathcal{PT}$-symmetric system $H_S$. According to Eq.\eqref{M_Map}, Eqs.\eqref{AD PD_Dep} and Eq.\eqref{P_N0}, we can calculate the renormalized population ${P_\mathrm{N}}_0(t)$ under AD, PD, Dep channel and no noise, respectively. Under the parameter settings of Fig.\ref{Magnus_expansion—figs}, Fig.\ref{Magnus2-fig} and Fig.\ref{Magnus_expansion—figs_mixed}, i.e., $\gamma=0.25,\theta=\pi/2,s=1,r=0.6$, the critical time of the legitimacy mentioned below Eq.\eqref{M_TD} $T_l=\infty$ because in the $\mathcal{PT}$-symmetry unbroken phase of $H_S$, all eigenvalues of $H_S$ are real. In addition, the $\max_{t\in[0, 8]}\|H_{AS}(t)\|_2\approx1.08$, $\max_{t\in[0, 8]}\|\overline{\overline{\mathcal{L}}}_{AD}(t)\|_2\approx2.19$, $\max_{t\in[0, 8]}\|\overline{\overline{\mathcal{L}}}_{PD}(t)\|_2\approx2.34$, $\max_{t\in[0, 8]}\|\overline{\overline{\mathcal{L}}}_{Dep}(t)\|_2\approx2.38$, so when we set time step $h=0.02$ (Fig.\ref{Magnus_expansion—figs}(b)(d) and Fig.\ref{Magnus_expansion—figs_mixed}) or $h=0.2$ (Fig.\ref{Magnus_expansion—figs}(a)(c) and Fig.\ref{Magnus2-fig}), the convergence condition, i.e., Eq.\eqref{convergence2}, will always be satisfied in every step because  $h<T_c$.

The relations between the renormalized population ${P_\mathrm{N}}_0$ and time $t$ in the case of quantum noise and no noise are given in Fig.\ref{Magnus_expansion—figs} and Fig.\ref{Magnus2-fig} respectively. At first, we focus on the cases of no noise, which are related to the green curves (the combination method) and blue dotted curves (the dilation method) in Fig.\ref{Magnus_expansion—figs}. In Fig.\ref{Magnus_expansion—figs}, the subfigures (a) and (b) are related to the relation between ${P_\mathrm{N}}_0$ and $t$ with Magnus series are calculated to the first term $\Omega_1$ in time steps $h=0.2$ and $h=0.02$ respectively according to Eqs.\eqref{H_hat_H_dyn-solutions}, Eqs.\eqref{AD PD_Dep} and Eq.\eqref{Magnus_product}, while in Fig.\ref{Magnus2-fig}(a), Magnus series is calculated to the second term $\Omega_2$ with time step $h=0.2$. In Fig.\ref{Magnus_expansion—figs}(a-b) and Fig.\ref{Magnus2-fig}(c), the green lines represent the results of no noise computed by the combination method (act as a theoretical result), which only involves calculations involving $H_S(t)$, not calculations involving $H_{AS}(t)$, while the blue doted lines represent the results of no noise computed by the dilation method (act as a simulation result), which involves calculations involving both $H_{S}(t)$ and $H_{AS}(t)$. The Fig.\ref{Magnus_expansion—figs}(c-d) and Fig.\ref{Magnus2-fig}(b) are errors between the green lines and the blue lines.
Compared Fig.\ref{Magnus_expansion—figs}(c) with (d), we can find that when the time step is increased from $h=0.2$ to $h=0.02$, the error is reduced by two orders of magnitude, which is duce to the error is $O(h^3)$ when the Magnus series is computed to the first term $\Omega_1$ according to Eq.\eqref{Magnus_product} \cite{Blanes2009}. Meanwhile, compared Fig.\ref{Magnus_expansion—figs}(c) with Fig.\ref{Magnus2-fig}(b), by calculating Magnus series to the second term $\Omega_2$, we can find the error in Fig.\ref{Magnus2-fig}(b) is also reduced by two orders of magnitude although they are obtained with the same time step $h=0.2$, which is due to the error is $O(h^5)$ when the Magnus series is computed to the second term $\Omega_2$ \cite{Blanes2009}. It is worth pointing that although computing Magnus series to high-order terms $\Omega_n (n\geqslant2)$ may lead to higher accuracy, $\Omega_n$ are usually difficult to be computed especially when $H_{AS}$ is high-dimensional multivariable symbolic matrix (see Eq.\eqref{Magnus series1-4} in Appendix \ref{chronological product}). On the contrary, the advantage of only computing $\Omega_1$ is very convenient, however, the improvement of accuracy has to be achieved by reducing the time step $h$, which means that more exponential matrices have to be calculated, which is usually computing-expensive especially when the matrix is high-dimensional. Therefore, considering the computational cost of achieving a specific accuracy, we usually have to make a compromise between the fewer computed terms of Magnus series and the bigger step size.

Now we focus on the effects of quantum noises on the renormalized population ${P_\mathrm{N}}_0$. From Fig.\ref{Magnus_expansion—figs}(a-b) and Fig.\ref{Magnus2-fig}, we can find that the curves related to Dep channel drops rapidly, and faster than the cases of AD, PD channels, which is caused by the tendency of Dep channel that transforms any quantum state into the maximal mixed state, where ${P_\mathrm{N}}_0$ will keep $1/2$. The similar phenomenon can also be seen in mixed-state density operator case in Fig.\ref{Magnus_expansion—figs_mixed}, where we have set the initial density operator of the mixed state as ${\rho_S}_{\mathrm{mixed}}(0)=\frac{1}{30}\begin{pmatrix}
              4 & 1 \\
              1 & 2
            \end{pmatrix}$.
Meanwhile, in the initial short time, the red curves related to AD channel almost coincides with the cyan curves related to PD channel, while after a long time, they separate from each other. The similar phenomenon can also be seen in Fig.\ref{Magnus_expansion—figs_mixed}. However, this phenomenon is accidental because the initial pure-state density operator $\rho_S(0)=\frac{1}{5}|0\rangle_S\langle0|$ is just located on the eigenstate (steady state, or the fixed point of superoperator of the noise channel) of the dissipation terms of AD channel given in Eq.\eqref{AD channel} and PD channel given in Eq.\eqref{PD channel} \cite{Minganti2019,andersson2007finding}, so their roles can be ignored in a short time, and only the left terms containing $\hat{H}_{AS}(t)$ play the roles; while in the long run, the evolution state $\rho_S(t)$ is far away from the initial pure state $|0\rangle_S$, so it can be affected by the dissipation terms where their roles cannot be ignored. Compared with the case of mixed state, we can understand this phenomenon more clearly. From Fig.\ref{Magnus_expansion—figs_mixed} we can see that in a short time, all the curves including noise and no noise cases rise, which are the results driven by Hamiltonian $\hat{H}_{AS}(t)$. However, the case of AD channel rises faster than all other curves, because the AD channel has a tendency to change all states to the state $|0\rangle_S$, which will contribute the renormalized population ${P_\mathrm{N}}_0$ (more strictly, $|0\rangle_S$ is the fixed point (steady state) of AD channel \cite{Minganti2019,andersson2007finding}.

\begin{figure*}
  \centering
    {\includegraphics[width=0.662\linewidth]{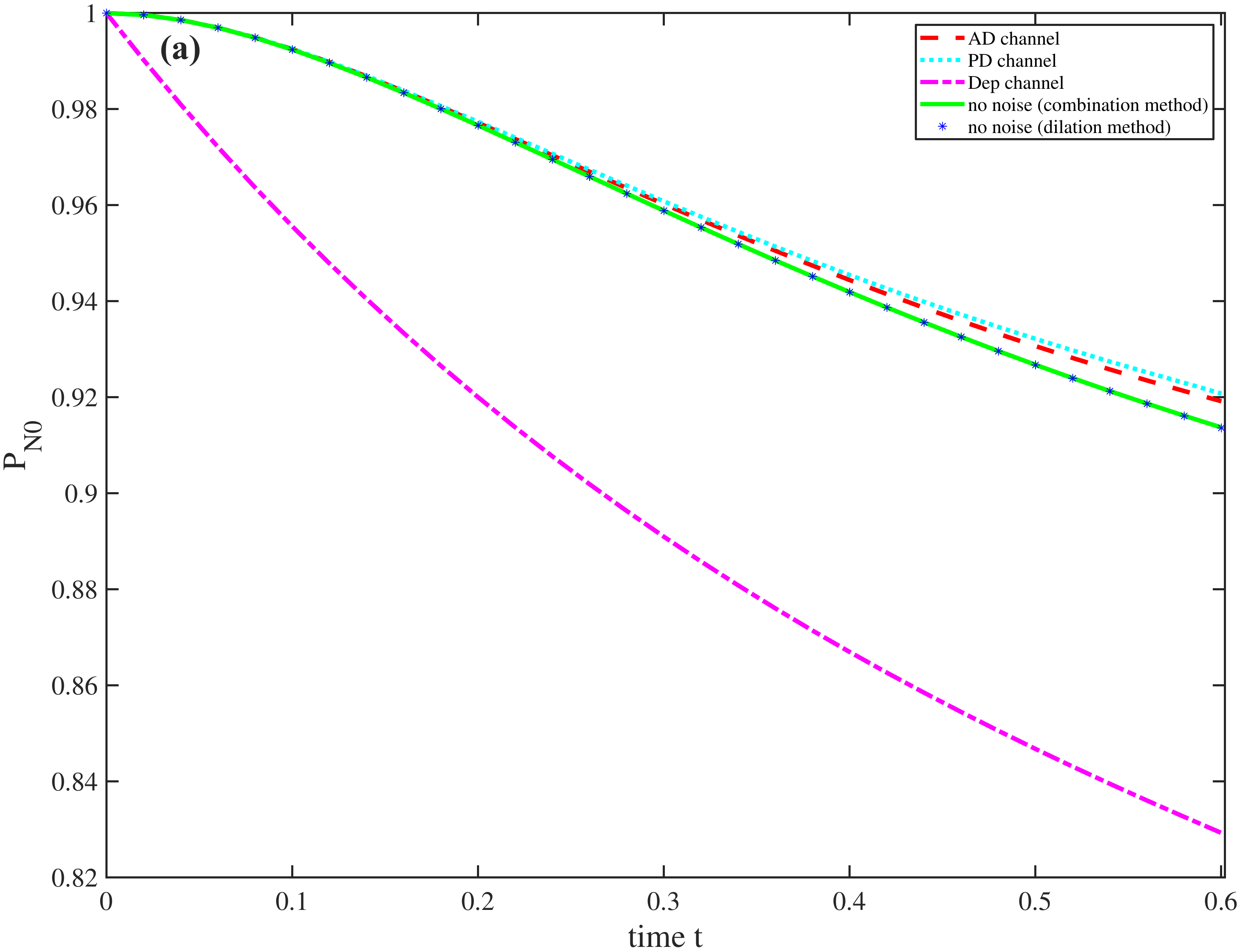}}\label{broken1_magnus1}
    {\includegraphics[width=0.280\linewidth]{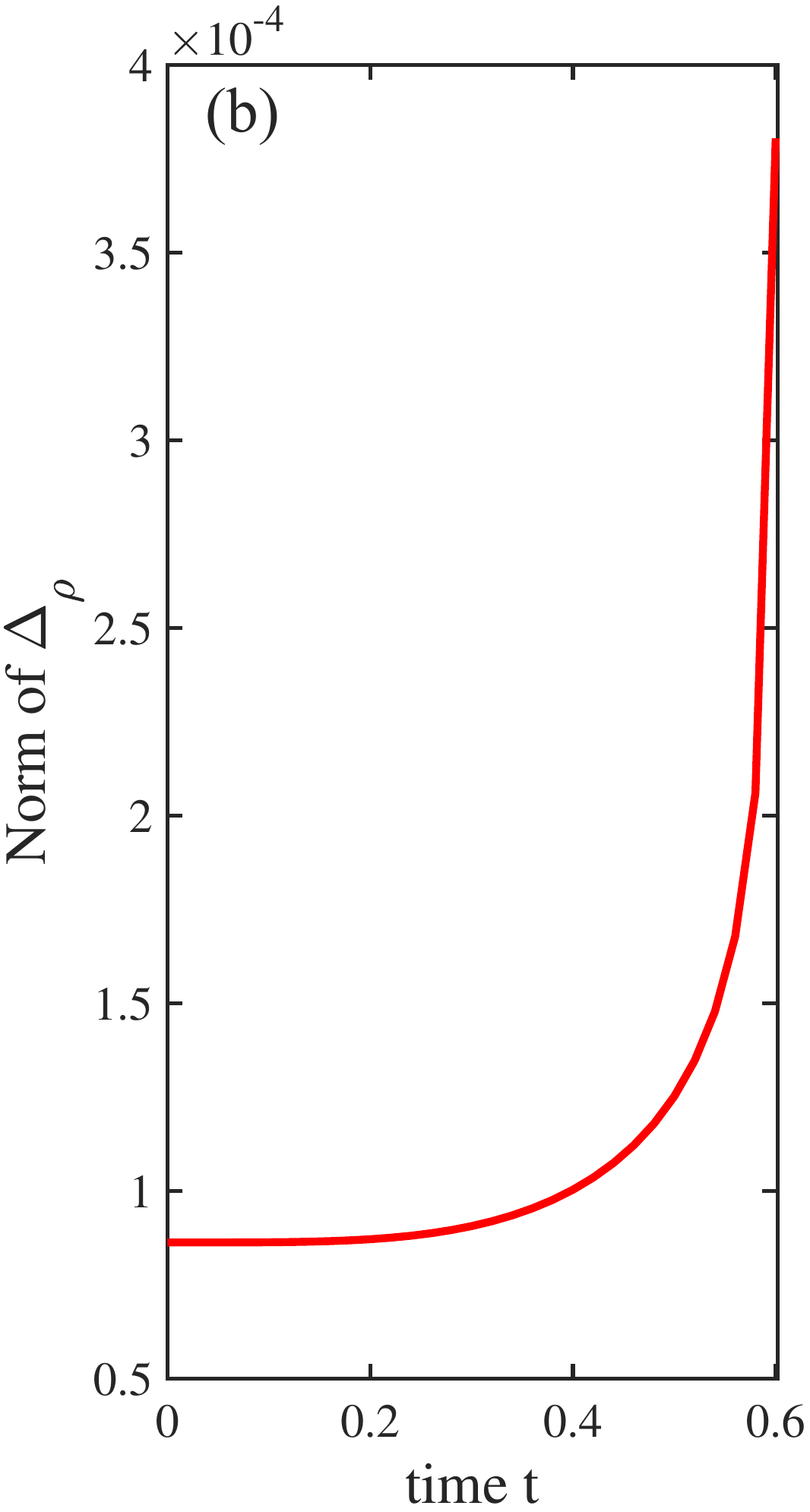}}\label{broken1_magnus1_norm}\\
  \caption{The effects of quantum noises on the dynamics simulation of $\mathcal{PT}$-symmetry broken system ($\gamma=0.25,\theta=\pi/2,s=1,r=1.4$, and set time step $h=0.02$) under the initial state of pure-state density operator. Both $X$ and $Y$ axes are dimensionless. (a)  Renormalized population ${P_\mathrm{N}}_0$ under the first term $\Omega_1$ of Magnus series according to Eq.\eqref{Magnus_product}. (b) The error between the green curve and the blue dotted curve corresponding to (a).}
\label{Magnus_expansion—figs_broken}
\end{figure*}

The effects of quantum noises on the dynamics simulation of $\mathcal{PT}$-symmetry broken system under the initial state of pure-state density operator $\rho_S=\frac{1}{5}|0\rangle_S\langle0|$ are given in Fig.\ref{Magnus_expansion—figs_broken}, and the renormalized population ${P_\mathrm{N}}_0$ is calculated to the first term of Magnus series $\Omega_1$. In this figure, we set the parameter as $\gamma=0.25,\theta=\pi/2,s=1,r=1.4$, then the legitimacy will be lost after the critical time $T_l\approx0.604$. Under this parameter configuration, $\max_{t\in[0, 0.6]}\|H_{AS}(t)\|_2\approx14.20$, $\max_{t\in[0, 0,6]}\|\overline{\overline{\mathcal{L}}}_{AD}(t)\|_2\approx28.40$, $\max_{t\in[0, 0.6]}\|\overline{\overline{\mathcal{L}}}_{PD}(t)\|_2\approx28.40$, $\max_{t\in[0, 0.6]}\|\overline{\overline{\mathcal{L}}}_{Dep}(t)\|_2\approx28.41$, so when we set time step $h=0.02$, the convergence condition, i.e., Eq.\eqref{convergence2}, will always be satisfied in every step because  $h<T_c$ (including noises case and no-noise case). From Fig.\ref{Magnus_expansion—figs_broken}(a) we can see that all curves decrease monotonically because the eigenvalues of $H_S$ will be complex numbers, which can be seen from Fig.\ref{magnus3_eigenvalue}, so the evolution operator $\mathbb{T}e^{-i\int_{0}^{t}H_S(\tau)d\tau}$ may cause the decay of model $|0\rangle_S$. In addition, the blue curve and the green curve almost completely coincide in the whole interval $t\in[0,T_l)$, which can be seen more clearly in the error diagram given in Fig.\ref{Magnus_expansion—figs_broken}(b), which shows high accuracy.

\section{Conclusions and discussions\label{conclusions}}

In this work, we generalized the results of Wu et al. in Ref.\cite{wu2019observation}, which are based on the dilation method, from the pure-state vectors case to the mixed-state case with the help of density operators, and provided a general theoretical framework based on density operators to analytically and numerically analyze the dynamics of TD arbitrary $\mathcal{PT}$-symmetric system and the influence of quantum noises. We make conclusions from the perspective of analytical analysis and numerical analysis, respectively.

At first, from the perspective of analytical analysis, more physical completeness was provided. In the process of derivation, we discussed the physical meaning of $M(t)$ ignored in Ref.\cite{wu2019observation}. Specifically, we proved that $M(t)$ is not the metric operator of $\mathcal{PT}$-symmetric system $H_S$, but the TD metric operator of $M(t)$-inner product space, which satisfies probability conservation. Meanwhile, we also gave a quantity $h_{S}(t)$ related to $M(t)$, and proved that it actually is a physical observable, because it can be mapped to the Hermitian quantity $H_{\mathrm{phys}}$, which has a real eigenspectrum, through a TD similarity transformation, i.e., the TD Dyson map. In addition, more mathematical completeness was also provided by us. Specifically, in the derivation, we obtained the dilated Hamiltonian $H_{AS}(t)$ by attaching a symmetric gauge instead of artificially assigning a quantity to the free variable as in Ref.\cite{wu2019observation}. As a result, the hidden symmetry of the eigenspectrum of the dilated Hamiltonian $H_{AS}(t)$ was able to be revealed. It is worth noting that when the system considered is time-independent and $\mathcal{PT}$-symmetry unbroken, the results of dynamics simulation in this work are consistent with our previous results in Ref.\cite{Li2022}, and when the state considered is pure state, the results of this work are consistent with the results given in Ref.\cite{wu2019observation}. Because the dilated system $\hat{H}_{AS}(t)$ is actually located in an open quantum system, the influence of environment will be inevitable, we introduced the tool of VDMS to solve the Lindblad master equation under three kinds of quantum noises, then the influence of quantum noises can be studied in the dynamics simulation of $\mathcal{PT}$-symmetric system.

Then, from the perspective of numerical analysis, we pointed out that on the premise of meeting the goal of calculation accuracy and saving computing resources, the time step $h$ of calculation and the cut-off term of Magnus series have to be carefully balanced. In addition, the time step $h$ of the numerical calculation should be restricted by the critical time  $T_c$ of convergence of Magnus series in every step, because the dilated higher-dimensional Hamiltonian $\hat{H}_{AS}(t)$ is usually time-dependent, the problem of chronological product has to be solved, and the Magnus series has to be calculated, which may diverge when $t>T_c$ so that the error may be amplified after Magnus series is truncated to high-order terms in calculation. Meanwhile, the implemented duration of experimental running is actually bounded by the critical time $T_l$ of the legitimacy of dilation method. This phenomenon occurs because when $t\rightarrow T_l$, at least one of eigenvalues of $M(t)$ given in Eq.\eqref{M_TD} will be close to one, then the corresponding eigenvalue of $\xi(t)$ given in Eq.\eqref{xi} will be close to zero, so the energy may diverge (see Fig.\ref{magnus3_eigenvalue}(b)). In fact, the problem of chronological product has to be solved not only in the numerical calculation, but also even in the experiment, because $\hat{H}_{AS}(t)$ has to be parameterized in advance by calculating it. When considering the influence of quantum noises, according to the results of the numerical calculation in our example, we know the depolarizing noise is the most fatal to the dynamics simulation of $\mathcal{PT}$-symmetric system among three kinds of quantum noises we considered and should be avoided as much as possible.

Finally, we have to mention that in Sec.\ref{example}, we actually give an example of time-independent Hamiltonian rather that TD Hamiltonian, just like in Ref.\cite{wu2019observation}, for the purpose of facilitate display and comparison. However, the method of analysis and calculation is universal, and what only needs additional attention is to be careful about the convergence of chronological product when calculating $M(t)$ given in Eq.\eqref{M_TD}, which has been avoid in our example.

\section{Acknowledgements}
We acknowledge the National Key Research and Development Program of China (Grant No. 2017YFA0303700), Beijing Advanced Innovation Center for Future Chip (ICFC), Tsinghua University Initiative Scientific Research Program, and the National Natural Science Foundation of China (Grant No. 11974205). C.Z. thanks the National Natural Science Foundation of China (Grants No. 12175002 and No. 11705004), the Beijing Natural Science Foundation (Grant No. 1222020), and NCUT Talents Project and Special Fund.

\appendix
\section{Derivations of $H_2(t)$ and $H_4(t)$ \label{appendix_derivation-H_24}}
Here we drive $H_2(t)$, $H_4(t)$ expressed by the operator $K(t)$ in the main text. First of all, we know
\begin{equation}
  M'(t)=(\xi^2(t)+I)'=\xi'(t)\xi(t)+\xi(t)\xi'(t),
\end{equation}
then according to Eq.\eqref{H_2},
\begin{widetext}
\begin{align}\label{H_2-app}
H_2(t)=&[-i\xi'(t)+H_S(t)\xi(t)-\xi(t)H_S(t)]M^{-1}(t)\xi(t) \nonumber \\
      =&-i\xi'(t)M^{-1}(t)+H_S(t)M^{-1}(t)\cdot\xi(t)-\xi(t)H_S(t)M^{-1}(t) \nonumber \\
      =&-i\xi'(t)M^{-1}(t)+[K(t)-\frac{i}{2}M^{-1}M'(t)M^{-1}(t)]\xi(t)-\xi(t)[K(t)-\frac{i}{2}M^{-1}M'(t)M^{-1}(t)] \nonumber \\
      =&K(t)\xi(t)-\xi(t)K(t)-\frac{i}{2}[2\xi'(t)M^{-1}(t)+M^{-1}(t)M'(t)\xi(t)M^{-1}(t)-M^{-1}\xi(t)M'(t)M^{-1}(t)] \nonumber \\
      =&K(t)\xi(t)\!-\!\xi(t)K(t)\!-\!\frac{i}{2}\{2\xi'(t)M^{-1}(t)\!+\!M^{-1}(t)[\xi'(t)\xi(t)\!+\!\xi(t)\xi'(t)]\xi(t)M^{-1}(t)-\nonumber\\
       &M^{-1}\xi(t)[\xi'(t)\xi(t)\!+\!\xi(t)\xi'(t)]M^{-1}(t)\}\nonumber\\
      =&K(t)\xi(t)-\xi(t)K(t)-\frac{i}{2}\{2\xi'(t)M^{-1}(t)+M^{-1}(t)[\xi'(t)(M(t)-I)M^{-1}(t)-M^{-1}(t)(M(t)-I)\xi'(t)M^{-1}\} \nonumber\\
      =&K(t)\xi(t)-\xi(t)K(t)-\frac{i}{2}[\xi'(t)M^{-1}(t)-M^{-1}(t)\xi'(t)].
\end{align}
\end{widetext}
In the similar way above, according to Eq.\eqref{H_4},
\begin{widetext}
\begin{align}\label{H_4-app}
  H_4(t)=&[i\xi'(t)\xi(t)+\xi(t)H_S(t)\xi(t)+H_S(t)]M^{-1}(t) \nonumber \\
        =&i\xi'(t)\xi(t)M^{-1}(t)+H_S(t)M^{-1}(t)+\xi(t)H_S(t)M^{-1}(t)\xi(t) \nonumber \\
        =&i\xi'(t)\xi(t)M^{-1}(t)+K(t)-\frac{i}{2}M^{-1}M'(t)M^{-1}(t)+\xi(t)[K(t)-\frac{i}{2}M^{-1}M'(t)M^{-1}(t)]\xi(t) \nonumber \\
        =&K(t)+\xi(t)K(t)\xi(t)+\frac{i}{2}M^{-1}(t)[2M(t)\xi'(t)\xi(t)-M'(t)-\xi(t)M'(t)\xi(t)]M^{-1}(t) \nonumber \\
        =&K(t)+\xi(t)K(t)\xi(t)+\frac{i}{2}M^{-1}(t)\{2M(t)\xi'(t)\xi(t)-[\xi'(t)\xi(t)+\xi(t)\xi'(t)]-\xi(t)[\xi'(t)\xi(t)+\xi(t)\xi'(t)]\xi(t)\}M^{-1}(t) \nonumber \\
        =&K(t)+\xi(t)K(t)\xi(t)\!+\!\frac{i}{2}M^{-1}(t)\{2M(t)\xi'(t)\xi(t)\!-\![\xi'(t)\xi(t)+\xi(t)\xi'(t)]\!-\!\xi(t)\xi'(t)[M(t)\!-\!I]-\nonumber\\
         &[M(t)\!-\!I]\xi'(t)\xi(t)\}M^{-1}(t) \nonumber \\
        =&K(t)+\xi(t)K(t)\xi(t)+\frac{i}{2}M^{-1}(t)[M(t)\xi'(t)\xi(t)-\xi(t)\xi'(t)M(t)]M^{-1}(t) \nonumber \\
        =&K(t)+\xi(t)K(t)\xi(t)+\frac{i}{2}[\xi'(t)\xi(t)M^{-1}-M^{-1}\xi(t)\xi'(t)].
\end{align}
\end{widetext}

\section{Problem of chronological product\label{chronological product}}

Considering a matrix differential equation \cite{Magnus1954}:
\begin{align}\label{matrix-differential-equation}
Y^{\prime}(t)=A(t) Y(t), \quad Y\left(t_{0}\right)=Y_{0},
\end{align}
where $A(t)$ is a known time-dependent matrix, $Y_{0}$ is the initial value of $Y(t)$, and $Y(t)$ is the matrix to be solved. The formal solution of the above equation is \cite{Blanes1998,Blanes2009}:
\begin{align}\label{formal solution}
Y(t)=\mathbb{T}\exp \left(\int_{t_{0}}^{t} A(s) d s\right) Y_{0},
\end{align}
where $\mathbb{T}$ is the time-ordering operator. For arbitrary two time $t_1$ and $t_2$ ($t_1\neq t_2$), in general, $[A(t_1), A(t_2)]\neq0$, then $e^{A(t_1)+A(t_2)}\neq e^{A(t_1)}\cdot e^{A(t_2)}$, the symbol $\mathbb{T}$ can not be ignored. When $[A(t_1), A(t_2)]=0$ for arbitrary two time $t_1$ and $t_2$, especially when $A$ is time-independent, the symbol $\mathbb{T}$ can be ignored.

The Eq.\eqref{formal solution} can be expressed as \cite{Blanes2009}:
\begin{align}
Y(t)=\exp \left(\Omega\left(t, t_{0}\right)\right) Y_{0},
\end{align}
where $\Omega(t)$ can be written as the sum of series:
\begin{align}\label{Magnus series}
\Omega(t)=\sum_{k=1}^{\infty} \Omega_{k}(t),
\end{align}
and $\Omega_n(t)$ is the $n$-th term of Magnus series.  Magnus points out that the differential of $\Omega$ with respect to $t$ can be written as:
\begin{align}
\Omega^{\prime}=\frac{\operatorname{ad}_{\Omega}}{\exp \left(\operatorname{ad}_{\Omega}\right)-1} A,
\end{align}
so the solutions of the above equation constitute Magnus expansion, or Magnus series.

The term $\Omega_{n}$ can be obtained by $S_{n}^{(j)}$, which can be obtained by the following recursive formula:
\begin{align}
S_{n}^{(j)}&=\sum_{m=1}^{n-j}\left[\Omega_{m}, S_{n-m}^{(j-1)}\right], \quad 2 \leq j \leq n-1 \nonumber\\
S_{n}^{(1)}&=\left[\Omega_{n-1}, A\right], \quad S_{n}^{(n-1)}=\operatorname{ad}_{\Omega_{1}}^{n-1}(A),
\end{align}
where $\operatorname{ad}_{\Omega}^{n}$ is a shorthand for an iterated commutator, and
\begin{align}
\operatorname{ad}_{\Omega}^{0} A=A, \quad \operatorname{ad}_{\Omega}^{k+1} A=\left[\Omega, \operatorname{ad}_{\Omega}^{k} A\right].
\end{align}
 For convenience, we set $t_0=0$. Therefore, we can get
\begin{align}
\Omega_{1} &=\int_{0}^{t} A(\tau) d \tau \nonumber \\
\Omega_{n} &=\sum_{j=1}^{n-1} \frac{B_{j}}{j !} \int_{0}^{t} S_{n}^{(j)}(\tau) d \tau, \quad n \geq 2,
\end{align}
where $B_j$ are the Bernoulli numbers, and $B_1=-1/2$. For the convenience of using, we write out the first four terms of $\Omega_{n}$ as follows:
\begin{align}\label{Magnus series1-4}
\begin{aligned}
\Omega_{1}(t)=& \int_{0}^{t} A\left(t_{1}\right) d t_{1} \\
\Omega_{2}(t)=& \frac{1}{2} \int_{0}^{t} d t_{1} \int_{0}^{t_{1}} d t_{2}\left[A\left(t_{1}\right), A\left(t_{2}\right)\right] \\
\Omega_{3}(t)=& \frac{1}{6} \int_{0}^{t} d t_{1} \int_{0}^{t_{1}} d t_{2} \int_{0}^{t_{2}} d t_{3}\cdot\\
&\left(\left[A\left(t_{1}\right),\left[A\left(t_{2}\right), A\left(t_{3}\right)\right]\right]+\left[A\left(t_{3}\right),\left[A\left(t_{2}\right), A\left(t_{1}\right)\right]\right]\right), \\
\Omega_{4}(t)=& \frac{1}{12} \int_{0}^{t} d t_{1} \int_{0}^{t_{1}} d t_{2} \int_{0}^{t_{2}} d t_{3} \int_{0}^{t_{3}} d t_{4}\left(\left[\left[\left[A_{1}, A_{2}\right], A_{3}\right], A_{4}\right]\right.\\
&+\left[A_{1},\left[\left[A_{2}, A_{3}\right], A_{4}\right]\right]+\left[A_{1},\left[A_{2},\left[A_{3}, A_{4}\right]\right]\right]+\\
&\left[A_{2},\left[A_{3},\left[A_{4}, A_{1}\right]\right]\right]).
\end{aligned}
\end{align}
It is worth noting that the Magnus series in Eq.\eqref{Magnus series} may diverge \cite{Blanes2009}, and a sufficient condition for it to converge in $t\in[0, T)$ is :
\begin{align}
\int_{0}^{T}\|A(s)\|_{2} \mathrm{d}s<\pi,
\end{align}
where $\|A\|_{2}$ denotes 2-norm of $A$.

\bibliography{PT_sim_TD.bib}

\begin{thebibliography}{65}%
\makeatletter
\providecommand \@ifxundefined [1]{%
 \@ifx{#1\undefined}
}%
\providecommand \@ifnum [1]{%
 \ifnum #1\expandafter \@firstoftwo
 \else \expandafter \@secondoftwo
 \fi
}%
\providecommand \@ifx [1]{%
 \ifx #1\expandafter \@firstoftwo
 \else \expandafter \@secondoftwo
 \fi
}%
\providecommand \natexlab [1]{#1}%
\providecommand \enquote  [1]{``#1''}%
\providecommand \bibnamefont  [1]{#1}%
\providecommand \bibfnamefont [1]{#1}%
\providecommand \citenamefont [1]{#1}%
\providecommand \href@noop [0]{\@secondoftwo}%
\providecommand \href [0]{\begingroup \@sanitize@url \@href}%
\providecommand \@href[1]{\@@startlink{#1}\@@href}%
\providecommand \@@href[1]{\endgroup#1\@@endlink}%
\providecommand \@sanitize@url [0]{\catcode `\\12\catcode `\$12\catcode
  `\&12\catcode `\#12\catcode `\^12\catcode `\_12\catcode `\%12\relax}%
\providecommand \@@startlink[1]{}%
\providecommand \@@endlink[0]{}%
\providecommand \url  [0]{\begingroup\@sanitize@url \@url }%
\providecommand \@url [1]{\endgroup\@href {#1}{\urlprefix }}%
\providecommand \urlprefix  [0]{URL }%
\providecommand \Eprint [0]{\href }%
\providecommand \doibase [0]{https://doi.org/}%
\providecommand \selectlanguage [0]{\@gobble}%
\providecommand \bibinfo  [0]{\@secondoftwo}%
\providecommand \bibfield  [0]{\@secondoftwo}%
\providecommand \translation [1]{[#1]}%
\providecommand \BibitemOpen [0]{}%
\providecommand \bibitemStop [0]{}%
\providecommand \bibitemNoStop [0]{.\EOS\space}%
\providecommand \EOS [0]{\spacefactor3000\relax}%
\providecommand \BibitemShut  [1]{\csname bibitem#1\endcsname}%
\let\auto@bib@innerbib\@empty
\bibitem [{\citenamefont {Griffiths}\ and\ \citenamefont
  {Schroeter}(2018)}]{Griffiths2018}%
  \BibitemOpen
  \bibfield  {author} {\bibinfo {author} {\bibfnamefont {D.~J.}\ \bibnamefont
  {Griffiths}}\ and\ \bibinfo {author} {\bibfnamefont {D.~F.}\ \bibnamefont
  {Schroeter}},\ }\href@noop {} {\emph {\bibinfo {title} {Introduction to
  quantum mechanics}}}\ (\bibinfo  {publisher} {Cambridge university press},\
  \bibinfo {year} {2018})\BibitemShut {NoStop}%
\bibitem [{\citenamefont {Bender}\ and\ \citenamefont
  {Boettcher}(1998)}]{Bender1998}%
  \BibitemOpen
  \bibfield  {author} {\bibinfo {author} {\bibfnamefont {C.~M.}\ \bibnamefont
  {Bender}}\ and\ \bibinfo {author} {\bibfnamefont {S.}~\bibnamefont
  {Boettcher}},\ }\bibfield  {title} {\bibinfo {title} {Real spectra in
  non-hermitian hamiltonians having $\mathcal{PT}$ symmetry},\ }\href
  {https://doi.org/10.1103/PhysRevLett.80.5243} {\bibfield  {journal} {\bibinfo
   {journal} {Phys. Rev. Lett.}\ }\textbf {\bibinfo {volume} {80}},\ \bibinfo
  {pages} {5243} (\bibinfo {year} {1998})}\BibitemShut {NoStop}%
\bibitem [{\citenamefont {Bender}\ \emph {et~al.}(1999)\citenamefont {Bender},
  \citenamefont {Boettcher},\ and\ \citenamefont {Meisinger}}]{Bender1999}%
  \BibitemOpen
  \bibfield  {author} {\bibinfo {author} {\bibfnamefont {C.~M.}\ \bibnamefont
  {Bender}}, \bibinfo {author} {\bibfnamefont {S.}~\bibnamefont {Boettcher}},\
  and\ \bibinfo {author} {\bibfnamefont {P.~N.}\ \bibnamefont {Meisinger}},\
  }\bibfield  {title} {\bibinfo {title} {Pt-symmetric quantum mechanics},\
  }\href {https://doi.org/10.1063/1.532860} {\bibfield  {journal} {\bibinfo
  {journal} {J. Math. Phys.}\ }\textbf {\bibinfo {volume} {40}},\ \bibinfo
  {pages} {2201} (\bibinfo {year} {1999})}\BibitemShut {NoStop}%
\bibitem [{\citenamefont {Bender}\ \emph {et~al.}(2002)\citenamefont {Bender},
  \citenamefont {Brody},\ and\ \citenamefont {Jones}}]{Bender2002}%
  \BibitemOpen
  \bibfield  {author} {\bibinfo {author} {\bibfnamefont {C.~M.}\ \bibnamefont
  {Bender}}, \bibinfo {author} {\bibfnamefont {D.~C.}\ \bibnamefont {Brody}},\
  and\ \bibinfo {author} {\bibfnamefont {H.~F.}\ \bibnamefont {Jones}},\
  }\bibfield  {title} {\bibinfo {title} {Complex extension of quantum
  mechanics},\ }\href {https://doi.org/10.1103/PhysRevLett.89.270401}
  {\bibfield  {journal} {\bibinfo  {journal} {Phys. Rev. Lett.}\ }\textbf
  {\bibinfo {volume} {89}},\ \bibinfo {pages} {270401} (\bibinfo {year}
  {2002})}\BibitemShut {NoStop}%
\bibitem [{\citenamefont {Bender}\ \emph {et~al.}(2004)\citenamefont {Bender},
  \citenamefont {Brody},\ and\ \citenamefont {Jones}}]{Bender2004}%
  \BibitemOpen
  \bibfield  {author} {\bibinfo {author} {\bibfnamefont {C.~M.}\ \bibnamefont
  {Bender}}, \bibinfo {author} {\bibfnamefont {D.~C.}\ \bibnamefont {Brody}},\
  and\ \bibinfo {author} {\bibfnamefont {H.~F.}\ \bibnamefont {Jones}},\
  }\bibfield  {title} {\bibinfo {title} {Erratum: Complex extension of quantum
  mechanics [phys. rev. lett. 89, 270401 (2002)]},\ }\href
  {https://doi.org/10.1103/PhysRevLett.92.119902} {\bibfield  {journal}
  {\bibinfo  {journal} {Phys. Rev. Lett.}\ }\textbf {\bibinfo {volume} {92}},\
  \bibinfo {pages} {119902} (\bibinfo {year} {2004})}\BibitemShut {NoStop}%
\bibitem [{\citenamefont
  {Mostafazadeh}(2002{\natexlab{a}})}]{Mostafazadeh2002b}%
  \BibitemOpen
  \bibfield  {author} {\bibinfo {author} {\bibfnamefont {A.}~\bibnamefont
  {Mostafazadeh}},\ }\bibfield  {title} {\bibinfo {title} {Pseudo-hermiticity
  versus pt symmetry: The necessary condition for the reality of the spectrum
  of a non-hermitian hamiltonian},\ }\href {https://doi.org/10.1063/1.1418246}
  {\bibfield  {journal} {\bibinfo  {journal} {J. Math. Phys.}\ }\textbf
  {\bibinfo {volume} {43}},\ \bibinfo {pages} {205} (\bibinfo {year}
  {2002}{\natexlab{a}})}\BibitemShut {NoStop}%
\bibitem [{\citenamefont
  {Mostafazadeh}(2002{\natexlab{b}})}]{Mostafazadeh2002a}%
  \BibitemOpen
  \bibfield  {author} {\bibinfo {author} {\bibfnamefont {A.}~\bibnamefont
  {Mostafazadeh}},\ }\bibfield  {title} {\bibinfo {title} {Pseudo-hermiticity
  versus pt-symmetry. ii. a complete characterization of non-hermitian
  hamiltonians with a real spectrum},\ }\href
  {https://doi.org/10.1063/1.1461427} {\bibfield  {journal} {\bibinfo
  {journal} {J. Math. Phys.}\ }\textbf {\bibinfo {volume} {43}},\ \bibinfo
  {pages} {2814} (\bibinfo {year} {2002}{\natexlab{b}})}\BibitemShut {NoStop}%
\bibitem [{\citenamefont
  {Mostafazadeh}(2002{\natexlab{c}})}]{Mostafazadeh2002}%
  \BibitemOpen
  \bibfield  {author} {\bibinfo {author} {\bibfnamefont {A.}~\bibnamefont
  {Mostafazadeh}},\ }\bibfield  {title} {\bibinfo {title} {Pseudo-hermiticity
  versus pt-symmetry iii: Equivalence of pseudo-hermiticity and the presence of
  antilinear symmetries},\ }\href {https://doi.org/10.1063/1.1489072}
  {\bibfield  {journal} {\bibinfo  {journal} {J. Math. Phys.}\ }\textbf
  {\bibinfo {volume} {43}},\ \bibinfo {pages} {3944} (\bibinfo {year}
  {2002}{\natexlab{c}})}\BibitemShut {NoStop}%
\bibitem [{\citenamefont {Mostafazadeh}(2003)}]{Mostafazadeh2003a}%
  \BibitemOpen
  \bibfield  {author} {\bibinfo {author} {\bibfnamefont {A.}~\bibnamefont
  {Mostafazadeh}},\ }\bibfield  {title} {\bibinfo {title} {Pseudo-hermiticity
  and generalized pt- and cpt-symmetries},\ }\href
  {https://doi.org/10.1063/1.1539304} {\bibfield  {journal} {\bibinfo
  {journal} {J. Math. Phys.}\ }\textbf {\bibinfo {volume} {44}},\ \bibinfo
  {pages} {974} (\bibinfo {year} {2003})}\BibitemShut {NoStop}%
\bibitem [{\citenamefont
  {Mostafazadeh}(2007{\natexlab{a}})}]{Mostafazadeh2007a}%
  \BibitemOpen
  \bibfield  {author} {\bibinfo {author} {\bibfnamefont {A.}~\bibnamefont
  {Mostafazadeh}},\ }\bibfield  {title} {\bibinfo {title} {Time-dependent
  pseudo-hermitian hamiltonians defining a unitary quantum system and
  uniqueness of the metric operator},\ }\href
  {https://www.sciencedirect.com/science/article/abs/pii/S0370269307005485}
  {\bibfield  {journal} {\bibinfo  {journal} {Phys. Lett. B}\ }\textbf
  {\bibinfo {volume} {650}},\ \bibinfo {pages} {208} (\bibinfo {year}
  {2007}{\natexlab{a}})}\BibitemShut {NoStop}%
\bibitem [{\citenamefont {Curtright}\ and\ \citenamefont
  {Mezincescu}(2007)}]{Curtright2007}%
  \BibitemOpen
  \bibfield  {author} {\bibinfo {author} {\bibfnamefont {T.}~\bibnamefont
  {Curtright}}\ and\ \bibinfo {author} {\bibfnamefont {L.}~\bibnamefont
  {Mezincescu}},\ }\bibfield  {title} {\bibinfo {title} {Biorthogonal quantum
  systems},\ }\href {https://doi.org/10.1063/1.2196243} {\bibfield  {journal}
  {\bibinfo  {journal} {J. Math. Phys.}\ }\textbf {\bibinfo {volume} {48}},\
  \bibinfo {pages} {092106} (\bibinfo {year} {2007})}\BibitemShut {NoStop}%
\bibitem [{\citenamefont {Brody}(2013)}]{Brody2013}%
  \BibitemOpen
  \bibfield  {author} {\bibinfo {author} {\bibfnamefont {D.~C.}\ \bibnamefont
  {Brody}},\ }\bibfield  {title} {\bibinfo {title} {Biorthogonal quantum
  mechanics},\ }\href {https://doi.org/10.1088/1751-8113/47/3/035305}
  {\bibfield  {journal} {\bibinfo  {journal} {J. Phys. A: Math. Theor.}\
  }\textbf {\bibinfo {volume} {47}},\ \bibinfo {pages} {035305} (\bibinfo
  {year} {2013})}\BibitemShut {NoStop}%
\bibitem [{\citenamefont {Zhang}\ \emph {et~al.}(2020)\citenamefont {Zhang},
  \citenamefont {Qin},\ and\ \citenamefont {Xiao}}]{Zhang2020}%
  \BibitemOpen
  \bibfield  {author} {\bibinfo {author} {\bibfnamefont {R.}~\bibnamefont
  {Zhang}}, \bibinfo {author} {\bibfnamefont {H.}~\bibnamefont {Qin}},\ and\
  \bibinfo {author} {\bibfnamefont {J.}~\bibnamefont {Xiao}},\ }\bibfield
  {title} {\bibinfo {title} {Pt-symmetry entails pseudo-hermiticity regardless
  of diagonalizability},\ }\href {https://doi.org/10.1063/1.5117211} {\bibfield
   {journal} {\bibinfo  {journal} {J. Math. Phys.}\ }\textbf {\bibinfo {volume}
  {61}},\ \bibinfo {pages} {012101} (\bibinfo {year} {2020})}\BibitemShut
  {NoStop}%
\bibitem [{\citenamefont {Bender}(2007)}]{Bender2007}%
  \BibitemOpen
  \bibfield  {author} {\bibinfo {author} {\bibfnamefont {C.~M.}\ \bibnamefont
  {Bender}},\ }\bibfield  {title} {\bibinfo {title} {Making sense of
  non-hermitian hamiltonians},\ }\href
  {https://doi.org/10.1088/0034-4885/70/6/r03} {\bibfield  {journal} {\bibinfo
  {journal} {Rep. Prog. Phys.}\ }\textbf {\bibinfo {volume} {70}},\ \bibinfo
  {pages} {947} (\bibinfo {year} {2007})}\BibitemShut {NoStop}%
\bibitem [{\citenamefont {Mostafazadeh}(2010)}]{Mostafazadeh2010}%
  \BibitemOpen
  \bibfield  {author} {\bibinfo {author} {\bibfnamefont {A.}~\bibnamefont
  {Mostafazadeh}},\ }\bibfield  {title} {\bibinfo {title} {Pseudo-hermitian
  representation of quantum mechanics},\ }\href
  {https://doi.org/10.1142/S0219887810004816} {\bibfield  {journal} {\bibinfo
  {journal} {Int J Geom Methods Mod Phys}\ }\textbf {\bibinfo {volume} {07}},\
  \bibinfo {pages} {1191} (\bibinfo {year} {2010})}\BibitemShut {NoStop}%
\bibitem [{\citenamefont {Makris}\ \emph {et~al.}(2008)\citenamefont {Makris},
  \citenamefont {El-Ganainy}, \citenamefont {Christodoulides},\ and\
  \citenamefont {Musslimani}}]{Makris2008}%
  \BibitemOpen
  \bibfield  {author} {\bibinfo {author} {\bibfnamefont {K.~G.}\ \bibnamefont
  {Makris}}, \bibinfo {author} {\bibfnamefont {R.}~\bibnamefont {El-Ganainy}},
  \bibinfo {author} {\bibfnamefont {D.~N.}\ \bibnamefont {Christodoulides}},\
  and\ \bibinfo {author} {\bibfnamefont {Z.~H.}\ \bibnamefont {Musslimani}},\
  }\bibfield  {title} {\bibinfo {title} {Beam dynamics in
  $\mathcal{P}\mathcal{T}$ symmetric optical lattices},\ }\href
  {https://doi.org/10.1103/PhysRevLett.100.103904} {\bibfield  {journal}
  {\bibinfo  {journal} {Phys. Rev. Lett.}\ }\textbf {\bibinfo {volume} {100}},\
  \bibinfo {pages} {103904} (\bibinfo {year} {2008})}\BibitemShut {NoStop}%
\bibitem [{\citenamefont {R{\"u}ter}\ \emph {et~al.}(2010)\citenamefont
  {R{\"u}ter}, \citenamefont {Makris}, \citenamefont {El-Ganainy},
  \citenamefont {Christodoulides}, \citenamefont {Segev},\ and\ \citenamefont
  {Kip}}]{Rueter2010}%
  \BibitemOpen
  \bibfield  {author} {\bibinfo {author} {\bibfnamefont {C.~E.}\ \bibnamefont
  {R{\"u}ter}}, \bibinfo {author} {\bibfnamefont {K.~G.}\ \bibnamefont
  {Makris}}, \bibinfo {author} {\bibfnamefont {R.}~\bibnamefont {El-Ganainy}},
  \bibinfo {author} {\bibfnamefont {D.~N.}\ \bibnamefont {Christodoulides}},
  \bibinfo {author} {\bibfnamefont {M.}~\bibnamefont {Segev}},\ and\ \bibinfo
  {author} {\bibfnamefont {D.}~\bibnamefont {Kip}},\ }\bibfield  {title}
  {\bibinfo {title} {Observation of parity--time symmetry in optics},\
  }\href@noop {} {\bibfield  {journal} {\bibinfo  {journal} {Nature physics}\
  }\textbf {\bibinfo {volume} {6}},\ \bibinfo {pages} {192} (\bibinfo {year}
  {2010})}\BibitemShut {NoStop}%
\bibitem [{\citenamefont {Peng}\ \emph {et~al.}(2014)\citenamefont {Peng},
  \citenamefont {Ozdemir}, \citenamefont {Lei}, \citenamefont {Monifi},
  \citenamefont {Gianfreda}, \citenamefont {Long}, \citenamefont {Fan},
  \citenamefont {Nori}, \citenamefont {Bender},\ and\ \citenamefont
  {Yang}}]{Peng2014}%
  \BibitemOpen
  \bibfield  {author} {\bibinfo {author} {\bibfnamefont {B.}~\bibnamefont
  {Peng}}, \bibinfo {author} {\bibfnamefont {S.~K.}\ \bibnamefont {Ozdemir}},
  \bibinfo {author} {\bibfnamefont {F.}~\bibnamefont {Lei}}, \bibinfo {author}
  {\bibfnamefont {F.}~\bibnamefont {Monifi}}, \bibinfo {author} {\bibfnamefont
  {M.}~\bibnamefont {Gianfreda}}, \bibinfo {author} {\bibfnamefont {G.~L.}\
  \bibnamefont {Long}}, \bibinfo {author} {\bibfnamefont {S.}~\bibnamefont
  {Fan}}, \bibinfo {author} {\bibfnamefont {F.}~\bibnamefont {Nori}}, \bibinfo
  {author} {\bibfnamefont {C.~M.}\ \bibnamefont {Bender}},\ and\ \bibinfo
  {author} {\bibfnamefont {L.}~\bibnamefont {Yang}},\ }\bibfield  {title}
  {\bibinfo {title} {Parity-time-symmetric whispering-gallery microcavities},\
  }\href {https://doi.org/10.1038/nphys2927} {\bibfield  {journal} {\bibinfo
  {journal} {Nature Physics}\ }\textbf {\bibinfo {volume} {10}},\ \bibinfo
  {pages} {394} (\bibinfo {year} {2014})}\BibitemShut {NoStop}%
\bibitem [{\citenamefont {Assawaworrarit}\ \emph {et~al.}(2017)\citenamefont
  {Assawaworrarit}, \citenamefont {Yu},\ and\ \citenamefont
  {Fan}}]{Assawaworrarit2017}%
  \BibitemOpen
  \bibfield  {author} {\bibinfo {author} {\bibfnamefont {S.}~\bibnamefont
  {Assawaworrarit}}, \bibinfo {author} {\bibfnamefont {X.}~\bibnamefont {Yu}},\
  and\ \bibinfo {author} {\bibfnamefont {S.}~\bibnamefont {Fan}},\ }\bibfield
  {title} {\bibinfo {title} {Robust wireless power transfer using a nonlinear
  parity-time-symmetric circuit},\ }\href {https://doi.org/10.1038/nature22404}
  {\bibfield  {journal} {\bibinfo  {journal} {Nature}\ }\textbf {\bibinfo
  {volume} {546}},\ \bibinfo {pages} {387} (\bibinfo {year}
  {2017})}\BibitemShut {NoStop}%
\bibitem [{\citenamefont {Ashida}\ \emph {et~al.}(2017)\citenamefont {Ashida},
  \citenamefont {Furukawa},\ and\ \citenamefont {Ueda}}]{Ashida2017}%
  \BibitemOpen
  \bibfield  {author} {\bibinfo {author} {\bibfnamefont {Y.}~\bibnamefont
  {Ashida}}, \bibinfo {author} {\bibfnamefont {S.}~\bibnamefont {Furukawa}},\
  and\ \bibinfo {author} {\bibfnamefont {M.}~\bibnamefont {Ueda}},\ }\bibfield
  {title} {\bibinfo {title} {Parity-time-symmetric quantum critical
  phenomena},\ }\href {https://doi.org/10.1038/ncomms15791} {\bibfield
  {journal} {\bibinfo  {journal} {Nature Communications}\ }\textbf {\bibinfo
  {volume} {8}},\ \bibinfo {pages} {15791} (\bibinfo {year}
  {2017})}\BibitemShut {NoStop}%
\bibitem [{\citenamefont {Couvreur}\ \emph {et~al.}(2017)\citenamefont
  {Couvreur}, \citenamefont {Jacobsen},\ and\ \citenamefont
  {Saleur}}]{Couvreur2017}%
  \BibitemOpen
  \bibfield  {author} {\bibinfo {author} {\bibfnamefont {R.}~\bibnamefont
  {Couvreur}}, \bibinfo {author} {\bibfnamefont {J.~L.}\ \bibnamefont
  {Jacobsen}},\ and\ \bibinfo {author} {\bibfnamefont {H.}~\bibnamefont
  {Saleur}},\ }\bibfield  {title} {\bibinfo {title} {Entanglement in nonunitary
  quantum critical spin chains},\ }\href
  {https://doi.org/10.1103/PhysRevLett.119.040601} {\bibfield  {journal}
  {\bibinfo  {journal} {Phys. Rev. Lett.}\ }\textbf {\bibinfo {volume} {119}},\
  \bibinfo {pages} {040601} (\bibinfo {year} {2017})}\BibitemShut {NoStop}%
\bibitem [{\citenamefont {Li}\ \emph {et~al.}(2019)\citenamefont {Li},
  \citenamefont {Harter}, \citenamefont {Liu}, \citenamefont {de~Melo},
  \citenamefont {Joglekar},\ and\ \citenamefont {Luo}}]{li2019observation}%
  \BibitemOpen
  \bibfield  {author} {\bibinfo {author} {\bibfnamefont {J.}~\bibnamefont
  {Li}}, \bibinfo {author} {\bibfnamefont {A.~K.}\ \bibnamefont {Harter}},
  \bibinfo {author} {\bibfnamefont {J.}~\bibnamefont {Liu}}, \bibinfo {author}
  {\bibfnamefont {L.}~\bibnamefont {de~Melo}}, \bibinfo {author} {\bibfnamefont
  {Y.~N.}\ \bibnamefont {Joglekar}},\ and\ \bibinfo {author} {\bibfnamefont
  {L.}~\bibnamefont {Luo}},\ }\bibfield  {title} {\bibinfo {title} {Observation
  of parity-time symmetry breaking transitions in a dissipative floquet system
  of ultracold atoms},\ }\href
  {https://www.nature.com/articles/s41467-019-08596-1} {\bibfield  {journal}
  {\bibinfo  {journal} {Nature communications}\ }\textbf {\bibinfo {volume}
  {10}},\ \bibinfo {pages} {1} (\bibinfo {year} {2019})}\BibitemShut {NoStop}%
\bibitem [{\citenamefont {Zhang}\ \emph
  {et~al.}(2019{\natexlab{a}})\citenamefont {Zhang}, \citenamefont {Sweeney},
  \citenamefont {Hsu}, \citenamefont {Yang}, \citenamefont {Stone},\ and\
  \citenamefont {Jiang}}]{Zhang2019a}%
  \BibitemOpen
  \bibfield  {author} {\bibinfo {author} {\bibfnamefont {M.}~\bibnamefont
  {Zhang}}, \bibinfo {author} {\bibfnamefont {W.}~\bibnamefont {Sweeney}},
  \bibinfo {author} {\bibfnamefont {C.~W.}\ \bibnamefont {Hsu}}, \bibinfo
  {author} {\bibfnamefont {L.}~\bibnamefont {Yang}}, \bibinfo {author}
  {\bibfnamefont {A.~D.}\ \bibnamefont {Stone}},\ and\ \bibinfo {author}
  {\bibfnamefont {L.}~\bibnamefont {Jiang}},\ }\bibfield  {title} {\bibinfo
  {title} {Quantum noise theory of exceptional point amplifying sensors},\
  }\href {https://doi.org/10.1103/PhysRevLett.123.180501} {\bibfield  {journal}
  {\bibinfo  {journal} {Phys. Rev. Lett.}\ }\textbf {\bibinfo {volume} {123}},\
  \bibinfo {pages} {180501} (\bibinfo {year} {2019}{\natexlab{a}})}\BibitemShut
  {NoStop}%
\bibitem [{\citenamefont {Chu}\ \emph {et~al.}(2020)\citenamefont {Chu},
  \citenamefont {Liu}, \citenamefont {Liu},\ and\ \citenamefont
  {Cai}}]{Chu2020}%
  \BibitemOpen
  \bibfield  {author} {\bibinfo {author} {\bibfnamefont {Y.}~\bibnamefont
  {Chu}}, \bibinfo {author} {\bibfnamefont {Y.}~\bibnamefont {Liu}}, \bibinfo
  {author} {\bibfnamefont {H.}~\bibnamefont {Liu}},\ and\ \bibinfo {author}
  {\bibfnamefont {J.}~\bibnamefont {Cai}},\ }\bibfield  {title} {\bibinfo
  {title} {Quantum sensing with a single-qubit pseudo-hermitian system},\
  }\href {https://doi.org/10.1103/PhysRevLett.124.020501} {\bibfield  {journal}
  {\bibinfo  {journal} {Phys. Rev. Lett.}\ }\textbf {\bibinfo {volume} {124}},\
  \bibinfo {pages} {020501} (\bibinfo {year} {2020})}\BibitemShut {NoStop}%
\bibitem [{\citenamefont {Yu}\ \emph {et~al.}(2020)\citenamefont {Yu},
  \citenamefont {Meng}, \citenamefont {Tang}, \citenamefont {Xu}, \citenamefont
  {Wang}, \citenamefont {Yin}, \citenamefont {Ke}, \citenamefont {Liu},
  \citenamefont {Li}, \citenamefont {Yang}, \citenamefont {Chen}, \citenamefont
  {Han}, \citenamefont {Li},\ and\ \citenamefont {Guo}}]{Yu2020}%
  \BibitemOpen
  \bibfield  {author} {\bibinfo {author} {\bibfnamefont {S.}~\bibnamefont
  {Yu}}, \bibinfo {author} {\bibfnamefont {Y.}~\bibnamefont {Meng}}, \bibinfo
  {author} {\bibfnamefont {J.-S.}\ \bibnamefont {Tang}}, \bibinfo {author}
  {\bibfnamefont {X.-Y.}\ \bibnamefont {Xu}}, \bibinfo {author} {\bibfnamefont
  {Y.-T.}\ \bibnamefont {Wang}}, \bibinfo {author} {\bibfnamefont
  {P.}~\bibnamefont {Yin}}, \bibinfo {author} {\bibfnamefont {Z.-J.}\
  \bibnamefont {Ke}}, \bibinfo {author} {\bibfnamefont {W.}~\bibnamefont
  {Liu}}, \bibinfo {author} {\bibfnamefont {Z.-P.}\ \bibnamefont {Li}},
  \bibinfo {author} {\bibfnamefont {Y.-Z.}\ \bibnamefont {Yang}}, \bibinfo
  {author} {\bibfnamefont {G.}~\bibnamefont {Chen}}, \bibinfo {author}
  {\bibfnamefont {Y.-J.}\ \bibnamefont {Han}}, \bibinfo {author} {\bibfnamefont
  {C.-F.}\ \bibnamefont {Li}},\ and\ \bibinfo {author} {\bibfnamefont {G.-C.}\
  \bibnamefont {Guo}},\ }\bibfield  {title} {\bibinfo {title} {Experimental
  investigation of quantum $\mathcal{P}\mathcal{T}$-enhanced sensor},\ }\href
  {https://doi.org/10.1103/PhysRevLett.125.240506} {\bibfield  {journal}
  {\bibinfo  {journal} {Phys. Rev. Lett.}\ }\textbf {\bibinfo {volume} {125}},\
  \bibinfo {pages} {240506} (\bibinfo {year} {2020})}\BibitemShut {NoStop}%
\bibitem [{\citenamefont {Croke}(2015)}]{Croke2015}%
  \BibitemOpen
  \bibfield  {author} {\bibinfo {author} {\bibfnamefont {S.}~\bibnamefont
  {Croke}},\ }\bibfield  {title} {\bibinfo {title} {$\mathcal{PT}$-symmetric
  hamiltonians and their application in quantum information},\ }\href
  {https://doi.org/10.1103/PhysRevA.91.052113} {\bibfield  {journal} {\bibinfo
  {journal} {Phys. Rev. A}\ }\textbf {\bibinfo {volume} {91}},\ \bibinfo
  {pages} {052113} (\bibinfo {year} {2015})}\BibitemShut {NoStop}%
\bibitem [{\citenamefont {Kawabata}\ \emph {et~al.}(2017)\citenamefont
  {Kawabata}, \citenamefont {Ashida},\ and\ \citenamefont
  {Ueda}}]{Kawabata2017}%
  \BibitemOpen
  \bibfield  {author} {\bibinfo {author} {\bibfnamefont {K.}~\bibnamefont
  {Kawabata}}, \bibinfo {author} {\bibfnamefont {Y.}~\bibnamefont {Ashida}},\
  and\ \bibinfo {author} {\bibfnamefont {M.}~\bibnamefont {Ueda}},\ }\bibfield
  {title} {\bibinfo {title} {Information retrieval and criticality in
  parity-time-symmetric systems},\ }\href
  {https://doi.org/10.1103/PhysRevLett.119.190401} {\bibfield  {journal}
  {\bibinfo  {journal} {Phys. Rev. Lett.}\ }\textbf {\bibinfo {volume} {119}},\
  \bibinfo {pages} {190401} (\bibinfo {year} {2017})}\BibitemShut {NoStop}%
\bibitem [{\citenamefont
  {Mostafazadeh}(2007{\natexlab{b}})}]{Mostafazadeh2007}%
  \BibitemOpen
  \bibfield  {author} {\bibinfo {author} {\bibfnamefont {A.}~\bibnamefont
  {Mostafazadeh}},\ }\bibfield  {title} {\bibinfo {title} {Quantum
  brachistochrone problem and the geometry of the state space in
  pseudo-hermitian quantum mechanics},\ }\href
  {https://doi.org/10.1103/PhysRevLett.99.130502} {\bibfield  {journal}
  {\bibinfo  {journal} {Phys. Rev. Lett.}\ }\textbf {\bibinfo {volume} {99}},\
  \bibinfo {pages} {130502} (\bibinfo {year} {2007}{\natexlab{b}})}\BibitemShut
  {NoStop}%
\bibitem [{\citenamefont {Bender}\ \emph {et~al.}(2007)\citenamefont {Bender},
  \citenamefont {Brody}, \citenamefont {Jones},\ and\ \citenamefont
  {Meister}}]{Bender2007a}%
  \BibitemOpen
  \bibfield  {author} {\bibinfo {author} {\bibfnamefont {C.~M.}\ \bibnamefont
  {Bender}}, \bibinfo {author} {\bibfnamefont {D.~C.}\ \bibnamefont {Brody}},
  \bibinfo {author} {\bibfnamefont {H.~F.}\ \bibnamefont {Jones}},\ and\
  \bibinfo {author} {\bibfnamefont {B.~K.}\ \bibnamefont {Meister}},\
  }\bibfield  {title} {\bibinfo {title} {Faster than hermitian quantum
  mechanics},\ }\href {https://doi.org/10.1103/PhysRevLett.98.040403}
  {\bibfield  {journal} {\bibinfo  {journal} {Phys. Rev. Lett.}\ }\textbf
  {\bibinfo {volume} {98}},\ \bibinfo {pages} {040403} (\bibinfo {year}
  {2007})}\BibitemShut {NoStop}%
\bibitem [{\citenamefont {G\"unther}\ and\ \citenamefont
  {Samsonov}(2008{\natexlab{a}})}]{Guenther2008}%
  \BibitemOpen
  \bibfield  {author} {\bibinfo {author} {\bibfnamefont {U.}~\bibnamefont
  {G\"unther}}\ and\ \bibinfo {author} {\bibfnamefont {B.~F.}\ \bibnamefont
  {Samsonov}},\ }\bibfield  {title} {\bibinfo {title}
  {$\mathcal{P}\mathcal{T}$-symmetric brachistochrone problem, lorentz boosts,
  and nonunitary operator equivalence classes},\ }\href
  {https://doi.org/10.1103/PhysRevA.78.042115} {\bibfield  {journal} {\bibinfo
  {journal} {Phys. Rev. A}\ }\textbf {\bibinfo {volume} {78}},\ \bibinfo
  {pages} {042115} (\bibinfo {year} {2008}{\natexlab{a}})}\BibitemShut
  {NoStop}%
\bibitem [{\citenamefont {G\"unther}\ and\ \citenamefont
  {Samsonov}(2008{\natexlab{b}})}]{Guenther2008a}%
  \BibitemOpen
  \bibfield  {author} {\bibinfo {author} {\bibfnamefont {U.}~\bibnamefont
  {G\"unther}}\ and\ \bibinfo {author} {\bibfnamefont {B.~F.}\ \bibnamefont
  {Samsonov}},\ }\bibfield  {title} {\bibinfo {title} {Naimark-dilated
  $\mathcal{P}\mathcal{T}$-symmetric brachistochrone},\ }\href
  {https://doi.org/10.1103/PhysRevLett.101.230404} {\bibfield  {journal}
  {\bibinfo  {journal} {Phys. Rev. Lett.}\ }\textbf {\bibinfo {volume} {101}},\
  \bibinfo {pages} {230404} (\bibinfo {year} {2008}{\natexlab{b}})}\BibitemShut
  {NoStop}%
\bibitem [{\citenamefont {Ramezani}\ \emph {et~al.}(2012)\citenamefont
  {Ramezani}, \citenamefont {Schindler}, \citenamefont {Ellis}, \citenamefont
  {G\"unther},\ and\ \citenamefont {Kottos}}]{Ramezani2012}%
  \BibitemOpen
  \bibfield  {author} {\bibinfo {author} {\bibfnamefont {H.}~\bibnamefont
  {Ramezani}}, \bibinfo {author} {\bibfnamefont {J.}~\bibnamefont {Schindler}},
  \bibinfo {author} {\bibfnamefont {F.~M.}\ \bibnamefont {Ellis}}, \bibinfo
  {author} {\bibfnamefont {U.}~\bibnamefont {G\"unther}},\ and\ \bibinfo
  {author} {\bibfnamefont {T.}~\bibnamefont {Kottos}},\ }\bibfield  {title}
  {\bibinfo {title} {Bypassing the bandwidth theorem with $\mathcal{PT}$
  symmetry},\ }\href {https://doi.org/10.1103/PhysRevA.85.062122} {\bibfield
  {journal} {\bibinfo  {journal} {Phys. Rev. A}\ }\textbf {\bibinfo {volume}
  {85}},\ \bibinfo {pages} {062122} (\bibinfo {year} {2012})}\BibitemShut
  {NoStop}%
\bibitem [{\citenamefont {Zheng}\ \emph {et~al.}(2013)\citenamefont {Zheng},
  \citenamefont {Hao},\ and\ \citenamefont {Long}}]{ZhengChao2013}%
  \BibitemOpen
  \bibfield  {author} {\bibinfo {author} {\bibfnamefont {C.}~\bibnamefont
  {Zheng}}, \bibinfo {author} {\bibfnamefont {L.}~\bibnamefont {Hao}},\ and\
  \bibinfo {author} {\bibfnamefont {G.~L.}\ \bibnamefont {Long}},\ }\bibfield
  {title} {\bibinfo {title} {Observation of a fast evolution in a
  parity-time-symmetric system},\ }\href
  {https://doi.org/10.1098/rsta.2012.0053} {\bibfield  {journal} {\bibinfo
  {journal} {Phil. Trans. R. Soc. A}\ }\textbf {\bibinfo {volume} {371}},\
  \bibinfo {pages} {20120053} (\bibinfo {year} {2013})}\BibitemShut {NoStop}%
\bibitem [{\citenamefont {Beygi}\ and\ \citenamefont
  {Klevansky}(2018)}]{Beygi2018}%
  \BibitemOpen
  \bibfield  {author} {\bibinfo {author} {\bibfnamefont {A.}~\bibnamefont
  {Beygi}}\ and\ \bibinfo {author} {\bibfnamefont {S.~P.}\ \bibnamefont
  {Klevansky}},\ }\bibfield  {title} {\bibinfo {title} {No-signaling principle
  and quantum brachistochrone problem in $\mathcal{PT}$-symmetric fermionic
  two- and four-dimensional models},\ }\href
  {https://doi.org/10.1103/PhysRevA.98.022105} {\bibfield  {journal} {\bibinfo
  {journal} {Phys. Rev. A}\ }\textbf {\bibinfo {volume} {98}},\ \bibinfo
  {pages} {022105} (\bibinfo {year} {2018})}\BibitemShut {NoStop}%
\bibitem [{\citenamefont {Brody}(2021)}]{Brody2021}%
  \BibitemOpen
  \bibfield  {author} {\bibinfo {author} {\bibfnamefont {D.~C.}\ \bibnamefont
  {Brody}},\ }\bibfield  {title} {\bibinfo {title} {Pt symmetry and the
  evolution speed in open quantum systems},\ }in\ \href
  {https://iopscience.iop.org/article/10.1088/1742-6596/2038/1/012005/meta}
  {\emph {\bibinfo {booktitle} {Journal of Physics: Conference Series}}},\
  Vol.\ \bibinfo {volume} {2038}\ (\bibinfo {organization} {IOP Publishing},\
  \bibinfo {year} {2021})\ p.\ \bibinfo {pages} {012005}\BibitemShut {NoStop}%
\bibitem [{\citenamefont {Bender}\ \emph {et~al.}(2013)\citenamefont {Bender},
  \citenamefont {Brody}, \citenamefont {Caldeira}, \citenamefont {Günther},
  \citenamefont {Meister},\ and\ \citenamefont {Samsonov}}]{Bender2013}%
  \BibitemOpen
  \bibfield  {author} {\bibinfo {author} {\bibfnamefont {C.~M.}\ \bibnamefont
  {Bender}}, \bibinfo {author} {\bibfnamefont {D.~C.}\ \bibnamefont {Brody}},
  \bibinfo {author} {\bibfnamefont {J.}~\bibnamefont {Caldeira}}, \bibinfo
  {author} {\bibfnamefont {U.}~\bibnamefont {Günther}}, \bibinfo {author}
  {\bibfnamefont {B.~K.}\ \bibnamefont {Meister}},\ and\ \bibinfo {author}
  {\bibfnamefont {B.~F.}\ \bibnamefont {Samsonov}},\ }\bibfield  {title}
  {\bibinfo {title} {Pt-symmetric quantum state discrimination},\ }\href
  {https://doi.org/10.1098/rsta.2012.0160} {\bibfield  {journal} {\bibinfo
  {journal} {Phil. Trans. R. Soc. A}\ }\textbf {\bibinfo {volume} {371}},\
  \bibinfo {pages} {20120160} (\bibinfo {year} {2013})}\BibitemShut {NoStop}%
\bibitem [{\citenamefont {Wang}\ \emph {et~al.}(2020)\citenamefont {Wang},
  \citenamefont {Li}, \citenamefont {Yu}, \citenamefont {Ke}, \citenamefont
  {Liu}, \citenamefont {Meng}, \citenamefont {Yang}, \citenamefont {Tang},
  \citenamefont {Li},\ and\ \citenamefont {Guo}}]{Wang2020}%
  \BibitemOpen
  \bibfield  {author} {\bibinfo {author} {\bibfnamefont {Y.-T.}\ \bibnamefont
  {Wang}}, \bibinfo {author} {\bibfnamefont {Z.-P.}\ \bibnamefont {Li}},
  \bibinfo {author} {\bibfnamefont {S.}~\bibnamefont {Yu}}, \bibinfo {author}
  {\bibfnamefont {Z.-J.}\ \bibnamefont {Ke}}, \bibinfo {author} {\bibfnamefont
  {W.}~\bibnamefont {Liu}}, \bibinfo {author} {\bibfnamefont {Y.}~\bibnamefont
  {Meng}}, \bibinfo {author} {\bibfnamefont {Y.-Z.}\ \bibnamefont {Yang}},
  \bibinfo {author} {\bibfnamefont {J.-S.}\ \bibnamefont {Tang}}, \bibinfo
  {author} {\bibfnamefont {C.-F.}\ \bibnamefont {Li}},\ and\ \bibinfo {author}
  {\bibfnamefont {G.-C.}\ \bibnamefont {Guo}},\ }\bibfield  {title} {\bibinfo
  {title} {Experimental investigation of state distinguishability in
  parity-time symmetric quantum dynamics},\ }\href
  {https://doi.org/10.1103/PhysRevLett.124.230402} {\bibfield  {journal}
  {\bibinfo  {journal} {Phys. Rev. Lett.}\ }\textbf {\bibinfo {volume} {124}},\
  \bibinfo {pages} {230402} (\bibinfo {year} {2020})}\BibitemShut {NoStop}%
\bibitem [{\citenamefont {Barnett}\ and\ \citenamefont
  {Andersson}(2002)}]{Barnett2002}%
  \BibitemOpen
  \bibfield  {author} {\bibinfo {author} {\bibfnamefont {S.~M.}\ \bibnamefont
  {Barnett}}\ and\ \bibinfo {author} {\bibfnamefont {E.}~\bibnamefont
  {Andersson}},\ }\bibfield  {title} {\bibinfo {title} {Bound on measurement
  based on the no-signaling condition},\ }\href
  {https://doi.org/10.1103/PhysRevA.65.044307} {\bibfield  {journal} {\bibinfo
  {journal} {Phys. Rev. A}\ }\textbf {\bibinfo {volume} {65}},\ \bibinfo
  {pages} {044307} (\bibinfo {year} {2002})}\BibitemShut {NoStop}%
\bibitem [{\citenamefont {Lee}\ \emph {et~al.}(2014)\citenamefont {Lee},
  \citenamefont {Hsieh}, \citenamefont {Flammia},\ and\ \citenamefont
  {Lee}}]{Lee2014}%
  \BibitemOpen
  \bibfield  {author} {\bibinfo {author} {\bibfnamefont {Y.-C.}\ \bibnamefont
  {Lee}}, \bibinfo {author} {\bibfnamefont {M.-H.}\ \bibnamefont {Hsieh}},
  \bibinfo {author} {\bibfnamefont {S.~T.}\ \bibnamefont {Flammia}},\ and\
  \bibinfo {author} {\bibfnamefont {R.-K.}\ \bibnamefont {Lee}},\ }\bibfield
  {title} {\bibinfo {title} {Local $\mathcal{P}\mathcal{T}$ symmetry violates
  the no-signaling principle},\ }\href
  {https://doi.org/10.1103/PhysRevLett.112.130404} {\bibfield  {journal}
  {\bibinfo  {journal} {Phys. Rev. Lett.}\ }\textbf {\bibinfo {volume} {112}},\
  \bibinfo {pages} {130404} (\bibinfo {year} {2014})}\BibitemShut {NoStop}%
\bibitem [{\citenamefont {Brody}(2016)}]{Brody2016}%
  \BibitemOpen
  \bibfield  {author} {\bibinfo {author} {\bibfnamefont {D.~C.}\ \bibnamefont
  {Brody}},\ }\bibfield  {title} {\bibinfo {title} {Consistency of
  {PT}-symmetric quantum mechanics},\ }\href
  {https://doi.org/10.1088/1751-8113/49/10/10lt03} {\bibfield  {journal}
  {\bibinfo  {journal} {J. Phys. A: Math. Theor.}\ }\textbf {\bibinfo {volume}
  {49}},\ \bibinfo {pages} {10LT03} (\bibinfo {year} {2016})}\BibitemShut
  {NoStop}%
\bibitem [{\citenamefont {Tang}\ \emph {et~al.}(2016)\citenamefont {Tang},
  \citenamefont {Wang}, \citenamefont {Yu}, \citenamefont {He}, \citenamefont
  {Xu}, \citenamefont {Liu}, \citenamefont {Chen}, \citenamefont {Sun},
  \citenamefont {Sun}, \citenamefont {Han} \emph {et~al.}}]{Tang2016}%
  \BibitemOpen
  \bibfield  {author} {\bibinfo {author} {\bibfnamefont {J.-S.}\ \bibnamefont
  {Tang}}, \bibinfo {author} {\bibfnamefont {Y.-T.}\ \bibnamefont {Wang}},
  \bibinfo {author} {\bibfnamefont {S.}~\bibnamefont {Yu}}, \bibinfo {author}
  {\bibfnamefont {D.-Y.}\ \bibnamefont {He}}, \bibinfo {author} {\bibfnamefont
  {J.-S.}\ \bibnamefont {Xu}}, \bibinfo {author} {\bibfnamefont {B.-H.}\
  \bibnamefont {Liu}}, \bibinfo {author} {\bibfnamefont {G.}~\bibnamefont
  {Chen}}, \bibinfo {author} {\bibfnamefont {Y.-N.}\ \bibnamefont {Sun}},
  \bibinfo {author} {\bibfnamefont {K.}~\bibnamefont {Sun}}, \bibinfo {author}
  {\bibfnamefont {Y.-J.}\ \bibnamefont {Han}}, \emph {et~al.},\ }\bibfield
  {title} {\bibinfo {title} {Experimental investigation of the no-signalling
  principle in parity--time symmetric theory using an open quantum system},\
  }\href {https://doi.org/10.1038/nphoton.2016.144} {\bibfield  {journal}
  {\bibinfo  {journal} {Nat. Photonics}\ }\textbf {\bibinfo {volume} {10}},\
  \bibinfo {pages} {642} (\bibinfo {year} {2016})}\BibitemShut {NoStop}%
\bibitem [{\citenamefont {Feng}\ \emph {et~al.}(2017)\citenamefont {Feng},
  \citenamefont {El-Ganainy},\ and\ \citenamefont {Ge}}]{feng2017non}%
  \BibitemOpen
  \bibfield  {author} {\bibinfo {author} {\bibfnamefont {L.}~\bibnamefont
  {Feng}}, \bibinfo {author} {\bibfnamefont {R.}~\bibnamefont {El-Ganainy}},\
  and\ \bibinfo {author} {\bibfnamefont {L.}~\bibnamefont {Ge}},\ }\bibfield
  {title} {\bibinfo {title} {Non-hermitian photonics based on parity--time
  symmetry},\ }\href
  {https://www.nature.com/articles/s41566-017-0031-1/briefing/signup/}
  {\bibfield  {journal} {\bibinfo  {journal} {Nat. Photonics}\ }\textbf
  {\bibinfo {volume} {11}},\ \bibinfo {pages} {752} (\bibinfo {year}
  {2017})}\BibitemShut {NoStop}%
\bibitem [{\citenamefont {Bagchi}\ and\ \citenamefont
  {Barik}(2020)}]{Bagchi2020}%
  \BibitemOpen
  \bibfield  {author} {\bibinfo {author} {\bibfnamefont {B.}~\bibnamefont
  {Bagchi}}\ and\ \bibinfo {author} {\bibfnamefont {S.}~\bibnamefont {Barik}},\
  }\bibfield  {title} {\bibinfo {title} {Remarks on the preservation of
  no-signaling principle in parity-time-symmetric quantum mechanics},\ }\href
  {https://doi.org/10.1142/S021773232050090X} {\bibfield  {journal} {\bibinfo
  {journal} {Mod. Phys. Lett. A}\ }\textbf {\bibinfo {volume} {35}},\ \bibinfo
  {pages} {2050090} (\bibinfo {year} {2020})}\BibitemShut {NoStop}%
\bibitem [{\citenamefont {Huang}\ \emph {et~al.}(2018)\citenamefont {Huang},
  \citenamefont {Kumar},\ and\ \citenamefont {Wu}}]{Huang2018}%
  \BibitemOpen
  \bibfield  {author} {\bibinfo {author} {\bibfnamefont {M.}~\bibnamefont
  {Huang}}, \bibinfo {author} {\bibfnamefont {A.}~\bibnamefont {Kumar}},\ and\
  \bibinfo {author} {\bibfnamefont {J.}~\bibnamefont {Wu}},\ }\bibfield
  {title} {\bibinfo {title} {Embedding, simulation and consistency of
  pt-symmetric quantum theory},\ }\href
  {https://doi.org/https://doi.org/10.1016/j.physleta.2018.06.047} {\bibfield
  {journal} {\bibinfo  {journal} {Phys. Lett. A}\ }\textbf {\bibinfo {volume}
  {382}},\ \bibinfo {pages} {2578} (\bibinfo {year} {2018})}\BibitemShut
  {NoStop}%
\bibitem [{\citenamefont {Wu}\ \emph {et~al.}(2019)\citenamefont {Wu},
  \citenamefont {Liu}, \citenamefont {Geng}, \citenamefont {Song},
  \citenamefont {Ye}, \citenamefont {Duan}, \citenamefont {Rong},\ and\
  \citenamefont {Du}}]{wu2019observation}%
  \BibitemOpen
  \bibfield  {author} {\bibinfo {author} {\bibfnamefont {Y.}~\bibnamefont
  {Wu}}, \bibinfo {author} {\bibfnamefont {W.}~\bibnamefont {Liu}}, \bibinfo
  {author} {\bibfnamefont {J.}~\bibnamefont {Geng}}, \bibinfo {author}
  {\bibfnamefont {X.}~\bibnamefont {Song}}, \bibinfo {author} {\bibfnamefont
  {X.}~\bibnamefont {Ye}}, \bibinfo {author} {\bibfnamefont {C.-K.}\
  \bibnamefont {Duan}}, \bibinfo {author} {\bibfnamefont {X.}~\bibnamefont
  {Rong}},\ and\ \bibinfo {author} {\bibfnamefont {J.}~\bibnamefont {Du}},\
  }\bibfield  {title} {\bibinfo {title} {Observation of parity-time symmetry
  breaking in a single-spin system},\ }\href
  {https://www.science.org/doi/abs/10.1126/science.aaw8205} {\bibfield
  {journal} {\bibinfo  {journal} {Science}\ }\textbf {\bibinfo {volume}
  {364}},\ \bibinfo {pages} {878} (\bibinfo {year} {2019})}\BibitemShut
  {NoStop}%
\bibitem [{\citenamefont {Gui-Lu}(2006)}]{GuiLu2006}%
  \BibitemOpen
  \bibfield  {author} {\bibinfo {author} {\bibfnamefont {L.}~\bibnamefont
  {Gui-Lu}},\ }\bibfield  {title} {\bibinfo {title} {General quantum
  interference principle and duality computer},\ }\href
  {https://doi.org/10.1088/0253-6102/45/5/013} {\bibfield  {journal} {\bibinfo
  {journal} {Commun. Theor. Phys.}\ }\textbf {\bibinfo {volume} {45}},\
  \bibinfo {pages} {825} (\bibinfo {year} {2006})}\BibitemShut {NoStop}%
\bibitem [{\citenamefont {Zheng}(2018)}]{Zheng2018}%
  \BibitemOpen
  \bibfield  {author} {\bibinfo {author} {\bibfnamefont {C.}~\bibnamefont
  {Zheng}},\ }\bibfield  {title} {\bibinfo {title} {Duality quantum simulation
  of a general parity-time-symmetric two-level system},\ }\href
  {https://doi.org/10.1209/0295-5075/123/40002} {\bibfield  {journal} {\bibinfo
   {journal} {{EPL} (Europhysics Letters)}\ }\textbf {\bibinfo {volume}
  {123}},\ \bibinfo {pages} {40002} (\bibinfo {year} {2018})}\BibitemShut
  {NoStop}%
\bibitem [{\citenamefont {Gao}\ \emph {et~al.}(2021)\citenamefont {Gao},
  \citenamefont {Zheng}, \citenamefont {Liu}, \citenamefont {Wang},\ and\
  \citenamefont {Wang}}]{Gao2021}%
  \BibitemOpen
  \bibfield  {author} {\bibinfo {author} {\bibfnamefont {W.-C.}\ \bibnamefont
  {Gao}}, \bibinfo {author} {\bibfnamefont {C.}~\bibnamefont {Zheng}}, \bibinfo
  {author} {\bibfnamefont {L.}~\bibnamefont {Liu}}, \bibinfo {author}
  {\bibfnamefont {T.-J.}\ \bibnamefont {Wang}},\ and\ \bibinfo {author}
  {\bibfnamefont {C.}~\bibnamefont {Wang}},\ }\bibfield  {title} {\bibinfo
  {title} {Experimental simulation of the parity-time symmetric dynamics using
  photonic qubits},\ }\href {https://doi.org/10.1364/OE.405815} {\bibfield
  {journal} {\bibinfo  {journal} {Opt. Express}\ }\textbf {\bibinfo {volume}
  {29}},\ \bibinfo {pages} {517} (\bibinfo {year} {2021})}\BibitemShut
  {NoStop}%
\bibitem [{\citenamefont {Zheng}(2019)}]{Zheng2019a}%
  \BibitemOpen
  \bibfield  {author} {\bibinfo {author} {\bibfnamefont {C.}~\bibnamefont
  {Zheng}},\ }\bibfield  {title} {\bibinfo {title} {Duality quantum simulation
  of a generalized anti-{PT}-symmetric two-level system},\ }\href
  {https://doi.org/10.1209/0295-5075/126/30005} {\bibfield  {journal} {\bibinfo
   {journal} {Europhys. Lett.}\ }\textbf {\bibinfo {volume} {126}},\ \bibinfo
  {pages} {30005} (\bibinfo {year} {2019})}\BibitemShut {NoStop}%
\bibitem [{\citenamefont {Huang}\ \emph {et~al.}(2019)\citenamefont {Huang},
  \citenamefont {Lee}, \citenamefont {Zhang}, \citenamefont {Fei},\ and\
  \citenamefont {Wu}}]{Huang2019}%
  \BibitemOpen
  \bibfield  {author} {\bibinfo {author} {\bibfnamefont {M.}~\bibnamefont
  {Huang}}, \bibinfo {author} {\bibfnamefont {R.-K.}\ \bibnamefont {Lee}},
  \bibinfo {author} {\bibfnamefont {L.}~\bibnamefont {Zhang}}, \bibinfo
  {author} {\bibfnamefont {S.-M.}\ \bibnamefont {Fei}},\ and\ \bibinfo {author}
  {\bibfnamefont {J.}~\bibnamefont {Wu}},\ }\bibfield  {title} {\bibinfo
  {title} {Simulating broken $\mathcal{PT}$-symmetric hamiltonian systems by
  weak measurement},\ }\href {https://doi.org/10.1103/PhysRevLett.123.080404}
  {\bibfield  {journal} {\bibinfo  {journal} {Phys. Rev. Lett.}\ }\textbf
  {\bibinfo {volume} {123}},\ \bibinfo {pages} {080404} (\bibinfo {year}
  {2019})}\BibitemShut {NoStop}%
\bibitem [{\citenamefont {Li}\ \emph {et~al.}(2022)\citenamefont {Li},
  \citenamefont {Zheng}, \citenamefont {Gao},\ and\ \citenamefont
  {Long}}]{Li2022}%
  \BibitemOpen
  \bibfield  {author} {\bibinfo {author} {\bibfnamefont {X.}~\bibnamefont
  {Li}}, \bibinfo {author} {\bibfnamefont {C.}~\bibnamefont {Zheng}}, \bibinfo
  {author} {\bibfnamefont {J.}~\bibnamefont {Gao}},\ and\ \bibinfo {author}
  {\bibfnamefont {G.}~\bibnamefont {Long}},\ }\bibfield  {title} {\bibinfo
  {title} {Efficient simulation of the dynamics of an $n$-dimensional
  $\mathcal{PT}$-symmetric system with a
  local-operations-and-classical-communication protocol based on an embedding
  scheme},\ }\href {https://doi.org/10.1103/PhysRevA.105.032405} {\bibfield
  {journal} {\bibinfo  {journal} {Phys. Rev. A}\ }\textbf {\bibinfo {volume}
  {105}},\ \bibinfo {pages} {032405} (\bibinfo {year} {2022})}\BibitemShut
  {NoStop}%
\bibitem [{\citenamefont {Nielsen}\ and\ \citenamefont
  {Chuang}(2002)}]{nielsen2002quantum}%
  \BibitemOpen
  \bibfield  {author} {\bibinfo {author} {\bibfnamefont {M.~A.}\ \bibnamefont
  {Nielsen}}\ and\ \bibinfo {author} {\bibfnamefont {I.}~\bibnamefont
  {Chuang}},\ }\href@noop {} {\bibinfo {title} {Quantum computation and quantum
  information}} (\bibinfo {year} {2002})\BibitemShut {NoStop}%
\bibitem [{\citenamefont {Minganti}\ \emph {et~al.}(2019)\citenamefont
  {Minganti}, \citenamefont {Miranowicz}, \citenamefont {Chhajlany},\ and\
  \citenamefont {Nori}}]{Minganti2019}%
  \BibitemOpen
  \bibfield  {author} {\bibinfo {author} {\bibfnamefont {F.}~\bibnamefont
  {Minganti}}, \bibinfo {author} {\bibfnamefont {A.}~\bibnamefont
  {Miranowicz}}, \bibinfo {author} {\bibfnamefont {R.~W.}\ \bibnamefont
  {Chhajlany}},\ and\ \bibinfo {author} {\bibfnamefont {F.}~\bibnamefont
  {Nori}},\ }\bibfield  {title} {\bibinfo {title} {Quantum exceptional points
  of non-hermitian hamiltonians and liouvillians: The effects of quantum
  jumps},\ }\href {https://doi.org/10.1103/PhysRevA.100.062131} {\bibfield
  {journal} {\bibinfo  {journal} {Phys. Rev. A}\ }\textbf {\bibinfo {volume}
  {100}},\ \bibinfo {pages} {062131} (\bibinfo {year} {2019})}\BibitemShut
  {NoStop}%
\bibitem [{\citenamefont {Ohlsson}\ and\ \citenamefont
  {Zhou}(2021)}]{Ohlsson2021}%
  \BibitemOpen
  \bibfield  {author} {\bibinfo {author} {\bibfnamefont {T.}~\bibnamefont
  {Ohlsson}}\ and\ \bibinfo {author} {\bibfnamefont {S.}~\bibnamefont {Zhou}},\
  }\bibfield  {title} {\bibinfo {title} {Density-matrix formalism for
  $\mathcal{PT}$-symmetric non-hermitian hamiltonians with the lindblad
  equation},\ }\href {https://doi.org/10.1103/PhysRevA.103.022218} {\bibfield
  {journal} {\bibinfo  {journal} {Phys. Rev. A}\ }\textbf {\bibinfo {volume}
  {103}},\ \bibinfo {pages} {022218} (\bibinfo {year} {2021})}\BibitemShut
  {NoStop}%
\bibitem [{\citenamefont {Magnus}(1954)}]{Magnus1954}%
  \BibitemOpen
  \bibfield  {author} {\bibinfo {author} {\bibfnamefont {W.}~\bibnamefont
  {Magnus}},\ }\bibfield  {title} {\bibinfo {title} {On the exponential
  solution of differential equations for a linear operator},\ }\href
  {https://doi.org/https://doi.org/10.1002/cpa.3160070404} {\bibfield
  {journal} {\bibinfo  {journal} {Commun. Pure Appl. Math.}\ }\textbf {\bibinfo
  {volume} {7}},\ \bibinfo {pages} {649} (\bibinfo {year} {1954})}\BibitemShut
  {NoStop}%
\bibitem [{\citenamefont {Blanes}\ \emph {et~al.}(1998)\citenamefont {Blanes},
  \citenamefont {Casas}, \citenamefont {Oteo},\ and\ \citenamefont
  {Ros}}]{Blanes1998}%
  \BibitemOpen
  \bibfield  {author} {\bibinfo {author} {\bibfnamefont {S.}~\bibnamefont
  {Blanes}}, \bibinfo {author} {\bibfnamefont {F.}~\bibnamefont {Casas}},
  \bibinfo {author} {\bibfnamefont {J.}~\bibnamefont {Oteo}},\ and\ \bibinfo
  {author} {\bibfnamefont {J.}~\bibnamefont {Ros}},\ }\bibfield  {title}
  {\bibinfo {title} {Magnus and fer expansions for matrix differential
  equations: the convergence problem},\ }\href
  {https://iopscience.iop.org/article/10.1088/0305-4470/31/1/023/meta}
  {\bibfield  {journal} {\bibinfo  {journal} {J. Phys. A: Math. Gen.}\ }\textbf
  {\bibinfo {volume} {31}},\ \bibinfo {pages} {259} (\bibinfo {year}
  {1998})}\BibitemShut {NoStop}%
\bibitem [{\citenamefont {Blanes}\ \emph {et~al.}(2009)\citenamefont {Blanes},
  \citenamefont {Casas}, \citenamefont {Oteo},\ and\ \citenamefont
  {Ros}}]{Blanes2009}%
  \BibitemOpen
  \bibfield  {author} {\bibinfo {author} {\bibfnamefont {S.}~\bibnamefont
  {Blanes}}, \bibinfo {author} {\bibfnamefont {F.}~\bibnamefont {Casas}},
  \bibinfo {author} {\bibfnamefont {J.-A.}\ \bibnamefont {Oteo}},\ and\
  \bibinfo {author} {\bibfnamefont {J.}~\bibnamefont {Ros}},\ }\bibfield
  {title} {\bibinfo {title} {The magnus expansion and some of its
  applications},\ }\href
  {https://www.sciencedirect.com/science/article/pii/S0370157308004092}
  {\bibfield  {journal} {\bibinfo  {journal} {Physics reports}\ }\textbf
  {\bibinfo {volume} {470}},\ \bibinfo {pages} {151} (\bibinfo {year}
  {2009})}\BibitemShut {NoStop}%
\bibitem [{\citenamefont {Huang}\ \emph {et~al.}(2021)\citenamefont {Huang},
  \citenamefont {Lee}, \citenamefont {Zhang},\ and\ \citenamefont
  {Wu}}]{huang2021solvable}%
  \BibitemOpen
  \bibfield  {author} {\bibinfo {author} {\bibfnamefont {M.}~\bibnamefont
  {Huang}}, \bibinfo {author} {\bibfnamefont {R.-K.}\ \bibnamefont {Lee}},
  \bibinfo {author} {\bibfnamefont {G.-Q.}\ \bibnamefont {Zhang}},\ and\
  \bibinfo {author} {\bibfnamefont {J.}~\bibnamefont {Wu}},\ }\bibfield
  {title} {\bibinfo {title} {A solvable dilation model of pt-symmetric
  systems},\ }\href {https://arxiv.org/abs/2104.05039} {\bibfield  {journal}
  {\bibinfo  {journal} {arXiv:2104.05039 [quant-ph]}\ } (\bibinfo {year}
  {2021})}\BibitemShut {NoStop}%
\bibitem [{\citenamefont {Zhang}\ \emph
  {et~al.}(2019{\natexlab{b}})\citenamefont {Zhang}, \citenamefont {Wang},\
  and\ \citenamefont {Gong}}]{Zhang2019}%
  \BibitemOpen
  \bibfield  {author} {\bibinfo {author} {\bibfnamefont {D.-J.}\ \bibnamefont
  {Zhang}}, \bibinfo {author} {\bibfnamefont {Q.-h.}\ \bibnamefont {Wang}},\
  and\ \bibinfo {author} {\bibfnamefont {J.}~\bibnamefont {Gong}},\ }\bibfield
  {title} {\bibinfo {title} {Time-dependent $\mathcal{PT}$-symmetric quantum
  mechanics in generic non-hermitian systems},\ }\href
  {https://doi.org/10.1103/PhysRevA.100.062121} {\bibfield  {journal} {\bibinfo
   {journal} {Phys. Rev. A}\ }\textbf {\bibinfo {volume} {100}},\ \bibinfo
  {pages} {062121} (\bibinfo {year} {2019}{\natexlab{b}})}\BibitemShut
  {NoStop}%
\bibitem [{\citenamefont {Fring}\ and\ \citenamefont
  {Moussa}(2016)}]{Fring2016}%
  \BibitemOpen
  \bibfield  {author} {\bibinfo {author} {\bibfnamefont {A.}~\bibnamefont
  {Fring}}\ and\ \bibinfo {author} {\bibfnamefont {M.~H.~Y.}\ \bibnamefont
  {Moussa}},\ }\bibfield  {title} {\bibinfo {title} {Unitary quantum evolution
  for time-dependent quasi-hermitian systems with nonobservable hamiltonians},\
  }\href {https://doi.org/10.1103/PhysRevA.93.042114} {\bibfield  {journal}
  {\bibinfo  {journal} {Phys. Rev. A}\ }\textbf {\bibinfo {volume} {93}},\
  \bibinfo {pages} {042114} (\bibinfo {year} {2016})}\BibitemShut {NoStop}%
\bibitem [{\citenamefont {Luiz}\ \emph {et~al.}(2020)\citenamefont {Luiz},
  \citenamefont {de~Ponte},\ and\ \citenamefont {Moussa}}]{Luiz2020}%
  \BibitemOpen
  \bibfield  {author} {\bibinfo {author} {\bibfnamefont {F.~S.}\ \bibnamefont
  {Luiz}}, \bibinfo {author} {\bibfnamefont {M.~A.}\ \bibnamefont {de~Ponte}},\
  and\ \bibinfo {author} {\bibfnamefont {M.~H.~Y.}\ \bibnamefont {Moussa}},\
  }\bibfield  {title} {\bibinfo {title} {Unitarity of the time-evolution and
  observability of non-hermitian hamiltonians for time-dependent dyson maps},\
  }\href {https://doi.org/10.1088/1402-4896/ab80e5} {\bibfield  {journal}
  {\bibinfo  {journal} {Physica Scripta}\ }\textbf {\bibinfo {volume} {95}},\
  \bibinfo {pages} {065211} (\bibinfo {year} {2020})}\BibitemShut {NoStop}%
\bibitem [{\citenamefont {Brody}\ and\ \citenamefont
  {Graefe}(2012)}]{Brody2012}%
  \BibitemOpen
  \bibfield  {author} {\bibinfo {author} {\bibfnamefont {D.~C.}\ \bibnamefont
  {Brody}}\ and\ \bibinfo {author} {\bibfnamefont {E.-M.}\ \bibnamefont
  {Graefe}},\ }\bibfield  {title} {\bibinfo {title} {Mixed-state evolution in
  the presence of gain and loss},\ }\href
  {https://doi.org/10.1103/PhysRevLett.109.230405} {\bibfield  {journal}
  {\bibinfo  {journal} {Phys. Rev. Lett.}\ }\textbf {\bibinfo {volume} {109}},\
  \bibinfo {pages} {230405} (\bibinfo {year} {2012})}\BibitemShut {NoStop}%
\bibitem [{\citenamefont {Xiao}\ \emph {et~al.}(2019)\citenamefont {Xiao},
  \citenamefont {Wang}, \citenamefont {Zhan}, \citenamefont {Bian},
  \citenamefont {Kawabata}, \citenamefont {Ueda}, \citenamefont {Yi},\ and\
  \citenamefont {Xue}}]{Xiao2019}%
  \BibitemOpen
  \bibfield  {author} {\bibinfo {author} {\bibfnamefont {L.}~\bibnamefont
  {Xiao}}, \bibinfo {author} {\bibfnamefont {K.}~\bibnamefont {Wang}}, \bibinfo
  {author} {\bibfnamefont {X.}~\bibnamefont {Zhan}}, \bibinfo {author}
  {\bibfnamefont {Z.}~\bibnamefont {Bian}}, \bibinfo {author} {\bibfnamefont
  {K.}~\bibnamefont {Kawabata}}, \bibinfo {author} {\bibfnamefont
  {M.}~\bibnamefont {Ueda}}, \bibinfo {author} {\bibfnamefont {W.}~\bibnamefont
  {Yi}},\ and\ \bibinfo {author} {\bibfnamefont {P.}~\bibnamefont {Xue}},\
  }\bibfield  {title} {\bibinfo {title} {Observation of critical phenomena in
  parity-time-symmetric quantum dynamics},\ }\href
  {https://doi.org/10.1103/PhysRevLett.123.230401} {\bibfield  {journal}
  {\bibinfo  {journal} {Phys. Rev. Lett.}\ }\textbf {\bibinfo {volume} {123}},\
  \bibinfo {pages} {230401} (\bibinfo {year} {2019})}\BibitemShut {NoStop}%
\bibitem [{\citenamefont {Bender}\ \emph {et~al.}(2003)\citenamefont {Bender},
  \citenamefont {Brody},\ and\ \citenamefont {Jones}}]{Bender2003}%
  \BibitemOpen
  \bibfield  {author} {\bibinfo {author} {\bibfnamefont {C.~M.}\ \bibnamefont
  {Bender}}, \bibinfo {author} {\bibfnamefont {D.~C.}\ \bibnamefont {Brody}},\
  and\ \bibinfo {author} {\bibfnamefont {H.~F.}\ \bibnamefont {Jones}},\
  }\bibfield  {title} {\bibinfo {title} {Must a hamiltonian be hermitian?},\
  }\href {https://doi.org/10.1119/1.1574043} {\bibfield  {journal} {\bibinfo
  {journal} {Am. J. Phys.}\ }\textbf {\bibinfo {volume} {71}},\ \bibinfo
  {pages} {1095} (\bibinfo {year} {2003})}\BibitemShut {NoStop}%
\bibitem [{\citenamefont {Andersson}\ \emph {et~al.}(2007)\citenamefont
  {Andersson}, \citenamefont {Cresser},\ and\ \citenamefont
  {Hall}}]{andersson2007finding}%
  \BibitemOpen
  \bibfield  {author} {\bibinfo {author} {\bibfnamefont {E.}~\bibnamefont
  {Andersson}}, \bibinfo {author} {\bibfnamefont {J.~D.}\ \bibnamefont
  {Cresser}},\ and\ \bibinfo {author} {\bibfnamefont {M.~J.}\ \bibnamefont
  {Hall}},\ }\bibfield  {title} {\bibinfo {title} {Finding the kraus
  decomposition from a master equation and vice versa},\ }\href
  {https://www.tandfonline.com/doi/abs/10.1080/09500340701352581} {\bibfield
  {journal} {\bibinfo  {journal} {J. Mod. Opt.}\ }\textbf {\bibinfo {volume}
  {54}},\ \bibinfo {pages} {1695} (\bibinfo {year} {2007})}\BibitemShut
  {NoStop}%
\end{thebibliography}%
\end{document}